\documentclass[11pt]{article}
\pdfoutput=1
\usepackage{jheppub}
\usepackage{amsmath}
\usepackage{bm}
\usepackage{slashed}
\usepackage{graphicx}

\newcommand{\tr}{\mathrm{tr}\,}

\newlength{\dummysp}
\newcommand{\beq}{\begin{eqnarray}}
\newcommand{\eeq}{\end{eqnarray}}

\newcommand{\gappeq}{\mathrel{\rlap {\raise.5ex\hbox{$>$}}
{\lower.5ex\hbox{$\sim$}}}}
\newcommand{\lappeq}{\mathrel{\rlap{\raise.5ex\hbox{$<$}}
{\lower.5ex\hbox{$\sim$}}}}

\newcommand{\ben}{\begin{enumerate}}
\newcommand{\een}{\end{enumerate}}

\newcommand{\bit}{\begin{itemize}}
\newcommand{\eit}{\end{itemize}}

\def\[{\left [}
\def\]{\right ]}
\def\({\left (}
\def\){\right )}
\def\R{{\mathbb R}}
\def\S{{\mathbb S}}
\def\Z{{\mathbb Z}}

\title{On the Global Structure of Deformed Yang-Mills Theory and QCD(adj) on $\mathbf{\R^3 \times\S^1}$\\
 }
   
\author[a]{Mohamed M. Anber,}\author[b]{Erich Poppitz} 
  
\affiliation[a]{Institute de Th\' eorie des Ph\' enomen\` es Physiques, \' Ecole Polytechnique F\' ed\' erale de Lausanne, CH-1015 Lausanne, Switzerland}
\affiliation[b]{Department of Physics,   University of Toronto, 
Toronto, ON M5S 1A7, Canada}
\emailAdd{mohamed.anber@epfl.ch}\emailAdd{poppitz@physics.utoronto.ca}    
\abstract{ 
Spatial compactification  on $\R^{3} \times \S^1_L$  at small $\S^1$-size $L$ often leads to a calculable vacuum structure, where various ``topological molecules" are responsible for confinement and the realization of the center and discrete chiral symmetries. Within this semiclassically calculable framework, we study how distinct theories  with the same $SU(N_c)/\Z_k$ gauge group (labeled by ``discrete $\theta$-angles")  arise upon  gauging  of appropriate $\Z_k$ subgroups of the one-form global  center symmetry of an $SU(N_c)$ gauge theory. We determine the possible $\Z_k$ actions on the local electric and magnetic effective degrees of freedom, find  the ground states, and  use domain walls and confining strings  to give a physical picture of the  vacuum structure of the different  $SU(N_c)/\Z_k$ theories.    Some of our results reproduce ones from earlier supersymmetric studies, but most are new and do not invoke supersymmetry.  We also study  a further finite-temperature  compactification to $\R^{2}\times\S^1_\beta\times\S^1_L$. We argue that,  in deformed Yang-Mills theory,  the effective theory near the deconfinement temperature  $\beta_c \gg L$ exhibits an emergent  Kramers-Wannier duality and that it exchanges  high- and low-temperature theories with  different global structure, sharing features with both the Ising model and $S$-duality in ${\cal N}$$=$$4$ supersymmetric Yang-Mills theory. }

\begin{document}

\maketitle

\flushbottom

\section{Introduction}

Gauge theories are usually formulated in terms of their Lie algebra, which determines the interactions and Lagrangian. While it is well known that there are different Lie groups with the same algebra, e.g.~$SU(2)$ vs.~$SO(3)$, usually one goes without specifying the choice of gauge group. This is because the local dynamics of the theory is insensitive to the  global structure. However, it is also  known that dualities can interchange theories with the same algebra but different gauge groups. The most notable example is the electric-magnetic duality of ${\cal N}$$=$$4$ supersymmetric Yang-Mills (SYM) theory (whose origin is in \cite{Goddard:1976qe}; see \cite{Kapustin:2006pk} for a complete list of references). Lattice  gauge theories for different choice of gauge group with the same algebra have also been the  studied, see~e.g.~\cite{deForcrand:2002vs}  and references therein.

Interestingly, it was only recently realized   that even when the gauge group is chosen, there is a further set of discrete parameters, called ``discrete $\theta$-angles" in \cite{Aharony:2013hda}, that label different theories with the same gauge group (we refer to the choice of gauge group and discrete $\theta$-angle parameters as ``global structure"). One way\footnote{We note that while the terminology in   the recent works sometimes differs from that in the lattice literature, there is a relation between the electric and magnetic flux (or ``twist")  sectors of 't Hooft \cite{'tHooft:1979uj}  and the discrete $\theta$-parameters, explained in \cite{Aharony:2013hda,Gaiotto:2014kfa} (see also 
\cite{Witten:2000nv,Kapustin:2014gua,Tachikawa:2014mna,Amariti:2015dxa}).}
 to describe the meaning of these discrete parameters is that they label the different choices of sets of mutually-local line (Wilson and 't Hooft) operators for a given choice of gauge group, while the sets corresponding to different discrete $\theta$ angles are  not mutually local with respect to each other. Since  Wilson and 't Hooft line operators  characterize the phases of gauge theories,  a physical picture of their behavior in   theories with different  global structure was given in \cite{Aharony:2013hda} using confinement in  softly broken Seiberg-Witten theory  as an example. The action of $S$-duality in  ${\cal N}$$=$$4$ SYM was also refined to include the new discrete parameters, leading to an intricate  consistent web of dualities  \cite{Aharony:2013hda}.

In this paper, we study the behavior of  theories with different global structure in a setting where the nonperturbative dynamics of the theory is understood in an analytically controlled way. Our aim is to provide a physical picture of their ground states  using the understood  confining dynamics, in a more general set   of  theories (not necessarily supersymmetric). We study two classes of theories, deformed Yang-Mills theory (dYM) and Yang-Mills theory with adjoint fermions (QCD(adj)), compactified on a spatial circle,  $\R^3 \times \S_L^1$, with periodic boundary conditions for the fermions,   whose study began in \cite{Shifman:2008ja,Unsal:2008ch,Unsal:2007jx}. We focus on theories with $su(N_c)$ Lie algebra  in the $\Lambda_{QCD} L N_c$$\ll$$1$ semiclassically calculable regime, where $\Lambda_{QCD}$ is the strong coupling scale.
In addition to ensuring semiclassical calculability, compactification  makes the  different global structures both more straightforward to study and more dramatic in their effect. This is because the line operators that distinguish the various theories can now wrap around  $\S^1_L$, becoming local operators in the long distance theory \cite{Aharony:2013hda,Aharony:2013kma}. Thus   theories with different global structure on $\R^3 \times \S_L^1$  can have different vacuum structure, labeled by the expectation value of these wrapped line operators. 

The first  original contribution of this paper is to systematically study the global structure in the calculable regime on $\R^3 \times \S^1_L$ in dYM and QCD(adj). We  determine the vacuum structure in theories with different global structure and give it a physical interpretation  using the interplay between domain walls and confining strings on $\R^3 \times \S_L^1$ recently discussed in \cite{Anber:2015kea}.
 The main technical tool we work out   is  the action of the zero-form part of the (to-be-gauged) center symmetry on the local electric and magnetic degrees of freedom in the effective  theory on $\R^3 \times \S^1_L$. We use it to study the vacuum structure and to explicitly construct the mutually local gauge invariant operators in each theory.
 
The second contribution of this paper is an observation regarding the role of the global structure upon  further thermal compactification on $\R^2 \times \S^1_\beta \times \S^1_L$. Previous work found that in the low-temperature
 $\beta$$\gg$$L$ regime, still at  $\Lambda_{\mbox{\scriptsize QCD}} L N_c$$\ll$$1$, there is a thermal deconfinement transition, both in  dYM \cite{Simic:2010sv} and QCD(adj) \cite{Anber:2011gn}. The effective theory near the transition is  a two-dimensional Coulomb gas of electrically and magnetically charged particles. For dYM, this Coulomb gas exhibits an emergent Kramers-Wannier (high-$T$/low-$T$) duality which simultaneously interchanges electric and magnetic charges.\footnote{In fact, puzzles related to the global structure in the thermal case were part of the original motivation for this study.}  We argue that this duality exchanges theories with different global structure and shares common features with  both the Kramers-Wannier duality in the Ising model, recently pointed out in \cite{Kapustin:2014gua},  and $S$-duality in $N=4$ SYM \cite{Aharony:2013hda}. To the best of our knowledge, the Kramers-Wannier duality of the effective theory is the only example of an  electric-magnetic duality in the framework of nonsupersymmetric pure YM theory.\footnote{Although phenomenological models  relevant for the deconfinement transition with some degree of electric-magnetic duality have been proposed in \cite{Liao:2006ry}.}
 
 \section{Summary and overview}
 
 \subsection{Summary, physical picture, and outlook}

The first broad conclusion from our study  of both dYM (Section~\ref{dYM}) and QCD(adj) (Section~\ref{adj}) is that the counting of vacua on $\R^3 \times \S^1$ via the ``splitting of vacua" mechanism of \cite{Aharony:2013hda} is more general than the particular confinement mechanism that was used to argue for it---monopole or dyon condensation in Seiberg-Witten theory on $\R^4$ with soft breaking to ${\cal N}=1$ or ${\cal N}=0$. It was argued in \cite{Aharony:2013hda} that  confining vacua in Seiberg-Witten theory on
$\R^4$  can have an emergent discrete magnetic gauge symmetry, whose nature depends on the global structure,  and that these vacua split after an $\R^3 \times \S^1$ compactification. As we show here, on $\R^3 \times \S^1$, vacua with broken discrete magnetic symmetries appear even in theories where the confinement mechanism on $\R^4$ is unknown. Indeed, while  dYM and SYM can  be thought of as being connected to broken Seiberg-Witten theory, by increasing the relevant supersymmetry breaking parameters and hoping for continuity, this is not so for non-supersymmetric QCD(adj)---in fact, for sufficiently large number of adjoint Weyl flavours, QCD(adj) on $\R^4$ may not even be confining, see discussion in~\cite{Poppitz:2009uq,Poppitz:2009tw,Anber:2011de}.

The confinement mechanism in the calculable regime on $\R^3 \times \S^1$ is quite different from that of Seiberg-Witten theory on $\R^4$ (they share  one broad feature---their abelian nature). In dYM and QCD(adj),  confinement is due to a generalization of the three-dimensional Polyakov mechanism, which arises  due to Debye screening in an instanton gas of magnetically charged objects. The magnetic charges (monopole-instantons) proliferate in the Euclidean $\R^3$  vacuum, rather than by a condensation of magnetically charged particles, as in Seiberg-Witten theory on $\R^4$.\footnote{For the relation between monopole-instantons on $\R^3 \times \S^1$ and monopole particles on $\R^4$, see \cite{Poppitz:2011wy,Poppitz:2012sw}.} Furthermore, there are  important differences between Polyakov's mechanism on $\R^3$ and confinement  on $\R^3 \times \S^1$.   In dYM there is  an extra contribution from a ``Kaluza-Klein" monopole-instanton  \cite{Shifman:2008ja,Unsal:2008ch}, thanks to the compact $\mathbb S_L^1$. In QCD(adj) the additional feature is that the gas is composed of topological molecules, magnetic bions \cite{Unsal:2007jx}, instead of fundamental monopole-instantons. 
 In both classes of theories we study, the broken magnetic discrete symmetries  
on $\R^3 \times \S^1$ manifest themselves in the existence of vacua with different expectation values of the dual photon fields  (or of 't Hooft loops wrapped around $\S^1$) in their respective fundamental domains. 

A second observation is that 
the abundance of vacua in theories with different global structure in the $\R^3 \times \S^1$ setup can be explained using the dichotomy between domain walls\footnote{On $\R^3\times \S^1$, a more precise term would be ``domain lines," but we  use the conventional terminology.} and confining strings. It is based on the idea that a domain wall-like object is either a domain wall  interpolating between different vacua or a confining string, but not both. This picture is simplest to argue for in dYM. There,  confining strings are domain wall-like configurations that carry appropriate electric fluxes. These objects are distinct from the genuine domain walls separating different  vacua; for example, if a theory has no confined local probes, all domain walls are genuine and all minima of the potential are distinct ground states, see  Section~\ref{dymprime} for more examples. 
This view of  theories with different  global structure is harder to explain in QCD(adj)  and SYM, since domain walls there are not confining strings, as they carry half the flux.
However,  the composite nature of confining strings in QCD(adj) found in \cite{Anber:2015kea} still allows  distinguishing  theories with different global structure via the confining string/domain-wall dichotomy (the rank-1 case is described in detail in Section~\ref{adjsu2}). 

Our final result is the curious observation of a Kramers-Wannier duality emerging in thermal dYM on $\R^2 \times \S^1_\beta \times \S^1_L$ in the $\beta \gg L$ calculable regime, see Section~\ref{thermal}, in particular its interplay with the global structure. We only discuss a rank-1 example in  detail, but have noted that the similarities to spin models and ${\cal N}=4$ SYM $S$-duality referred to earlier are more general. It may be of some interest to pursue this further.

 We also note that while there is no oblique confinement in the calculable regime on $\R^3 \times \S^1$,  the relation between theories with  different global structure  by $2 \pi$ shifts of $\theta$ \cite{Aharony:2013hda} arises here due to the ``topological interference" effect \cite{Unsal:2012zj}, where the Euclidean magnetic plasma exhibits $\theta$ dependence due to an analogue of the Witten effect for monopole-instantons. The $su(2)$ dYM case  is an example discussed in detail at the end of Section~\ref{thermal}.

We end with some comments for the future. An explicit way of defining theories with different global structure  was given in \cite{Kapustin:2014gua}: to construct gauge theories with an $SU(N_c)/\Z_k$ gauge group, one gauges a $\Z_k$  subgroup of the discrete $\Z_{N_c}$ global one-form center symmetry of a theory with an $SU(N_c)$ gauge group (we use the terminology of \cite{Gaiotto:2014kfa}, for a traditional lattice definition see e.g.~\cite{Greensite:2011zz}). The gauging proceeds via coupling the gauge theory to a discrete topological gauge theory (dTQFT). The action of the dTQFT, which  also has a lattice formulation  \cite{Kapustin:2014gua}, contains explicit discrete $\theta$-angle parameters labeling the  global structures.
 It might be an interesting future exercise to work out the details of the 
coupling of the dTQFT to the electric and magnetic degrees of freedom in the long-distance theory on $\R^3 \times \S^1_L$ and give it further physical interpretation, e.g.~along the lines of \cite{Dierigl:2014xta}. We also suspect that there are further interesting not-yet-uncovered consequences of the observations of \cite{Anber:2015kea} relating domain walls and confining strings in the classes of theories we discuss. 
 
\subsection{Organization of the paper}

 Section~\ref{symmetriesanddynamics} is devoted to a review and the development of our main tools---the fields, symmetries, and dynamics of  the low energy effective theory of dYM and QCD(adj) on $\R^3 \times \S^1_L$. 
Most of this Section is a review of known results. The exception is the discussion of the $\Z_{N_c}$ zero-form center-symmetry transformation of the dual photon fields (Section~\ref{weylchambersection}) for the general non-supersymmetric case, crucial for the study of Section~\ref{differenttheories}, and the explicit construction of the Wilson, 't Hooft, and dyonic line operators on $\mathbb R^3 \times \mathbb S_L^1$ (most of Section \ref{lineoperators}).
 
 In 
Section~\ref{abelianization} we give a brief  definition of dYM and QCD(adj). We do not review the dynamics that leads to their abelianization, $SU(N_c) \rightarrow U(1)^{N_c-1}$, as this has been done many times in the literature. We do, however, explain the structure of the perturbative abelian action both in terms of the original  electric  gauge fields, (\ref{main Lagrangian with chern term}), and  dual  magnetic  variables, (\ref{total action of the system}), as well as the relevant scale hierarchy. Section~\ref{weylchambersection} contains both a review of some old results and a detailed derivation of some new ones---the (zero-form) center symmetry transformations of the low-energy magnetic variables. For completeness,  in Section~\ref{sigmadomain}, we review the periodicity of the magnetic variables (``dual photons") for different choices of gauge group ($SU(N_c)/\Z_k$), giving two different derivations, one of which is in Appendix~\ref{appendixsigma}. The notion of the magnetic center symmetry is also reviewed there. 

Section~\ref{groundstates}  reviews the nonperturbative effective potentials for dYM and QCD(adj) and their minima. The nonperturbative dynamics leading to the potentials for the dual photons given there is quite rich and we do not do it justice, but simply refer to earlier work.

Section~\ref{lineoperators} studies the 't Hooft and Wilson operators in the $\R^3 \times \S^1$ long-distance theory. All  derivations are given in  Appendix \ref{operatorsappendix}. We  define the line operators in the  canonical formalism  and give a self-contained review of 't Hooft and Wilson operators in $\mathbb R^4$. Then, we give explicit expressions for these operators in $\mathbb R^3 \times \mathbb S^1$, their commutation relations, and the Witten effect within that formalism. We end Section~\ref{lineoperators}, the last  of Section~\ref{symmetriesanddynamics}, by reviewing the classification of the different choices of mutually local line operators for $SU(N_c)/\Z_k$ gauge groups of \cite{Aharony:2013hda}, i.e. the different global structures.

In Section~\ref{differenttheories}, we use the results from Section~\ref{symmetriesanddynamics} to study the vacua of dYM and QCD(adj) with different global structure, obtained by different gauging of  (subgroups of) the zero-form $\Z_{N_c}$ global symmetry. 
In Section~\ref{generalities},  we further specify the action of the to-be-gauged center symmetry on the long-distance magnetic degrees of freedom, Eq.~(\ref{centerk2}) being the most relevant.  

In Section~\ref{dYM}, we study dYM with $SU(N_c)/\Z_{N_c}$ and prime $N_c$ (Section~\ref{dymprime}), nonprime $N_c$  (Section~\ref{dymnonprime}), and $SU(N_c)/\Z_{k}$ with $kk'=N_c$ (Section~\ref{dymk}). The thermally compactified dYM and Kramers-Wannier duality are studied, from the point of view of the global structure, in Section~\ref{thermal}. The physical picture using domain walls and confining strings is also explained there. 

In Section~\ref{adj}, we study the vacua of QCD(adj) for different global structures with  $su(2), su(3)$ and $su(4)$ algebras in Sections~\ref{adjsu2}, \ref{adjsu3} and \ref{adjsu4}, respectively (the details of the latter case are in Appendix~\ref{su4appendix}), where some previous results for the supersymmetric case (a single adjoint flavor) are rederived.

%%%%%%%%%%%%%%%%%%%%%%%%%%%%%%%%%%%%%%%%%%%%%%%%%%%%%%%%%%
\section{Symmetries and dynamics of dYM and QCD(adj) on $\mathbf{\R^3 \times \S^1}$}
%%%%%%%%%%%%%%%%%%%%%%%%%%%%%%%%%%%%%%%%%%%%%%%%%%%%%%%%%%
\label{symmetriesanddynamics}

\subsection{Abelianization, duality, and long-distance theory} 
\label{abelianization}

We consider   four dimensional  Yang-Mills (YM) theory with a gauge Lie algebra $su(N_c)$. We compactify the theory on $\mathbb R^{1,2} \times \mathbb S_L^1$ and we take the compact direction along the  third spatial axis such that $x_3\sim x_3+2\pi R$, and $L\equiv 2\pi R$ is the circumference of the $\mathbb S^1_L$ circle. 

The two classes of theories we consider are:
\begin{enumerate}
\item {\bf dYM}: deformed Yang-Mills theory, i.e.~pure YM theory with the usual action plus a center-stabilizing double-trace deformation\footnote{If one is worried about adding a nonlocal term to the action, note that a center-stabilizing  effect equivalent to that of $\Delta S$ can be due to integrating out massive adjoint fermions with $m L N_c \le {\cal O}(1)$ \cite{Azeyanagi:2010ne,Misumi:2014raa,Bergner:2014dua}.}
\beq \label{deformation}
\Delta S= {1 \over  L^3} \int\limits_{\R^3} \sum\limits_{n=1}^{\lfloor N/2 \rfloor} a_n |\tr_{F} \;\Omega_{L}^n(x)|^2.
\eeq The trace  is taken in the fundamental representation $F$. 
$\Omega_L$ is the Polyakov loop operator, or $\S^1_L$ holonomy \begin{eqnarray}
\label{polyakovloop}
\Omega_{L} (x)=Pe^{i\int_{\mathbb S_L^1}v_3(x, x^3)}\,,
\end{eqnarray}
where $x \in \mathbb R^{1,2}$, $P$ denotes  path ordering and  $v_3$ is the gauge field component along the compact $x_3$ direction. 

The physics of YM theory with the  double trace deformation  (\ref{deformation}) has been studied  in the continuum \cite{Shifman:2008ja,Unsal:2008ch} (motivated in part by large-$N_c$ volume independence) and on the lattice \cite{Myers:2007vc}.
The double-trace deformation $\Delta S$ ensures that the vacuum is at the center-symmetric point,  see (\ref{center0})  below. This is easy to verify  at small $L$, the only regime that we shall study in this paper, where center stability occurs for $a_n \sim 1$. 

\item {\bf QCD(adj)}: YM theory with $n_f$ massless Weyl fermions in the adjoint representation. The $n_f=1$ case is   SYM. When the gauge group is $SU(N_c)$, QCD(adj) has an $ (SU(n_f)\times \mathbb Z_{2n_f N_c})/\mathbb Z_{n_f}$ global chiral symmetry. At small $L$, the $SU(n_f)$ chiral symmetry remains unbroken. The genuine discrete chiral symmetry\footnote{For  theories with an $SU(N_c)/\Z_{k}$ gauge group there is no discrete chiral $\Z_{N_c}$ symmetry. One way to see this, sufficient for us,  is to note   \cite{Aharony:2013hda} that a discrete chiral symmetry transformation shifts the $\theta$ angle by $2\pi$ and thus changes the spectrum of genuine line operators (by the Witten effect, see Appendix \ref{witteneffect} for discussion) mapping one theory to another. Equivalently, upon gauging the $\Z_k$ one-form symmetry \cite{Kapustin:2014gua}, one finds that a discrete chiral transformation shifts the discrete $\theta$-angle. This follows from the fact  that the theory with ungauged center has  a mixed [(discrete zero-form chiral) (one-form center)$^2$]  't Hooft anomaly \cite{Gaiotto:2014kfa}.  } is $\Z_{N_c}$ and is spontaneously broken, as we shall see further below. It is crucial for calculability of the dynamics that the fermions are taken periodic along the $\S^1_L$ circle.

The vacuum in QCD(adj) is also at the center symmetric point. Here, center stability is not due to a deformation (\ref{deformation}), as in dYM, but occurs for different  dynamical reasons, depending on  $n_f$ \cite{Unsal:2007jx}.
\end{enumerate}

We shall discuss the small-$L$ dynamics in   these two theories in parallel, as the bosonic sectors of their respective low-energy effective theories are quite similar, despite the different reasons for center stability and abelianization.
We already alluded to the fact that  
both dYM and QCD(adj) have a one-form $\Z_{N_c}$ global center symmetry acting on line operators. 
When the theory is compactified on $\R^3 \times \S^1_L$, the one-form center symmetry gives rise to a zero-form ``ordinary"  center symmetry  and a one-form symmetry. The former acts on line operators wrapping the $\S_L^1$, such as the Wilson or Polyakov loop. These become local operators in the long-distance theory on $\R^3 \times \S^1_L$. In this paper, we shall study in detail the action of the zero-form part of the center symmetry on the long-distance local observables in the $\R^3 \times \S^1_L$ theory.

The action of the $\Z_{N_c}$ center symmetry (from an $\R^3$ point of view, a zero-form symmetry)  on the  trace of the   $\S^1_L$ Wilson loop  in the fundamental representation is \begin{equation}
\label{center}
\mbox{tr}_{F} \Omega_L\rightarrow e^{i\frac{2\pi k}{N_c}}\mbox{tr}_{F} \Omega_L , \;\;\; k=1,2,...,N_c .
\end{equation} 
Without going into detailed dynamical explanation,\footnote{Briefly, in dYM, center-stability is due to the deformation $\Delta S$ overcoming the one-loop bosonic Gross-Pisarsky-Yaffe potential \cite{Gross:1980br}, which tends to break center symmetry. In QCD(adj) with $SU(N_c)$ gauge group and $n_f>1$ center symmetry is due to the combined one-loop Coleman-Weinberg potential of the bosons and periodic fermions (note that abelianization  at small $L$ is not a property of nonsupersymmetric QCD(adj) for all gauge groups, see \cite{Argyres:2012ka} for an extensive discussion). In SYM ($n_f=1$), where the Coleman-Weinberg potential vanishes due to supersymmetry, center stability holds for all gauge groups, due to the nonperturbative effects of neutral bions.}  the expectation value of 
  $\Omega_L$ (recall that the $\S^1_L$ Wilson loop eigenvalues are gauge invariant)  in both theories can be taken  %
\begin{eqnarray}
\label{center0}
\langle\Omega_{L} \rangle=\eta\; \mbox{diag}\left(1,\omega_{N_c},\omega_{N_c}^2,...,\omega_{N_c}^{N_c-1}\right)\,,\quad \omega_{N_c}\equiv e^{i\frac{2\pi}{N_c}}\,,
\end{eqnarray}
where $\eta=e^{\frac{i\pi}{N_c}}$ for even $N_c$, and $\eta=1$ for odd $N_c$. The Polyakov loop eigenvalues (\ref{center0}) are uniformly spread along the unit circle, $\mbox{tr}\langle\Omega_{L}^k \rangle=0$, $k=1,\ldots N_c -1$, and the $\mathbb Z_{N_c}$ center symmetry of the $SU(N_c)$ gauge theory is preserved. 

   From an $\R^3$ point of view, the Polyakov loop (\ref{polyakovloop}) is an adjoint scalar field, whose expectation value  (\ref{center0}) breaks $SU(N_c)$ to $U(1)^{N_c-1}$. The scale of the breaking is clearly related to the $\S^1_L$ size. Thus,  
by taking the spatial circle to be small, i.e.~$N_c L\Lambda_{\mbox{\scriptsize QCD}}\ll 1$,\footnote{The weak-coupling condition demands that   the mass of the lightest nonabelian gauge boson ($W$-boson), which is $1/(N_c L)$, be larger than the strong scale.} the coupling constant $g^2$ at the scale $1/L$ remains small so that we can perform reliable perturbative analysis at weak coupling.  
We  integrate out the tower of $W$-bosons, the corresponding fermion components, and their Kaluza-Klein modes, remembering that both gauge bosons and fermions obey periodic boundary conditions along $\mathbb S_L^1$.  We shall not do this explicitly in this paper. In order to introduce notation, however, we note that any gauge field $v_m$ or fermion $\lambda_I$ component (denoted by $X$)  are decomposed as 
$
X=X^At^A=\pmb X\cdot \pmb H+\sum_{\pmb \beta_+}X_{\pmb \beta}E_{\pmb \beta}+\sum_{\pmb \beta_+}X^*_{\pmb \beta}E_{-\pmb \beta}\,,
$ where $\pmb X=(X_1,X_2,...,X_r)$ denotes the Cartan components of the field, and $\{H^i\}$, $i=1,2,...,r$, is the set of the Cartan generators (the rank $r=N_c-1$ for $su(N_c)$). The components along the generators $\pmb E_{\beta_\pm}$ (they obey $\left[H^i, E_{\pm \pmb \beta}\right]=\pm \pmb\beta^i E_{\pm \pmb \beta}$, where $\pmb\beta \in \{\pmb \beta_+\}$, the set of all positive roots) are the heavy $W$-bosons. 
The Lagrangian of the long-distance theory, see (\ref{main Lagrangian with chern term}) below, valid at energies smaller than the lightest $W$-boson mass, is  written only in terms of the Cartan components of the fields. 

In what follows we shall write the bosonic part of the effective Lagrangian for both dYM and QCD(adj). To this end, we   use $\pmb v_3$ and $\pmb v_{\mu=0,1,2}$ to denote the $r$-dimensional vectors of Cartan components of the gauge field in the $\S^1_L$ and $\R^{1,2}$ directions, respectively. 
We shall further introduce a dimensionless field 
\beq
\label{phi}
\pmb \phi = \pmb v_3 L~.
\eeq
 Notice that in terms of $\pmb \phi$, the  eigenvalues of $\Omega_L$ in the fundamental representation are $e^{i \;\pmb \nu_k \cdot \pmb \phi}$, where $\pmb \nu_k$, $k=1,...,N_c$, are the weights of the fundamental representation (i.e.~the eigenvalues of the fundamental Cartan generators). The expectation value (\ref{center0})  can be written in terms of 
  the field $\pmb\phi$ as
\begin{eqnarray}
\label{vacuumcenter}
\langle \pmb \phi \rangle = \pmb \phi_0=\frac{2\pi \pmb \rho}{N_c}\, ,
\end{eqnarray}
modulo shifts by $2 \pi$ times vectors in the co-root lattice (see the discussion around equation (\ref{phiperiodicity1})).\footnote{\label{basis1}A useful basis of weights for $su(N_c)$ is as follows.
Let $\pmb e_i$, $i$$=$$1,...N_c$,  denote the $i$-th unit  Cartesian basis vector of $\R^{N_c}$. All roots and weights are then orthogonal to the vector $\pmb e_1 + \pmb e_2 + \ldots + \pmb e_{N_c}$$=$$(1,1,1,...1)$. 
The simple roots are $\pmb \alpha_i$$=$$\pmb e_i - \pmb e_{i+1}$, $1 \le i \le N_c -1$, and the affine (or lowest) root is $\pmb \alpha_0$$=$$\pmb\alpha_{N_c}$$=$$\pmb e_{N_c} - \pmb e_1$. The co-weights $\pmb w_k^*$, obeying $\pmb w_k^* \cdot \pmb \alpha_j$$=$$\delta_{kj}$ ($k,j$$=$$1,...N_c -1$), are then $\pmb w_k^*$$=$$\sum\limits_{1\le i \le k} \pmb e_i - {k \over N_c} \sum\limits_{1\le i \le N_c} \pmb e_i$. Since we use a normalization where $\pmb \alpha_i^2=2$, the roots and co-roots as well as weights and co-weights are naturally identified for the $su(N_c)$ algebra. The $N_c$ weights of the fundamental are $\pmb \nu_1 = \pmb w_1$, $\pmb \nu_2 = \pmb w_1 - \pmb \alpha_1, \ldots, \pmb \nu_{N_c} = \pmb w_1 - (\pmb\alpha_1 + \ldots \pmb\alpha_{N_c - 1})$.}
Here $\pmb\rho$ is the Weyl vector defined as $\pmb\rho=\sum_{a=1}^{N_c-1}\pmb w_a$,  $\pmb w_a$ are the fundamental weight vectors, which satisfy $\pmb \alpha_a^*\cdot \pmb w_b=\delta_{ab}$, $a=1,2,...,N_c-1$, and $\pmb \alpha_a^*$ are the dual simple roots.
As already mentioned, for a generic expectation value of $\pmb \phi$ (or $\Omega$), the gauge group $G$ is broken down to $U(1)^r$. The dimensionally reduced effective action of the theory reads:
\begin{eqnarray}
S&=&\frac{L}{g^2}\int_{\mathbb R^{1,2}}  d^3 x \left\{ -\frac{\partial_\mu \pmb \phi\cdot\partial^\mu \pmb \phi}{L^2}-\frac{1}{2}\pmb v_{\mu\nu}^2 +\frac{\theta}{8\pi^2}  \epsilon_{\mu\nu\rho}\partial^\mu \pmb \phi\cdot \pmb v^{\nu\rho} - g^2V_{\mbox{\scriptsize eff}}(\pmb \phi)\right\} \,,
\label{main Lagrangian with chern term}
\end{eqnarray} 
where $\pmb v_{\mu\nu} = \partial_\mu \pmb v_\nu -  \partial_\nu \pmb v_\mu$, we have kept the $\theta$-angle dependence, and have denoted the perturbative  potential for $\pmb \phi$ by $V_{\mbox{\scriptsize eff}}$. We stress again that the difference between  QCD(adj) and dYM  is in the dynamics generating this potential; in particular it vanishes for SYM, is given by (\ref{deformation}) plus loop correction for dYM, and is loop-generated in QCD(adj).

In (\ref{main Lagrangian with chern term}) and further in this paper, $g$ denotes the four-dimensional gauge coupling at the scale $1/L$. $\pmb \phi$-dependent loop corrections to the moduli space metric (the kinetic terms in (\ref{main Lagrangian with chern term})) have been omitted; these will be important at one point in our discussion and shall be reintroduced.
The action (\ref{main Lagrangian with chern term}),  valid for $N_c L\Lambda_{\mbox{\scriptsize QCD}}\ll 1$, describes $r$ free massless photons $\pmb v_\mu$ and $r$ scalars (the $rn_f$ free massless  Weyl fermions in QCD(adj) are omitted). For QCD(adj) with $n_f>1$ and dYM, the scalars $\pmb \phi$ have masses of order $\frac{\sqrt{N_c}g}{L}$. The special case $n_f=1$ corresponds to a supersymmetric theory where  the scalars are massless.

Next, we can write a dual description of the three dimensional photons by introducing the auxiliary Lagrangian
\begin{eqnarray}
S_{\mbox{\scriptsize aux}}=\frac{1}{4\pi}\int d^3x \epsilon_{\mu\nu\rho}\partial^\mu \pmb \sigma\cdot \pmb v^{\nu\rho}\,.
\label{auxiliary S}
\end{eqnarray}
Varying $S_{\mbox{\scriptsize aux}}$ with respect to $\pmb \sigma$ we obtain the Bianchi identity $\epsilon_{\mu\nu\rho} \pmb v^{\nu\rho}=0$. Further, by varying  $S+S_{\mbox{\scriptsize aux}}$ with respect to $\pmb v^{\alpha\beta}$ we find
\begin{eqnarray}
\pmb v^{\nu\rho}=\frac{g^2}{4\pi L}\left(\partial_\mu \pmb \sigma+\frac{\theta}{2\pi}\partial_\mu \pmb \phi\right)\epsilon^{\mu \nu\rho}\,.
\label{v in terms of sigma}
\end{eqnarray}
Substituting (\ref{v in terms of sigma}) into  $S+S_{\mbox{\scriptsize aux}}$  we find
\begin{eqnarray}\label{total action of the system} %main Lagrangian with chern term
S_{eff}  
 =\frac{1}{L}\int d^3 x \left\{-\frac{1}{g^2}\left(\partial_\mu \pmb \phi\right)^2-\frac{g^2}{16\pi^2}\left(\partial_\mu \pmb\sigma+\frac{\theta}{2\pi}\partial_\mu\pmb \phi \right)^2- V_{\scriptsize\mbox{eff}}(\pmb \phi)  \right\}\,,
\end{eqnarray}
i.e.~the action in terms of electric ($\pmb \phi$) and magnetic ($\pmb \sigma$) variables. We stress that for $n_f = 1$ QCD(adj) all fields in (\ref{total action of the system}) are massless, while in the nonsupersymmetric QCD(adj) $n_f>1$ and dYM, there is a scale hierarchy among the fields in (\ref{total action of the system}):
\begin{equation}
\label{perturbativescalehierarchy}
m_W = {1 \over L N_c}\; \; \;  \; \gg \; \; \;\; m_{\pmb \phi} \sim {\sqrt{N_c} g \over L} \; \; \; \; \gg \; \;  \;\; m_{\pmb \sigma} = 0~.
\end{equation}
This perturbative hierarchy justifies the validity of the effective theory (\ref{total action of the system}), allowing us to keep  the fields $\pmb \phi$ and $\pmb \sigma$ (and the corresponding fermion components, when present) while integrating out the heavy $W$-bosons and fermions. 

The action (\ref{main Lagrangian with chern term}) and   its dual (\ref{total action of the system}) will be the basis for our study of the theories with different  global structure of the gauge group. To this end, we need to understand the action of the zero-form  $\Z_{N_c}$ center symmetry on the fields in the effective long-distance theory.

\subsection{The Weyl chamber and the action of  center symmetry on the electric and magnetic degrees of freedom}
\label{weylchambersection}

We begin with a description of the Weyl chamber of $\pmb \phi$. This is the region of physically inequivalent values of $\pmb \phi$---equivalence under large gauge transformations periodic in $SU(N_c)$ and under discrete Weyl reflections is imposed. Since this is important for us, we dwell on the structure of the Weyl chamber in some detail. The field $\pmb \phi$ can be shifted by large gauge transformations, $\pmb \phi \rightarrow\pmb \phi + 2 \pi \pmb a$, generated by $U= e^{ i {x^3 \over R} \pmb a \cdot \pmb H_{\cal R}}$, where $\pmb H_{\cal R}$ denotes the Cartan generator in a representation ${\cal R}$. Periodicity of $U$ for all electric representations allowed by the global choice of gauge group  requires $\pmb a \cdot \pmb \nu_{\cal R}\in \Z$, 
where $\pmb \nu_{\cal R}$ is any allowed weight of $G$, i.e. a vector in its group lattice $\Gamma_G$.\footnote{A review of some useful terminology follows.
The group lattice is spanned by the weights of the faithful representations of $G=SU(N_c)/\Z_k$. One extreme example  is where the gauge group is the covering group $\tilde{G}=SU(N_c$), where $\Gamma_{\tilde G} = \Gamma_w$, the weight lattice of $SU(N_c)$.  Another  case is when the gauge group is the adjoint group, $G=SU(N_c)/\Z_{N_c}$, when $\Gamma_G = \Gamma_r$, the root lattice of $SU(N_c)$ and no charges with ``smaller" electric representations are allowed. In the intermediate cases when $\Z_k \subset \Z_{N_c}$, the group lattice is intermediate between the coarse root lattice and the fine weight lattice $\Gamma_r \subset \Gamma_G \subset \Gamma_w$. The basis $\{{\pmb g_k}\}$ of the group lattice $\Gamma_G$, for $G=SU(N_c)/\Z_k$ with $k k'=N_c$  can in each particular case be constructed from appropriate combinations of the weight-lattice basis vectors, such that the weight of any representation of $N$-ality $k p$, $p=1,...,k'$, can be written as their linear combination, see example in Appendix~\ref{su4appendix}. Finally, the dual to the group lattice for  $\tilde{G}=SU(N_c)$  is  the co-root lattice $\Gamma_r^*$ (dual to the weight lattice $\Gamma_w$), while for $G=SU(N_c)/\Z_{N_c}$ it is the co-weight lattice $\Gamma_w^*$  dual to the root lattice $\Gamma_r$. 
} 
 This  implies that $\pmb a$ is an element of the lattice   $\Gamma_G^*$  dual to $\Gamma_G$. 
 
 Equivalently, the fact that shifts of $\pmb \phi$ by $2 \pi$ times $\pmb a \in \Gamma_G^*$ are not observable can also be seen by noting that the gauge invariant eigenvalues of  Wilson loops around $\S^1_L$ in all allowed representations, i.e.~$e^{i \pmb \nu \cdot \pmb \phi}$ for arbitrary $\pmb \nu \in \Gamma_G$, do not change under a shift of $\pmb \phi$ by $2\pi$ times $\Gamma_G^*$ vectors.

We conclude that for an $SU(N_c)$ gauge group, where $\Gamma_G^*$ is the dual to the weight lattice, the co-root lattice $\Gamma_r^*$, we have that the fundamental domain of $\pmb\phi$ is the unit cell of $\Gamma_r^*$, i.e.
\beq\label{phiperiodicity1}
\pmb \phi \equiv \pmb \phi + 2 \pi \pmb \alpha^*_k~, ~~ k=1,\ldots N_c-1~.
\eeq
Imposing further identifications under Weyl reflections, the Weyl chamber for $SU(N_c)$, see  \cite{Argyres:2012ka}, is given by  $\pmb \phi$ obeying the inequalities \begin{equation}
\label{weylchamber}
\pmb \alpha_a \cdot \pmb \phi > 0, a = 1,...,r, \; {\rm and } \; - \pmb \alpha_0 \cdot \pmb \phi < 2 \pi~,
\end{equation} where $\pmb \alpha_0$ is the affine or lowest root. The result (\ref{weylchamber}) can also be derived physically as the smallest connected region in $\pmb \phi$ space, containing $\pmb \phi = 0$ (where all $W$ bosons are massless) such that no massless $W$ bosons appear  anywhere except at its boundaries, including any Kaluza-Klein modes.  This follows by studying the $W$-boson spectrum, given by $|{2\pi p\over L} + { \pmb\beta \cdot \pmb \phi  \over L}|$, where $p$ is the integer Kaluza-Klein number and $\pmb \beta$ is any root.

\begin{figure}[htbp] %  figure placement: here, top, bottom, or page
   \centering
   \includegraphics[width=.5\textwidth]{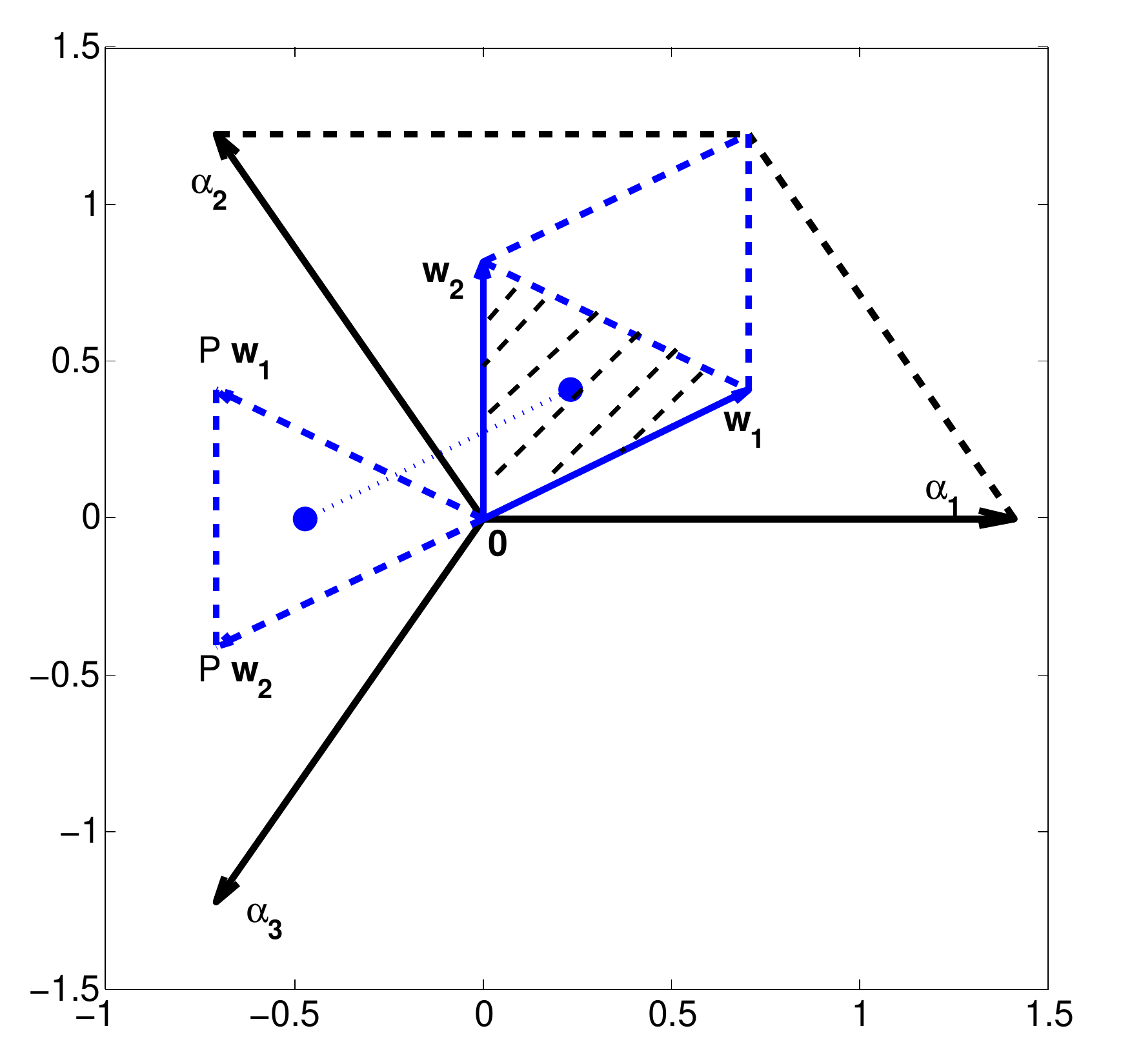} 
   \caption{The Weyl chamber of $\pmb\phi \over 2 \pi$ for $SU(3)$ is the shaded equilateral  triangle between the two fundamental weights $\pmb w_1$ and $\pmb w_2$. The dot in the center of the triangle is the center symmetric point $\pmb \rho \over 3$ (\ref{vacuumcenter}). The global $\Z_3$ center  transformation (\ref{zenphi}) acts as a counterclockwise $\pi/3$ rotation $\cal P$ (\ref{calP}) around the origin (the vectors ${\cal P} \pmb w_{1,2}$ are shown)  followed by a $\pmb w_1$ shift. In effect, this produces a $\pi/3$ rotation of the shaded triangle around its center. In the $SU(3)/\Z_3$ theory, the $\Z_3$ rotation of the Weyl chamber around its center is a gauge symmetry and the Weyl chamber is correspondingly reduced. }
   \label{fig:su3weyl}
\end{figure}

Geometrically, the Weyl chamber of $\pmb \phi$ can be  described as the region in an $r$-dimensional space, which is the inside of the volume whose  boundary is given by the  $r+1$ hyperplanes where the inequalities (\ref{weylchamber}) become an identity---a triangle for $r=2$, a tetrahedron for $r=3$, etc.; see Figure~\ref{fig:su3weyl} for the rank two case (notice also, as per the discussion in the paragraph after Eq.~(\ref{zenphi}), that when the gauge group is reduced upon modding $SU(N_c)$ by a subgroup of the center, the fundamental domain of $\pmb \phi$ is correspondingly reduced).

Now, we are ready to study the action of the zero-form $\Z_{N_c}$ center symmetry. This is a transformation of the $\pmb \phi$ fields that: {\it a.})~maps the Weyl chamber to itself and {\it b.})~acts on the $\S_L^1$ Wilson loop eigenvalues by a $\Z_{N_c}$ phase, as in (\ref{center}).
It consists of a cyclic Weyl reflection ${\cal{P}}$ plus a fundamental weight-vector shift \cite{Argyres:2012ka}. In a basis-independent language, the cyclic Weyl reflection is defined as follows.\footnote{We notice that we do not always distinguish between weights and coweights, or roots and coroots, as they are naturally identified in $su(N_c)$.}
 Let $\pmb v$ be an arbitrary vector in weight space and $\hat{s}_{\pmb \alpha} \pmb v = \pmb v -  2 \pmb \alpha \; { \pmb \alpha \cdot \pmb v  \over  \pmb \alpha \cdot \pmb \alpha }$ be its reflection in a plane perpendicular to the root $\pmb \alpha$. Then,
\beq
\label{calP}
{\cal{P}} =  \hat{s}_{{\pmb \alpha}_{1}} \hat{s}_{{\pmb \alpha}_{2}}
 \ldots \hat{s}_{{\pmb \alpha}_{N_c -2}} \hat{s}_{{\pmb \alpha}_{N_c -1}}   \eeq
is the ordered product of the Weyl reflections with respect to all simple roots. In the $N_c$-dimensional basis, with all weight vectors  orthogonal to  ($1,1,1,...,1$), where $\pmb v = (v_1, ... , v_{N_c})$, we have $ {\cal{P}} \pmb v = (v_{N_c}, v_1,v_2 \ldots, v_{N_c-1})$. The action on the simple and affine roots is ${\cal{P}} \pmb \alpha_k = \pmb\alpha_{k+1  ({\rm mod}  N_c)}$---thus, in the $SU(3)$ weight diagram of Figure~\ref{fig:su3weyl}, this is  a counterclockwise $\pi/3$ rotation around the origin. The action of ${\cal P}$ on the fundamental weights is also  easily seen to be ${\cal P} \pmb w_k = \pmb w_k - (\pmb \alpha_1 + \ldots + \pmb \alpha_k)$.

 In terms of the cyclic Weyl reflection $\cal{P}$,  the zero-form $\Z_{N_c}$ center symmetry action on $\pmb \phi$ is
\beq
\label{zenphi}
 \hat\gamma \in \Z_{N_c}: ~\pmb\phi \rightarrow {\cal{P}} \pmb \phi + 2 \pi \pmb w_1^*~.
\eeq
It is straightforward to see that (\ref{zenphi}) maps the Weyl chamber (\ref{weylchamber}) to itself and that, since 
$\tr_{F} \Omega_L = \sum\limits_{k = 1}^{N_c}  e^{i \pmb \nu_k \cdot \pmb\phi}$, Eq.~(\ref{center}) is a consequence of (\ref{zenphi}) (notice that $e^{ 2 \pi i \pmb w_p \cdot \pmb w_k^*} = e^{- {2 \pi i p k\over N_c}}$). Clearly, the vev $\pmb \phi_0$ (\ref{vacuumcenter}) is a fixed point of (\ref{zenphi}). These features are illustrated for $SU(3)$ on Fig.~\ref{fig:su3weyl}.
 
We pause to stress that the reason for our detailed study of the action of the global center symmetry on the low-energy degrees of freedom  is that upon restricting the allowed electric representations, i.e. by taking  the gauge group to be a quotient of $SU(N_c)$  by a $\Z_k$ subgroup of its center, further large gauge transformations are allowed---for example, ones periodic in $SU(N_c)/\Z_k$, rather than just $SU(N_c)$, since the condition $\pmb a \cdot \pmb \nu \in \Z$ becomes less restrictive when the set of  allowed electric charges $\pmb \nu$ is reduced. This means that shifts of $\pmb\phi$, as in (\ref{phiperiodicity1}), by vectors in lattices finer than the dual root lattice  (e.g.~by $\pmb w_1^*$) become  gauge symmetries. Thus, depending on the choice of $\Z_k$, part of the global symmetry (\ref{zenphi})  becomes gauged. In particular, if we take $k k' = N_c$, then $\hat\gamma^{k'}$ generates $\Z_k$ and is gauged in the $G/\Z_k$ theory.

A further observation\footnote{Tangentially, this has important consequences   for  the confining strings in theories on $\R^3 \times \S^1$.} made in \cite{Anber:2015kea},  crucial to our study here, is that the $\Z_{N_c}$ generator $\hat\gamma$  has to also act on the dual photon field  $\pmb \sigma$. As we shall see, ultimately this follows from the fact that $\Z_{N_c}$ of (\ref{zenphi}) is a symmetry of the long-distance theory (\ref{total action of the system}), unbroken in the vacuum (\ref{center0}). The quickest argument makes use of supersymmetry. In SYM, $\pmb \phi + i \pmb \sigma$ is the lowest component of a chiral superfield and since $\hat\gamma$ should act on the entire superfield, we  have, along with (\ref{zenphi}),
\beq
\label{zensigma}
\hat\gamma \in \Z_{N_c}: \pmb\sigma \rightarrow {\cal{P}} \pmb \sigma~.
\eeq
In fact,  (\ref{zensigma}) holds independent of supersymmetry and applies also to dYM and QCD(adj) with $n_f>1$. Since Eq.~(\ref{zensigma})  is our main tool for studying vacua identified by the action of the $\Z_{N_c}$ zero-form symmetry, we now pause to give the general argument. The discussion in the following three paragraphs may appear  lengthy and technical, but in view of its importance we give it in detail.

The way to argue that $\hat\gamma$ should act as in (\ref{zenphi}, \ref{zensigma}) is to show that this is a symmetry of the full partition function of the long distance theory. In the following we show that this is true to one-loop order in  
the effective Lagrangian (the argument is, in fact, more general, see the comment at the end of this Section). Consider (\ref{main Lagrangian with chern term}) before the duality transformation, but now include the loop-corrected moduli space metric,\footnote{This was omitted in (\ref{main Lagrangian with chern term}) and in the rest of the paper as it is only relevant for the present argument.} $g_{ab}(\pmb\phi)$. It adds to the kinetic terms of both $\pmb\phi$ and $\pmb v_\mu$ from (\ref{main Lagrangian with chern term}) a loop contribution of the form 
\beq
\label{kinetic}
  {1 \over L} g_{ab}^{(1)}( \pmb\phi  )\; \partial_\mu  \phi^a \partial^\mu  \phi^b + L g_{ab}^{(2)}(  \pmb\phi )  \;v_{\mu\nu}^a v^{\mu\nu \; b} ~,
\eeq
were $a$ and $b$ run over the Cartan subalgebra. The one-loop correction to the metric was calculated  for SYM in \cite{Anber:2014lba}, via the $\R^3 \times \S^1$-index theorem  \cite{Poppitz:2008hr} in monopole-instanton backgrounds, and in Ref.~\cite{Anber:2014sda} via Feynman diagrams in QCD(adj) and dYM. The explicit form, including coefficients and details of renormalization, can be found there.

   It is convenient to shift $\pmb\phi$ around its vev $\pmb \phi_0$ (\ref{vacuumcenter}). For brevity, in the discussion below we use $\pmb \phi$ to denote the slowly-varying fluctuation around $\pmb \phi_0$. Since $\pmb \phi_0$ is invariant under (\ref{zenphi}), the fluctuation $\pmb \phi$ transforms homogeneously under $\hat\gamma$: $\pmb\phi \rightarrow {\cal P}\pmb \phi$.
In the next paragraph, we show that $\partial_\mu ({\cal P} \phi)^a \partial^\mu ({\cal P} \phi)^b g^{(1)}_{ab}({\cal P} \pmb \phi) = \partial_\mu   \phi ^a \partial^\mu    \phi ^b g^{(1)}_{ab}(  \pmb \phi)$, i.e.~the low-energy theory effective action of $\pmb \phi$ is invariant under $\cal P$ transformations. We also note  that  $g_{ab}^{(1)}$ and $g_{ab}^{(2)}$ transform in the same manner, as explicitly shown in Eq.~(\ref{oneloopmetric1}) below. This implies that the photon field $\pmb v_\mu$ should transform as $\pmb \phi$  in order to keep the long-distance lagrangian (\ref{kinetic}) invariant, i.e.~as $\pmb v_\mu \rightarrow {\cal P} \pmb v_\mu$. After the duality  (\ref{auxiliary S}, \ref{v in terms of sigma}),  the  $\Z_{N_c}$ transformation of $\pmb v_\mu$ induces the transformation (\ref{zensigma}) on the dual photon $\pmb\sigma$. 

To substantiate the conclusion from the above paragraph,  we consider the non-diagonal part  of the metric. Up to 
theory-dependent constants and a $\pmb\phi$-independent contribution renormalizing the gauge coupling, which can be found in \cite{Anber:2014lba,Anber:2014sda} for the various cases,
both one loop functions $g_{ab}^{(1, 2)}$ from (\ref{kinetic}) are of the form
\beq
\label{oneloopmetric}
g_{ab}^{1-loop}(\pmb\phi) = \sum_{\beta^+} \beta^a \beta^b \left[\Psi ({\pmb\rho\cdot \pmb\beta \over N_c}+{\pmb\phi\cdot \pmb\beta \over 2 \pi}) + \Psi(1 -{\pmb\rho\cdot \pmb\beta \over N_c}- {\pmb\phi\cdot \pmb\beta \over 2 \pi})\right]~,
\eeq where the  sum  is over all positive roots $\pmb\beta$ and $\Psi$ is the logarithmic derivative of the gamma function. 
Next, we recall that the $su(N_c)$ roots are $\pmb \beta_{ij} = \pmb e_{i} - \pmb e_{j}$ and that the set of positive roots that is summed over in (\ref{oneloopmetric}) corresponds to summing over $1 \le i < j \le N_c$. Below, we shall use $\pmb \beta^{+}_{ij}$  to denote roots for which $1 \le i < j \le N_c$, i.e. positive roots. We also have that   $ { ({\cal P} \pmb \phi) \cdot \pmb\beta_{ij}} ={  \pmb \phi \cdot  \pmb \beta_{i-1 \; j-1}}$ and thus
$   \Psi ({\pmb\rho\cdot \pmb\beta^+_{ij} \over N_c}+{ ({\cal P} \pmb\phi)\cdot \pmb\beta^+_{ij} \over 2 \pi}) =  \Psi ({\pmb\rho\cdot \pmb\beta^+_{i j } \over N_c}+{   \pmb\phi \cdot \pmb\beta_{i-1 \; j-1} \over 2 \pi}) 
.$ But $\pmb \beta_{i-1 \; j-1}$ is a positive root only for $i > 1$, while for $i=1$, we have $\pmb \beta_{0 \; j-1} = - \pmb \beta_{j-1 \; N_c}^+$. Thus, using $\pmb\rho \cdot \pmb \beta_{ij} = j-i$, we find that
\beq
 \label{psitransform01}
\Psi ({\pmb\rho\cdot \pmb\beta^+_{ij} \over N_c}+{ ({\cal P} \pmb\phi)\cdot \pmb\beta^+_{ij} \over 2 \pi}) = 
\left\{ \begin{array}{cc}  \Psi ({\pmb\rho\cdot \pmb\beta^+_{i-1 j-1 } \over N_c}+{   \pmb\phi \cdot \pmb\beta^+_{i-1 \; j-1} \over 2 \pi}) & {\rm for } \; i>1 ~,\\      \Psi (1- {\pmb\rho\cdot \pmb\beta^+_{j-1 \; N_c } \over N_c}- {   \pmb\phi \cdot \pmb\beta^+_{  j-1\; N_c} \over 2 \pi}) & {\rm for } \; i=1~ .\end{array} \right.
 \eeq
We can similarly work out the transformation of the second term in (\ref{oneloopmetric}), combine it with (\ref{psitransform01}),  introduce $\hat{\pmb\beta}_{ij}  \equiv {\pmb\beta}_{i-1 \; j-1} $ for $i>1$, and $\hat{\pmb\beta}_{1j}  \equiv {\pmb\beta}_{j-1 N_c}$, to conclude that  
 \begin{eqnarray}
 \label{psitransform}
 && \Psi ({\pmb\rho\cdot \pmb\beta^+_{ij} \over N_c}+{ ({\cal P} \pmb\phi)\cdot \pmb\beta^+_{ij} \over 2 \pi})+ \Psi(1 -{\pmb\rho\cdot \pmb\beta^+_{ij} \over N_c}- { ({\cal P} \pmb\phi)\cdot \pmb\beta^+_{ij} \over 2 \pi}) \\ &&\qquad \qquad=     \Psi ({\pmb\rho\cdot \hat{\pmb \beta}^+_{i  j } \over N_c}+{   \pmb\phi \cdot  \hat{\pmb\beta}^+_{i j } \over 2 \pi}) + \Psi(1 -{\pmb\rho\cdot \hat{\pmb\beta}^+_{i  j } \over N_c}- { \pmb\phi \cdot  \hat{\pmb\beta}^+_{i  j }\over 2 \pi})~ \nonumber~.
   \end{eqnarray}
   Then, using (\ref{psitransform}),    we deduce the transformation of (\ref{oneloopmetric}): 
\begin{eqnarray}  \label{oneloopmetric1}
   g_{ab}^{1-loop}({\cal P} \pmb \phi) = \sum\limits_{1 \le i < j \le N_c} \beta^a_{i  j}  \beta^b_{ij}  \left[ \Psi ({\pmb\rho\cdot \hat{\pmb\beta}_{i j} \over N_c}+{   \pmb\phi \cdot \hat{\pmb\beta}_{i j } \over 2 \pi}) + \Psi(1 -{\pmb\rho\cdot \hat{\pmb\beta}_{i j} \over N_c}- { \pmb\phi \cdot \hat{\pmb\beta}_{i j} \over 2 \pi})\right]~.
\end{eqnarray} 
Finally, we recall the transformation of the derivatives, $\partial_\mu ({\cal P} \pmb \phi) \cdot \pmb \beta_{ij} = \partial_\mu   \pmb \phi \cdot \pmb \beta_{i-1\; j-1}$. Together with (\ref{oneloopmetric1}), they imply\footnote{Again, we use  $\pmb\beta_{0 j-1} = - \pmb \beta_{j-1 N_c}$, noting that every root  appears squared in the derivative terms and that  $\pmb\beta \rightarrow \hat{\pmb\beta}$ is simply a relabeling of all the positive roots.} the already noted invariance $\partial_\mu ({\cal P} \phi)^a \partial^\mu ({\cal P} \phi)^b g_{ab}({\cal P} \pmb \phi) = \partial_\mu   \phi ^a \partial^\mu    \phi ^b g_{ab}(  \pmb \phi)$ as well as the transformation $\pmb v_\mu \rightarrow {\cal P} \pmb v_\mu$ required to keep the invariance of the long-distance theory (\ref{kinetic}).

We stress that the invariance  of the long-distance action under (\ref{zenphi}) is exact to all loop orders (and, as we shall see below in all cases we study, nonperturbatively\footnote{Indeed, finding the symmetry (\ref{zensigma}) from the nonperturbative potentials (\ref{dympotential}, \ref{qcdadjpotential}) is quick, but it is important to realize that it is an exact symmetry to all loop orders.})   despite our  use of the  one-loop corrected moduli space metric (\ref{oneloopmetric}) to illustrate it. In essence, $\cal P$ invariance of the long-distance theory holds because the interactions between the heavy and light modes, as well as the spectrum of heavy $W$ bosons is invariant under the $\cal P$ transformation of the light fields, provided the vacuum is center symmetric.\footnote{After some Kaluza-Klein frequency relabeling---responsible for the $i=1$ shift in (\ref{psitransform01})---which is inessential since the $W$ bosons and their Kaluza-Klein modes are integrated out  (this gave rise to the particular combination of $\Psi$ functions in (\ref{oneloopmetric})).}  Thus, $\cal P$ invariance  (\ref{zenphi},\ref{zensigma}) of the long-distance theory is a consequence of the unbroken  center symmetry of dYM and QCD(adj).

In summary, the main results of this Section to be used later are the description of the $SU(N_c)$ Weyl chamber (\ref{weylchamber}) and the action of the $\Z_{N_c}$ generator $\hat\gamma$ on $\pmb \phi$ and $\pmb \sigma$, (\ref{zenphi}, \ref{zensigma}).

 \subsection{The fundamental domain of the dual photon for different choices of gauge group}
 \label{sigmadomain}
 
 The fundamental domain of the dual photon field $\pmb \sigma$ is determined by the allowed electric charges in theory. 
 The allowed charges are, in turn,   determined by the global structure of the gauge group. 
For gauge theories with an $su(N_c)$ algebra, 
 the universal covering group is $\tilde G= SU(N_c)$ and the possible choices of the gauge group are $G=SU(N_c)/\Z_k$, with $\Z_k$ a subgroup of the $\Z_{N_c}$ center.   
The periodicity of $\pmb \sigma$  is determined by the group lattice 
 \begin{equation}
 \label{sigmaperiodicity1}
 \pmb \sigma \equiv \pmb \sigma + 2 \pi \pmb g_k~,
 \end{equation}
 where $\pmb g_k$, $k=1,...r$, form a basis of the group lattice $\Gamma_G$. 
A quick way to argue this is via the duality relation (\ref{v in terms of sigma}),\footnote{For  $\theta\ne 0$, notice that $\pmb \phi$ has no monodromy around electric charges. A  Hamiltonian derivation of (\ref{sigmaperiodicity1}), based on   further spatial compactification  on $\mathbb T_2$,  magnetic flux quantization, and the duality (\ref{v in terms of sigma}), is given in Appendix \ref{appendixsigma}.} which implies that the electric field is $\pmb v^{0 i} ={g^2 \over 4 \pi L} \epsilon^{ij} \partial_j \pmb\sigma$, where $i=1,2$, $\epsilon^{12} = 1$. Thus, the monodromy of $\pmb \sigma$ around a spatial loop $C \in \R^2$ measures the electric charge inside, $\oint_{C}d \pmb \sigma = {4 \pi L \over g^2} \; \pmb Q$ where $\pmb Q$ is the flux of $\pmb v^{0i}$ through $C$. In the normalization of (\ref{main Lagrangian with chern term}), Gau\ss' law for a static charge (weight) $\pmb \lambda$ at the origin is $\partial_i \pmb v^{0 i}(x) = {g^2 \over 2 L} \pmb \lambda \delta^{(2)}(x)$, hence $\pmb Q = {g^2 \over 2 L} \pmb \lambda$ and so the monodromy becomes $\oint_{C}d \pmb \sigma = 2 \pi \pmb \lambda$. The condition that the dual photon be single valued around all allowed charges, dynamical or probes, in a gauge theory with gauge group $G$, i.e.~for all $\pmb \lambda \in \Gamma_G$, implies the identification (\ref{sigmaperiodicity1}).

In particular, for $G= \tilde{G} = SU(N_c)$ (we denote by $\tilde{G}$ the covering group), the fundamental domain of $\pmb\sigma$ is the unit cell of the weight lattice $\Gamma_w$ (the finest lattice for $su(N_c)$), while for $SU(N_c)/\Z_{N_c}$ it is the unit cell of the root lattice $\Gamma_r$, with the group lattices $\Gamma_G$ for the intermediate cases. Thus, for gauge group $SU(N_c)/\Z_k$, 
  weight-lattice shifts of  $\pmb \sigma$ are meaningful. They represent global symmetries rather  identifications under (\ref{sigmaperiodicity1})---provided  $\Gamma_G$ is coarser than $\Gamma_w$.
  Recall that  $\Gamma_w/\Gamma_G = \pi_1(G)$ and  that the centers of $G$, $Z(G)$, and of $\tilde{G}$, $Z(\tilde{G})$, obey $Z(G) \ltimes \pi_1(G) = Z(\tilde{G})$.  For $G=SU(N_c)/\Z_k$, with $k k' = N_c$, we have  $Z(G) = Z(\tilde{G})/\Z_k = \Z_{k'}$. 
  Thus, for $G=SU(N_c)/\Z_{k}$, $\pi_1(G)$ is also a $\Z_{k}$ discrete symmetry,   called  the magnetic or dual center symmetry. This symmetry, being generated by shifts of $\pmb \sigma$ by weights in $\Gamma_w/\Gamma_G$,  naturally acts on 't Hooft operators (see Eq.~(\ref{operators1}) below).

 \begin{figure}[htbp] %  figure placement: here, top, bottom, or page
   \centering
   \includegraphics[width=.6\textwidth]{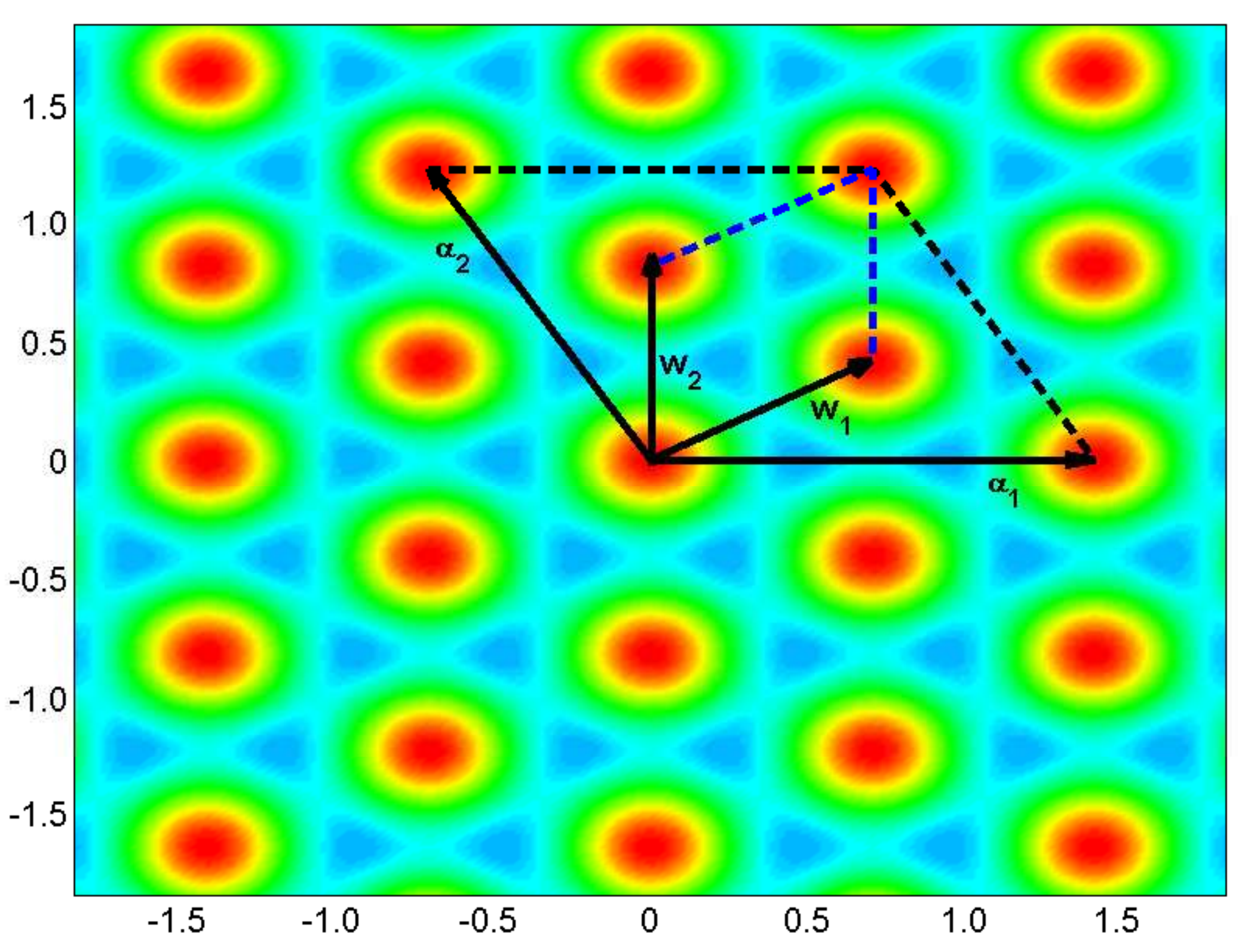} 
   \caption{ {\bf dYM}: The $\pmb\sigma\over 2\pi$ plane for $su(3)$. The $SU(3)$ fundamental domain is $\Gamma_w$, spanned by $\pmb w_{1,2}$. A contour plot of the potential (\ref{dympotential}) is overlaid with the minima (\ref{dymminima}) of the potential for dYM indicated by the dark (red) circles. There is a single ground state for dYM at $\pmb\sigma = 0$ within the  $SU(3)$ fundamental domain---but not within the larger domain, the root lattice $\Gamma_r$ spanned by $\pmb \alpha_{1,2}$, for $SU(3)/\Z_3$.    }
   \label{fig:dymsu3}
\end{figure}
    
 To summarize, in a theory with gauge group $G$,  nontrivial weight lattice shifts of $\pmb\sigma$, by vectors that   belong to $\Gamma_w/\Gamma_G$, act as global symmetries on the magnetic degrees of freedom. 
  We shall see below, when studying the action of the gauged center symmetry on the vacua and on the  Wilson, 't Hooft and dyonic  operators, that for $G= SU(N_c)/\Z_k$ there are $k$ inequivalent gaugings of the $\Z_k$ center. They differ by the choice of $\Gamma_w/\Gamma_G$ shifts in the gauged center symmetry transformation of $\pmb\sigma$.  
 
Before discussing the gauging, we next review the vacua  of the  $SU(N_c)$ theories on $\R^3 \times \S^1_L$.
     
 \subsection{The ground states of dYM and QCD(adj) for $\mathbf{su(N_c)}$ theories}
    \label{groundstates} 
    
\begin{figure}[htbp] %  figure placement: here, top, bottom, or page
   \centering
   \includegraphics[width=.6\textwidth]{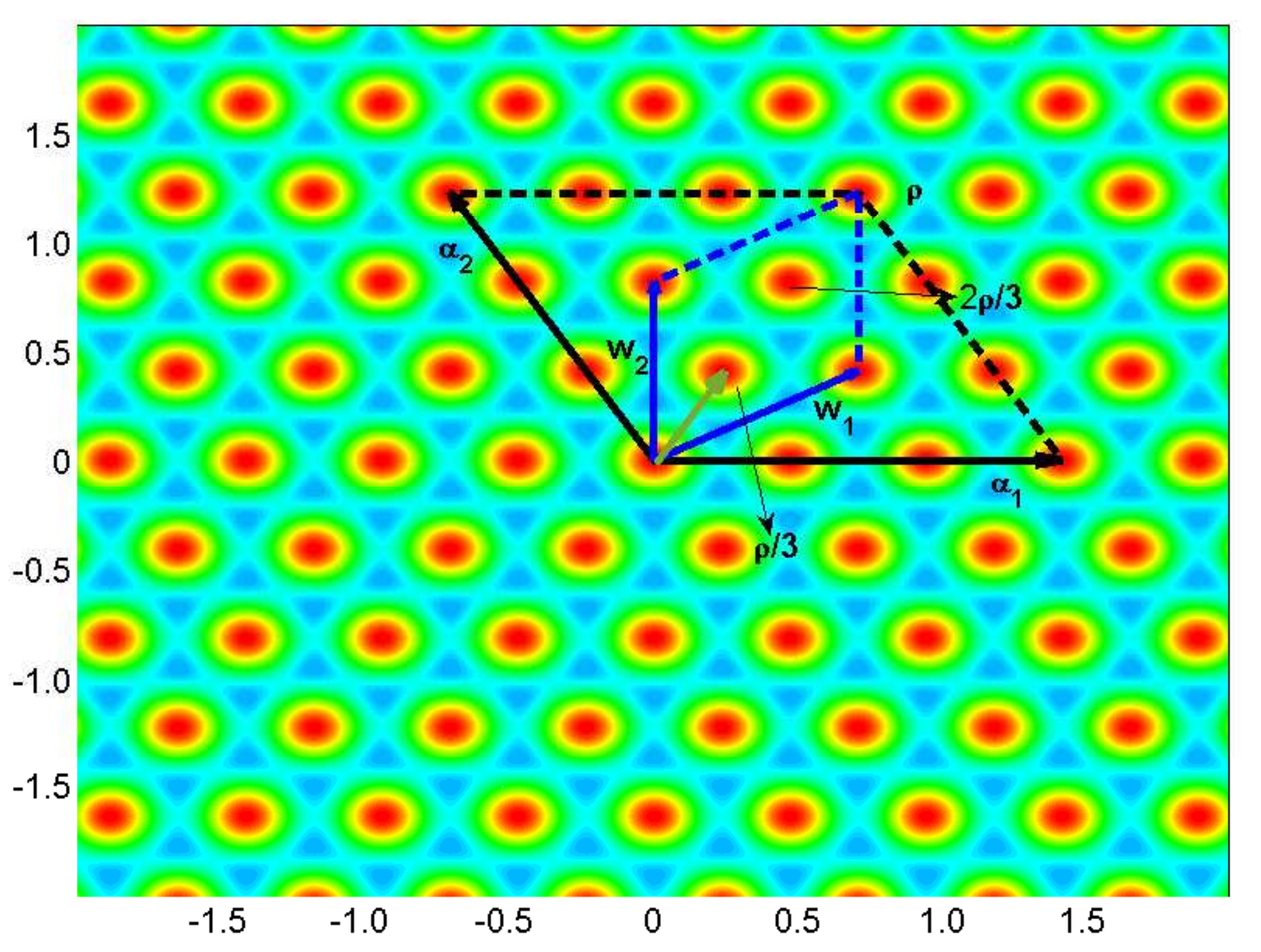} 
   \caption{ {\bf QCD(adj)}: The $\pmb\sigma\over 2\pi$ plane for $su(3)$ and the minima of the 
   potential---the extrema of (\ref{superpotential1})---indicated by dark (red) circles. There are three minima, at $\pmb \sigma = {2 \pi k \pmb \rho \over 3}$, $k=0,1,2$, within the $SU(3)$ fundamental domain, the weight lattice. As on Fig.~\ref{fig:dymsu3}, there are three times as many minima within the root lattice (used later in  finding the $[SU(3)/\Z_3]_p$ theory ground states).   }
   \label{fig:symsu3}
\end{figure}

In the following Sections, we shall describe how to study the vacua of $SU(N_c)/\Z_k$ gauge theories on $\R^3 \times \S^1$. At small $L$, the ground state is determined by calculable nonperturbative effects which generate potentials for $\pmb \sigma$.  The nonperturbative potentials in dYM and QCD(adj) have been derived before. We simply give them below and only mention their dynamical origin. The dynamical objects that are involved in their generation are the same, no matter what  choice of global structure is made---the dynamical objects   have root-lattice electric and co-root-lattice magnetic charges and are present for all choices of $G$.
\begin{enumerate}
\item {\bf dYM}: The potential is generated by $N_c$ magnetic monopole-instantons whose magnetic charges are labeled by the affine coroots of the  $su(N_c)$ algebra $\pmb \alpha^*_k$, $k=1,...N_c$. The potential can be written in the form \cite{Unsal:2008ch}
\beq
\label{dympotential}
V_{dYM}(\pmb \sigma) = {c \over L^3} e^{- {8 \pi^2 \over N_c g^2} } \sum\limits_{k=1}^{N_c} \left[ 1 - \cos \left( \pmb\alpha^*_k \cdot \pmb\sigma + {\theta \over N_c}\right) \right]~,
\eeq
where the overall constant $c$ has power law dependence on $g^2$ as well as numerical factors that are inessential for us. The $e^{- {8 \pi^2 \over N_c g^2} }$ factor and the  $\theta$-dependence reflect the fact that both the action and topological charge of these objects are $1/N_c$-th of the ones for BPST instantons. 

For further use, for $\theta=0$,\footnote{Nonzero-$\theta$ effects in dYM were studied in \cite{Thomas:2011ee,Unsal:2012zj}. We  mostly study $\theta=0$, except for remarks in Section~\ref{thermal} and Appendix \ref{witteneffect}.} the minima of (\ref{dympotential}) occur at 
\beq
\label{dymminima}
\langle \pmb\sigma \rangle = 2 \pi \pmb w_k, ~k=0,\ldots N_c-1\; ({\rm  mod}\; 2\pi \pmb w),~ \forall \pmb w \in \Gamma_w, {\rm with} \; \pmb w_0 \equiv \pmb 0.
\eeq
Notice in particular, that for the $G=SU(N_c)$ gauge group dYM has a single minimum, at $\pmb \sigma =0$, within the fundamental domain (the weight lattice $\Gamma_w$). See Fig.~\ref{fig:dymsu3} for an illustration for $su(3)$.

\item {\bf QCD(adj)}: The potential is generated by $N_c$ magnetic bions \cite{Unsal:2007jx}---correlated tunneling events composed of a monopole-instanton and an anti-monopole instanton, which are neighbors on the extended Dynkin diagram, i.e.~have magnetic charge $\pmb \alpha_k^* - \pmb \alpha_{k-1}^*$. The potential, see \cite{Anber:2011de,Argyres:2012ka}, evaluated at the center symmetric vev for $\pmb \phi$ (this is permitted by  the scale separation (\ref{perturbativescalehierarchy})), can  be cast in a ``supersymmetric" form, as already noted in \cite{Unsal:2007jx}. This reflects the similar nonperturbative origin of the potentials in SYM, see \cite{Davies:2000nw}, and QCD(adj) with $n_f>1$:
\beq \label{qcdadjpotential}
V_{QCD(adj)}(\pmb\sigma) = {c' \over L^3}\; e^{- {16 \pi^2 \over N_c g^2}}\;\sum\limits_{a=1}^{N_c-1} \;\frac{\partial {\cal W}(\pmb \sigma)}{\partial\sigma_a} \;\frac{\partial \bar{\cal W}(\pmb \sigma)}{\partial\sigma_a}\,,  
\eeq%
where the ``superpotential" ${\cal W}$ ($\bar{\cal W}$ is its complex conjugate) is given by
\begin{eqnarray}
\label{superpotential1}
{\cal W}(\pmb \sigma)=\sum_{a=0}^{N-1}e^{i\pmb\sigma\cdot\pmb \alpha_a^*}\,.
\end{eqnarray}   
The main difference between the ($n_f=1$) supersymmetric case and the nonsupersymmetric QCD(adj) is in the presence of light moduli $\pmb\phi$ in SYM, which mix with $\pmb \sigma$ in the potential (its full form   can be found in \cite{Poppitz:2012nz}). In both SYM and QCD(adj),   $\pmb \phi$ is stabilized at the center symmetric value $\pmb\phi_0$, while the minima for $\pmb \sigma$, given by the extrema of (\ref{superpotential1}), are \beq
\label{qcdadjminima}
\langle \pmb\sigma \rangle = {2 \pi k \pmb \rho \over N_c}, ~k=0,\ldots N_c-1\; ({\rm  mod}\; 2\pi \pmb w),~\forall \pmb w \in \Gamma_w ~.
\eeq
For  a $G=SU(N_c)$ gauge group, there are $N_c$ minima, $\pmb \sigma ={2\pi k \pmb \rho \over N_c} $, for QCD(adj) within the fundamental domain (the weight lattice $\Gamma_w$). These are  associated with the spontaneously broken discrete chiral symmetry, well known from past studies of SYM. See Fig.~\ref{fig:symsu3} for a contour plot of the potential for the $su(3)$ case.
 \end{enumerate}
Before we continue, recall the fact already alluded to---that the nonperturbative potentials (\ref{dympotential}, \ref{qcdadjpotential}) preserve the $\Z_{N_c}$ center symmetry  $\pmb \sigma \rightarrow {\cal P} \pmb \sigma$ (\ref{zensigma}). This follows upon inspection of the potentials and the fact that $\pmb \alpha_k \cdot ({\cal P} \pmb \sigma) = \pmb \alpha_{k-1 ({\rm mod} N_c)} \cdot  \pmb \sigma$. Clearly, the potentials also preserve the magnetic center symmetry (whenever present) as they are invariant under $2 \pi \pmb w_k$ shifts of $\pmb \sigma$.
 
Next, we are interested in finding the  ground states in dYM or QCD(adj) with a  $G=SU(N_c)/\Z_k$ gauge group. 
Thus, we shall begin with finding the minima of the $\pmb\sigma$ potential up to shifts by   $\Gamma_G$ (i.e. in the unit cell of $\Gamma_G$, the fundamental domain of $\pmb \sigma$).
As already discussed, in the theory with an $SU(N_c)/\Z_k$ gauge group, some of the global $\Z_{N_c}$ transformations (\ref{zenphi},\ref{zensigma})---the ones generated by $\hat\gamma^{k'}$---are now gauged. Thus, some  vacua within $\Gamma_G$ are identified.

In addition,  there is freedom to supplement  the $\hat\gamma^{k'}$ action on $\pmb\sigma$ by generators of   the magnetic $\Z_{k}$  symmetry, i.e. by shifts by basis vectors of $\Gamma_w/\Gamma_G$.  The different $[SU(N_c)/\Z_k]_r$ theories are distinguished by this action.  
  The genuine line operators are those that do not transform by a phase under the chosen $\Z_k$ shifts.
  The number of ground states in any given case is given by the number of minima within $\Gamma_G$ (given by Eq.~(\ref{dymminima}) for dYM), further identified by the action of  $\hat\gamma^k$ and the chosen shifts by  $\Gamma_w/\Gamma_G$ generators. We now review the classification of the different theories.

 \subsection{Wilson, 't Hooft, and dyonic line operators, and the classification of different $\mathbf{[SU(N_c)/\Z_k]_r}$ theories }
 \label{lineoperators}
 
 As discussed in the introduction, one way to distinguish different theories for a given choice of  gauge group is via their sets of mutually local genuine line operators. We thus begin with a short review of these operators in our setup.
We shall give a canonical (Hilbert space) definition of line operators in the low-energy effective theory on $\R^{1,2}$. 

To motivate the expressions that follow, we note that our long-distance theory is abelian, without light charged particles. Wilson ('t Hooft) loop operators create infinitely thin electric (magnetic) fluxes along their respective loops. Using 
Gau\ss' law, Wilson ('t Hooft) loops can be rewritten as operators measuring the magnetic (electric) flux through a surface $\Sigma$ bounded by the loop $C$.  A generic dyonic operator depends on both electric and magnetic fluxes
\beq
\label{Dgeneral}
{\cal D}(\pmb\nu_e, \pmb \nu_m, \Sigma) = e^{i \;2 \pi \pmb \nu_m \cdot \hat{\pmb \Phi}_e (\Sigma)   + i\; \pmb \nu_e \cdot \hat{\pmb \Phi}_m (\Sigma)}~.
\eeq 
Here, $\pmb\nu_{e,m}$  are electric and magnetic weights (see below) and $\hat{\pmb\Phi}_{e,m}$ are the operators of the electric or magnetic flux through the corresponding surface $\Sigma$. Explicitly,   $\hat{\pmb\Phi}_m(\Sigma) =  \int_\Sigma d^2 \sigma_i \hat{\pmb B}^i$ and $\hat{\pmb\Phi}_e(\Sigma) = \int_\Sigma  d^2 \sigma_i \hat{\pmb \Pi}^i$. Here, $i=1,2,3$ denotes spatial directions, $\hat{\pmb B}^i$ is the magnetic field operator, and $\hat{\pmb \Pi}^i$---the momentum operator conjugate to the gauge field $\hat{\pmb v}^i$ (for $\theta =0$ this is essentially the electric  field  operator).\footnote{See Appendix \ref{operatorsappendix} for normalizations and a short review of the Hilbert space definition of 't Hooft operators.} We also note that  no ordering issues arise in the long-distance abelian theory, as  evident from the final expressions (\ref{operators2},\ref{operators1}) below.

We already discussed that  the electric weights $\pmb \nu_e$ for a given choice of the gauge group $G$ take values in the group lattice $\Gamma_G$. Magnetic weights $\pmb \nu_m$ can, a priori, take values in the co-weight lattice, but    are restricted by the condition that operators in faithful representations of $G$ are single valued around $e^{i 2 \pi \pmb \nu_m \cdot \hat{\pmb \Phi}_e(\Sigma)}$. This leads to the condition that $\pmb \nu_m \cdot \pmb g \in \Z$, $\forall g \in \Gamma_G$, i.e.~the magnetic weights take values in the dual to the group lattice $\Gamma_G^*$; see Appendix \ref{operatorsappendix} for more detailed discussion.

\begin{figure}[htbp] %  figure placement: here, top, bottom, or page
   \centering
   \includegraphics[width=1.0\textwidth]{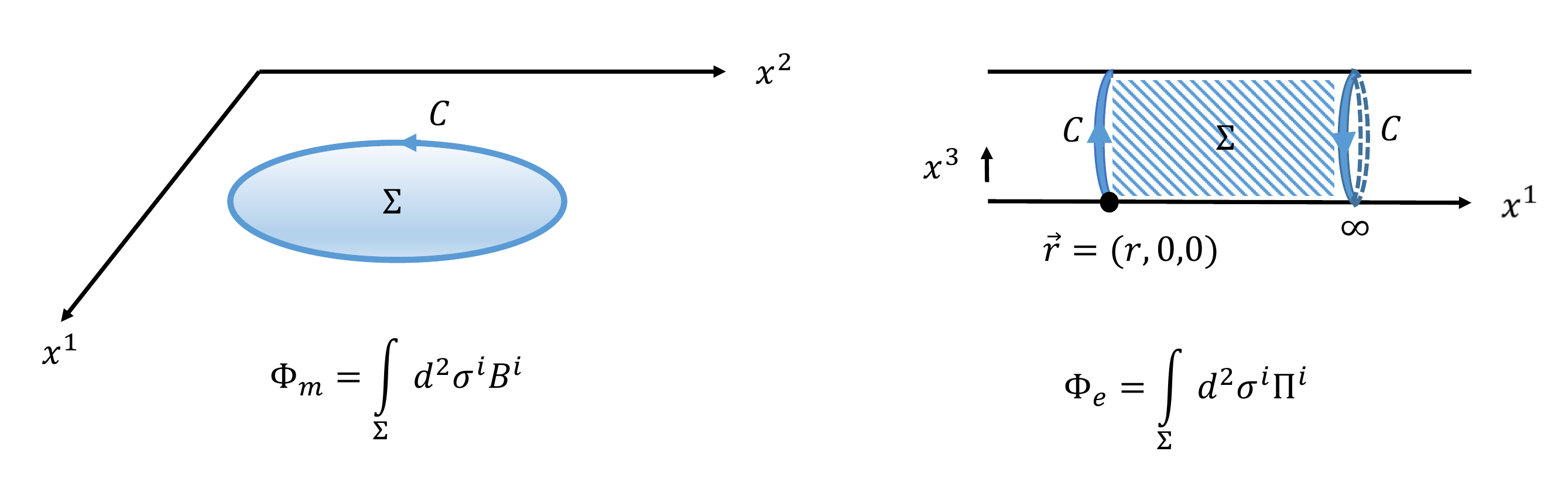} 
   \caption{Two kinds of loops $C$. The Wilson and  't Hooft (dyonic)  loop operators  measure the magnetic ($\Phi_m$) or electric ($\Phi_e$) flux (a combination thereof), respectively,  through surfaces $\Sigma$ spanning the contour $C$. The two kinds of surfaces shown give rise to the operators (\ref{operators2}), (\ref{operators1}), respectively.}
   \label{fig:loops}
\end{figure}

We next consider the  two kinds of loops   shown on Fig.~\ref{fig:loops}. One set  of loops  are boundaries of surfaces $\Sigma_{xy}$ in the noncompact $\R^2$, while others bound surfaces  wrapped around $\S^1_L$---where one end of the surface, i.e.~one of the two loops winding in opposite directions around $\S^1_L$ and spanning the surface, can be taken to infinity. 

Recalling that $\pmb \phi = L \pmb v_3$, so that $L \pmb F_{i3} = \partial_{i} \pmb \phi, L\pmb E_3 = \partial_t \pmb\phi$, using the duality (\ref{v in terms of sigma})  and the long-distance lagrangian (\ref{main Lagrangian with chern term}, \ref{total action of the system}), we find after some tedious but straightforward manipulations (see Appendix \ref{operatorsappendix})  that  Eq.~(\ref{Dgeneral}) becomes, for a surface $\Sigma_{xy}$  spanning the loop $C \in \R^2$ (the $xy$-plane)
     \beq
\label{operators2}
{\cal D}(\pmb \nu_e, \pmb \nu_m, \Sigma_{xy})=\exp\left[i 2 \pi  \int_{\Sigma_{xy}} d^2 s \left\{ \pmb\nu_m \cdot \pmb\Pi_\phi  +   \pmb\nu_e \cdot \pmb \Pi_\sigma \right\} \right]\,,
\eeq 
where $\pmb \Pi_{\phi, \sigma}$ are the conjugate momenta found upon quantizing (\ref{total action of the system}) (we omitted hats over operators). The dyonic operator corresponding to the loop winding around the $\S^1_L$ circle is labeled by the single point $\vec{r} \in \R^2$ and is given by
\beq\label{operators1}
{\cal D}(\pmb \nu_e, \pmb \nu_m, \vec{r})=\exp\left[-i \pmb\nu_m\cdot \pmb \sigma(\vec{r})  +i \pmb \nu_e \cdot \pmb \phi(\vec{r})  \right]\,.
\eeq
 From the canonical commutation relations, 
the nontrivial commutation relation of the dyonic operators (\ref{operators2},\ref{operators1}) is easily seen to be:
\begin{eqnarray}
\label{commutator}
{\cal D}(\pmb \nu_e^1, \pmb \nu_m^1, \Sigma_{xy})\; {\cal D}(\pmb \nu_e^2, \pmb \nu_m^2, \vec{r})&&= \\
&&
 e^{i 2\pi \ell(\Sigma_{xy}, \vec{r}) \left(\pmb \nu_{e}^1\cdot\pmb \nu_{m}^2-\pmb\nu_{m}^1\cdot\pmb \nu_{e}^2\right)} \; {\cal D}(\pmb \nu_e^2, \pmb \nu_m^2, \vec{r}) \; {\cal D}(\pmb \nu_e^1, \pmb \nu_m^1, \Sigma_{xy}) \nonumber~.
\end{eqnarray}
Here $\ell(\Sigma_{xy}, \vec{r})$ is unity if $\vec{r} \in \Sigma(xy)$ and zero otherwise. As expected, the dyonic operators (\ref{operators1}) are mutually local {\it provided} that 
\begin{eqnarray}
\label{diracrelation}\pmb \nu_{e}^1\cdot\pmb \nu_{m}^2-\pmb\nu_{m}^1\cdot\pmb \nu_{e}^2 \in \Z~,
\end{eqnarray}
i.e. the Dirac quantization condition  is satisfied (electric and magnetic weights obeying the conditions  discussed in the paragraph following (\ref{Dgeneral}) obey (\ref{diracrelation})).

As explained in \cite{Aharony:2013hda}, the different $SU(N_c)/\Z_k$ theories are distinguished by the possible choices of mutually local sets of line operators. There, these were called ``genuine line operators," as they do not involve observable surfaces (topological or otherwise) and we shall henceforth use this terminology as well.\footnote{Although our derivation of the line operators (\ref{operators2}, \ref{operators1}) involves surfaces,  in the end result the operators winding around $\S^1_L$ (\ref{operators1})
do not involve a surface (these could have been obtained more directly). Further, in (\ref{operators2}) the surface, while present using our low-energy variables,  is not observable if (\ref{diracrelation}) holds, i.e.~for genuine line operators. }
We are now in position to describe the classification  of the different $SU(N_c)/\Z_k$ theories. 

The remainder of this Section is a review of  observations of \cite{Aharony:2013hda}. 
 The dyonic operators  (\ref{operators1}) enable us to categorize the different theories for a given covering group $\tilde G$, as described in \cite{Aharony:2013hda}. We focus  on $SU(N_c)/\Z_k$ theories with $k k^\prime = N_c$. To this end, we denote a fundamental Wilson loop by $W$ and a fundamental 't Hooft loop by $H$. In particular,  we can think of $W$ and $H$ as   our operators 
 \beq
 \label{ouroperators}
 W = {\cal D}(\pmb w_1,0,  \vec{r})~,~~H= {\cal D}(0, \pmb w_1,  \vec{r})~,
 \eeq respectively, see  (\ref{operators1}), where we took both $\pmb \nu_e$ and $\pmb \nu_m$  to be the highest weight of the fundamental representation.\footnote{\label{s1only}For notational simplicity, we  refer to the $\S^1_L$-wrapped operators but shall remember that checking the mutual locality condition (\ref{diracrelation}) requires using the operators (\ref{operators2}). Once again, unless we have to, we do not distinguish between weights and co-weights.}   Similarly, we use  $W^p H^q$ to denote a dyonic operator with a Wilson loop in a representation of $N$-ality $p$ and 't Hooft loop with a magnetic weight of $N$-ality $q$.\footnote{The $N$-ality of a representation with  Dynkin labels $(q_1,...,q_{N_c-1})$, i.e.~of highest weight $\pmb \nu = \sum\limits_{i=1}^{N_c-1} q_i \pmb w_i$,  is given by $q_1 + 2 q_2 + ... + (N_c-1) q_{N_c -1} \; {\rm mod}(N)$. }

We begin by recalling that Wilson and 't Hooft loops with weights in the root lattice (or co-root lattice, which we identify with the root lattice for $su(N_c)$) are always allowed and play no role for distinguishing the global structure of the theories: they correspond to the dynamical fields ($W$-bosons) and  dynamical magnetic monopoles of the theory and occur irrespective of the global choice of gauge group. The operators that distinguish between the different theories are Wilson and 't Hooft loops with charges taking values in latices finer than the root lattice.

Consider first the purely electric probes.
Clearly, in an $SU(N_c)/\Z_{k}$ theory only electric probes of $N$-ality $k$ are allowed. Thus the lowest charge allowed for electric representations is, schematically, $W^k$, the $k$-th power of the fundamental Wilson loop; notice that if $k=N_c$, no  nontrivial $N$-ality electric probe is permitted.

Turning to magnetic line operators, note that the fundamental 't Hooft loop $H$ is not mutually local with respect to $W^k$. This follows, in our notation and using (\ref{diracrelation}), by noting that the weights of representations of $N$-ality $p$  and $k$  obey\footnote{Notice that (\ref{wproduct}) holds upon replacing $\pmb w_p$ or $\pmb w_k$ there by {\it any  weight} of an $N$-ality $p$ (or $k$) representation. For the purpose of classifying the different choices of mutually local line operators it suffices to consider  powers of the fundamental $W$ and $H$.
}
\begin{equation}
\label{wproduct}
e^{2 \pi i \pmb w_p \cdot \pmb w_k} = e^{ -2 \pi i \;{p k \over N_c}}~.
\end{equation}
Thus, for $p=1$ and $k < N_c$, the quantization condition (\ref{diracrelation}) does not hold and the operators do not commute, as per (\ref{commutator}). However,   (\ref{commutator}, \ref{wproduct}) also imply that the $k'$-th power of the 't Hooft loop $H^{k'}$, with a magnetic weight of $N$-ality $k'$ (e.g. $\pmb w_{k'}$ modulo roots),  is mutually local with respect to $W^k$ since $kk'=N_c$. This also implies that dyonic operators of the form $W^n H^{k'}$, for any $n$, are also mutually local with respect to $W^k$. However, $W^n H^{k'}$ is not  local with respect to $W^p H^{k'}$ with $n\ne p$. Thus one can choose the mutually local line operators for the $SU(N_c)/\Z_k$ theory to be in {\it one of} the following $N$-ality classes:
\begin{equation}
\label{choices}
(W^k, H^{k'})~,~~ (W^k, W H^{k'})~, ~~(W^k, W^2 H^{k'})~,...~~(W^k, W^n H^{k'}) ~,
\end{equation}
continuing (a priori) to arbitrary  $n$. We use the notation of \cite{Aharony:2013hda}, where the ordered pair, e.g.~$(W^k, W H^{k'})$,  denotes the mutually local purely electric ($W^k$) and magnetic or dyonic $(W H^{k'})$ operators in a given theory.
 Further, we note that only values of $n$ modulo $k$ lead to physically distinct choices of mutually local line operators, since $W^{p+k} H^{k'}$ has locality properties identical  to $W^p H^{k'}$, owing to $kk'  = N_c$ and (\ref{wproduct}). 
 
 The conclusion  \cite{Aharony:2013hda}  is that for $SU(N_c)/\Z_k$ there are $k$ possible choices of mutually local (or ``genuine")  line operators. These choices are listed in (\ref{choices}), with $n=k-1$. These $k$  choices label the different $[SU(N_c)/\Z_k]_r$, $r =0,...,k-1$, theories. The choice of genuine line operators is part of the definition of the theory. Their expectation values can be used to classify the  phases of the theories. One can also study how the theory behaves in the infrared after an ultraviolet perturbation by   various line operators.

After this review of \cite{Aharony:2013hda}, in the rest of the paper, we study  dYM and QCD(adj)  in the calculable regime on $\R^3 \times \S^1_L$ and show explicitly how the classification (\ref{choices}) arises naturally  when constructing the $SU(N_c)/\Z_k$ theories by gauging the $\Z_k$ subgroup of the center symmetry of the $SU(N_c)$ theories.
 In the partially compactified theory, the gauging can be worked out in a  straightforward manner for the zero-form part of the center, as it acts on the local degrees of freedom $\pmb \phi$, $\pmb\sigma$, in a way already determined in the previous Sections. The gauging  will  allow us to also determine the vacuum structure of the $[SU(N_c)/\Z_k]_r$ theories on $\R^3 \times \S^1_L$.

%%%%%%%%%%%%%%%%%%%%%%%%%%%%%%%%%%%%%%%%%%%%%%%%%%%%%%%%%%%%%%%%%%%%%%%%%%%%%%%%%%%%
\section{Theories with different global structure and their vacua on $\mathbf{\R^3 \times \S^1}$}
%%%%%%%%%%%%%%%%%%%%%%%%%%%%%%%%%%%%%%%%%%%%%%%%%%%%%%%%%%%%%%%%%%%%%%%%%%%%%%%%%%%%%
\label{differenttheories}

\subsection{Generalities}
\label{generalities}

As already explained, the different  choices of genuine line operators from (\ref{choices}) correspond to different gaugings of the $\Z_k$ symmetry: the electric $\Z_k$ symmetry, acting on $(\pmb \phi, \pmb\sigma)$ as in (\ref{zenphi}, \ref{zensigma}), is gauged in all cases, but can be supplemented by an action of the magnetic $\Z_k$ center on $\pmb \sigma$. 

To make this more explicit in our notation consider our example operators from the previous section. Let us explicitly write\footnote{Keeping footnote \ref{s1only} in mind.} the operators  (\ref{ouroperators})
\beq \label{WandT}
 W = e^{i \pmb w_1 \cdot \pmb \phi} ~,~~  H = e^{-i \pmb w_1 \cdot \pmb \sigma}~,  
 \eeq
as representatives of the fundamental representation Wilson and 't Hooft loops. Under the   $\Z_{N_c}$ center symmetry transformation $\hat\gamma$ (\ref{zenphi}, \ref{zensigma}), $\pmb \phi \rightarrow {\cal P} \pmb \phi + 2 \pi \pmb w_1$, we have,  up to Weyl reflections ${\cal P}$, $W \rightarrow e^{-   {2 \pi i \over N_c}} W$,\footnote{\label{weylfootnote1}More precisely, notice that under $\hat\gamma$, we have $e^{i \pmb \nu_k \cdot \pmb \phi} \rightarrow e^{-   {2 \pi i \over N_c}} e^{i \pmb \nu_{k-1 ({\rm mod} N_c)} \cdot \pmb \phi}$. Thus, if we were not modding by cyclic Weyl reflections (generated by $\cal P$ and part of the unbroken gauge group), using the fact that $\cal{P}$ cyclically permutes the $N_c$ weights $\pmb \nu_i$ of the fundamental representation,  we would have $\sum_{i=1}^{N_c} e^{i \pmb \nu_i \cdot \pmb \phi} \rightarrow e^{ - {2 \pi i \over N_c}}\sum_{i=1}^{N_c}  e^{i \pmb \nu_i \cdot \pmb \phi}$ instead of the shorthand $W \rightarrow e^{- {2 \pi i \over N_c}} W$. Similarly, $\sum_{i=1}^{N_c} e^{- i \pmb \nu_i \cdot \pmb \sigma} \rightarrow  \sum_{i=1}^{N_c}  e^{- i \pmb \nu_i \cdot \pmb \sigma}$ instead of $H\rightarrow H$. The same remark also applies to (\ref{centerk1}) and (\ref{centerk2}).} while under $\pmb \sigma \rightarrow {\cal P} \pmb \sigma$, the 't Hooft loop is invariant, $H \rightarrow H$, also up to Weyl reflections; note that we used   (\ref{wproduct}) again. Since $W$ is a gauge invariant operator and since it is not mutually local w.r.t. $H^k$ with $k<N_c$, the only choice of genuine line operators of nontrivial $N$-ality in the $SU(N_c)$ theory is in the $N$-ality class of $(W,1)\sim (W,H^{N_c})$ (in the notation of \cite{Aharony:2013hda}).

For gauge group $G = SU(N_c)/\Z_k$, the  $\Z_k$ center is generated by $\hat\gamma^{k'}$ and acts similarly: 
\beq
\label{centerk1}
\Z_k: ~ W \rightarrow e^{- {2\pi i k' \over N_c}} W~,~~ H  \rightarrow H ~, ~~ k'k= N_c.
\eeq
In other words, $W^{k}$ is invariant. It is also mutually local w.r.t.~$H^{k'}$.
The $N$-ality class $(W^k, H^{k'})$ is the first entry of (\ref{choices}). The theory with this choice of genuine line operators is denoted by $[SU(N_c)/\Z_k]_0$. This theory corresponds to  gauging the zero-form center symmetry acting on   $\pmb\phi$, $\pmb\sigma$ as $\hat\gamma^{k'}$, with $\hat\gamma$ from (\ref{zenphi},\ref{zensigma}).

 As explained earlier, theories with $G=SU(N_c)/\Z_k$ have a magnetic $\Z_k$ center acting on $\pmb\sigma$ as shifts by basis vectors in $\Gamma_w/\Gamma_G$. This is because, as opposed to $SU(N_c)$ theories, shifts by such weight vectors are not identifications, since the fundamental domain of $\pmb \sigma$ is now the larger (than the unit cell of the weight lattice) unit cell of $\Gamma_G$. The  shifts of $\pmb \sigma$  acting nontrivially on the fundamental 't Hooft loop $H$ are  generated by the $k-1$ fundamental weights   $\pmb w_q$, $q=1,\ldots,k-1$, i.e.~the highest weights of the  $q$-index antisymmetric tensor representations (of $N$-ality less than $k$).
We denote a $\hat\gamma^{k'}$ action modified by a $2 \pi \pmb w_q$ shift by $\hat\gamma^{(k', q)}$. We then have that 
\begin{eqnarray}
\label{centerk2}
\hat\gamma^{(k',q)}:~~~ \pmb \sigma &\rightarrow& {\cal P}^{k'} \pmb \sigma + 2 \pi \pmb w_q,  ~ q=1,\ldots k-1, \nonumber
\\
H^{k'} &\rightarrow& e^{ - i 2 \pi k' \pmb w_1 \cdot \pmb w_q} H^{k'} = e^{   i 2 \pi k' q \over N_c} H^{k'} ,
\\
W &\rightarrow&e^{- {2\pi i k'  \over N_c}} W,\nonumber \end{eqnarray}
i.e.~a $\Z_k$ action on the  operators $H^{k'}$ and $W$, where the action on $W$ is from (\ref{centerk1}) (recall footnote \ref{weylfootnote1}). We  extend the definition above to $q=0$ by understanding that $\hat\gamma^{(k,0)}$ does not involve a shift of $\pmb\sigma$. 
  Thus $W^q H^{k'}$ is invariant under $\hat\gamma^{(k',q)}$ for $0 \le q \le k-1$. Clearly, for the $k$ different values of   $q$ we have  the different (mutually nonlocal) choices, $(W^k, W^q H^{k'})$ of Eq.~(\ref{choices}), of sets of mutually local operators. The corresponding theories are called $[SU(N_c)/\Z_{k}]_{q}$, $q=0,...k-1$. This completes the classification  \cite{Aharony:2013hda} of theories with $G= SU(N_c)/\Z_{k}$.

 The main result from this Section that we use in what follows is the action of the (zero-form part of the) $\Z_k$ symmetry  whose gauging gives rise to the $[SU(N_c)/\Z_k]_q$ theory. The most relevant one is the $\hat\gamma^{(k',q)}$ action on $\pmb \sigma$ from Eq.~(\ref{centerk2}) with $q=0,...k-1$ (no shift for $q=0$)---this is because the vacuum structure of dYM and QCD(adj) is determined by the   potentials  for $\pmb \sigma$ that were already given in (\ref{dympotential},\ref{qcdadjpotential}). Our strategy now is to find their minima, already given in (\ref{dymminima},\ref{qcdadjminima}), that lie within the unit cell of $\Gamma_G$ and are left invariant under $\hat\gamma^{(k',q)}$.

%%%%%%%%%%%%%%%%%%%%%%%%
 \subsection{dYM}
%%%%%%%%%%%%%%%%%%%%%%%
 \label{dYM}
 
 According to (\ref{dymminima}), we have the minima $\langle \pmb \sigma \rangle_k = 2 \pi \pmb w_k, ~k=0,...N_c-1$ (modulo arbitrary $\Gamma_w$ shifts), where $\pmb w_0\equiv\pmb 0$. For an $SU(N_c)$ gauge group, the fundamental domain is $\Gamma_w$ itself, hence there is a unique minimum at the origin $\pmb w_0=\pmb 0$.
 
 \subsubsection{dYM for prime $\mathbf{N_c}$ and a physical picture}
 \label{dymprime}
 
  Apart from $SU(N_c)$, for prime $N_c$, one can only choose the gauge group  $SU(N_c)/\Z_{N_c}$. Then, the fundamental domain is $\Gamma_r$, where there are $N_c$ minima given by the origin and the $N_c-1$ fundamental weights (recall Fig.~\ref{fig:dymsu3}). The $[SU(N_c)/\Z_{N_c}]_q$ theories are distinguished by the action of $\hat\gamma^{(1,q)}$  of (\ref{centerk2}) which identifies various minima. 
 
For $q=0$, we can use $\pmb w_k - {\cal P} \pmb w_k = \pmb \beta_{1,k,+1} \in \Gamma_r$ (this follows from the ${\cal P}$ action on $\pmb w_k$ given earlier and $\pmb\beta_{ij}= \pmb e_i - \pmb e_j$).
 Thus the $q=0$ theory has $N_c$ vacua, as $\hat\gamma^{1,0}$ leaves each vacuum invariant (recall that the difference of two different fundamental weights is not a root). 
 
\begin{figure}[htbp] %  figure placement: here, top, bottom, or page
   \centering
   \includegraphics[width=1.01\textwidth]{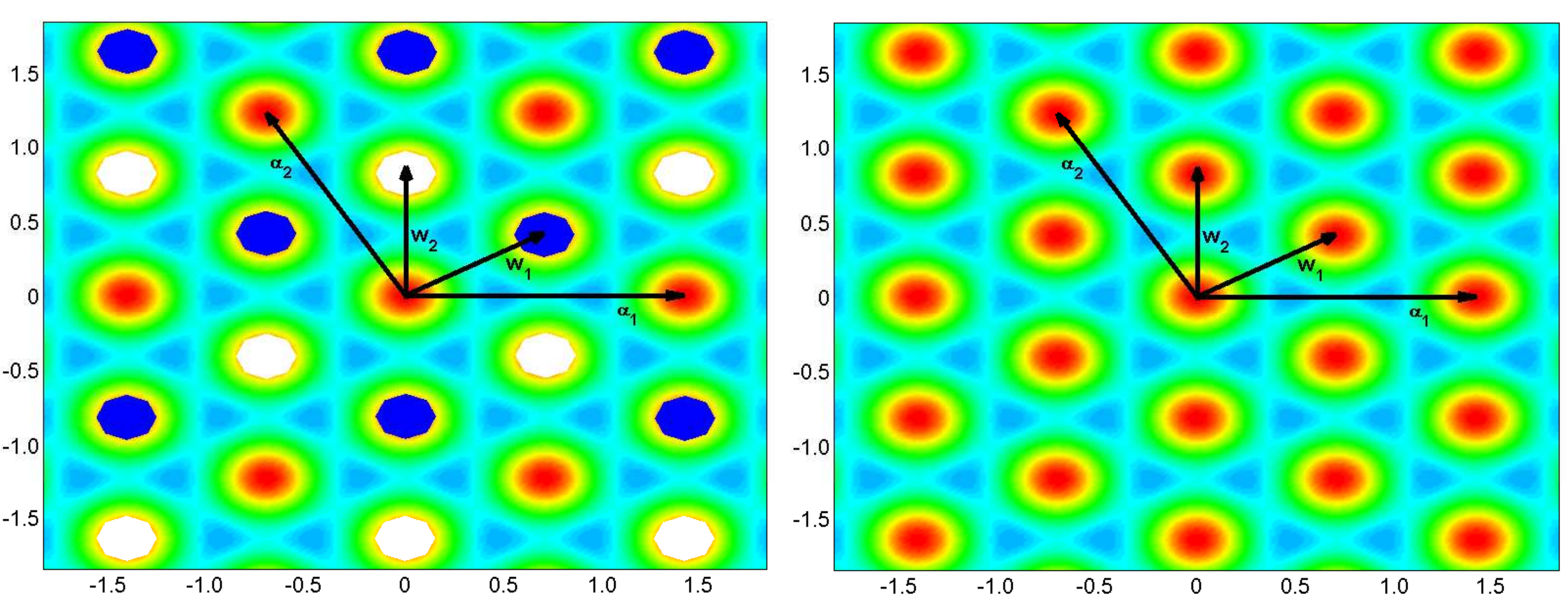} 
   \caption{ The identification of ground states by the action of $\hat\gamma^{(1,q)}$  for $[SU(3)/\Z_3]_q$. {\it Left:}~The $q=0$ theory has three vacua within the unit cell of the root lattice (shown by the solid ovals indicated by different colors).   There are true domain walls between them, consistent with the absence of confined local probes. {\it Right:}  The $q=1 \;(2)$ theories have all three vacua within $\Gamma_r$ identified, the ``domain walls" are now strings confining the $W H$ ($W^2 H$) local probes (and their powers).}
   \label{fig:psu3vacua}
\end{figure}

 For $q>0$, we notice that $ {\cal P}\pmb w_{k } +\pmb w_q = \pmb w_{k+q({\rm mod} N_c)} ({\rm mod}\; \Gamma_r)$. This implies that all $N_c$ minima within $\Gamma_r$ are identified under the action of $\hat\gamma^{(1,q)}$ with $q>0$ and thus the $[SU(N_c)/\Z_{N_c}]_{q>0}$ theories have unique ground states, as shown on the right panel of Fig.~\ref{fig:psu3vacua}  for $SU(3)$.

\begin{figure}[htbp] %  figure placement: here, top, bottom, or page
   \centering
   \includegraphics[width=.4\textwidth]{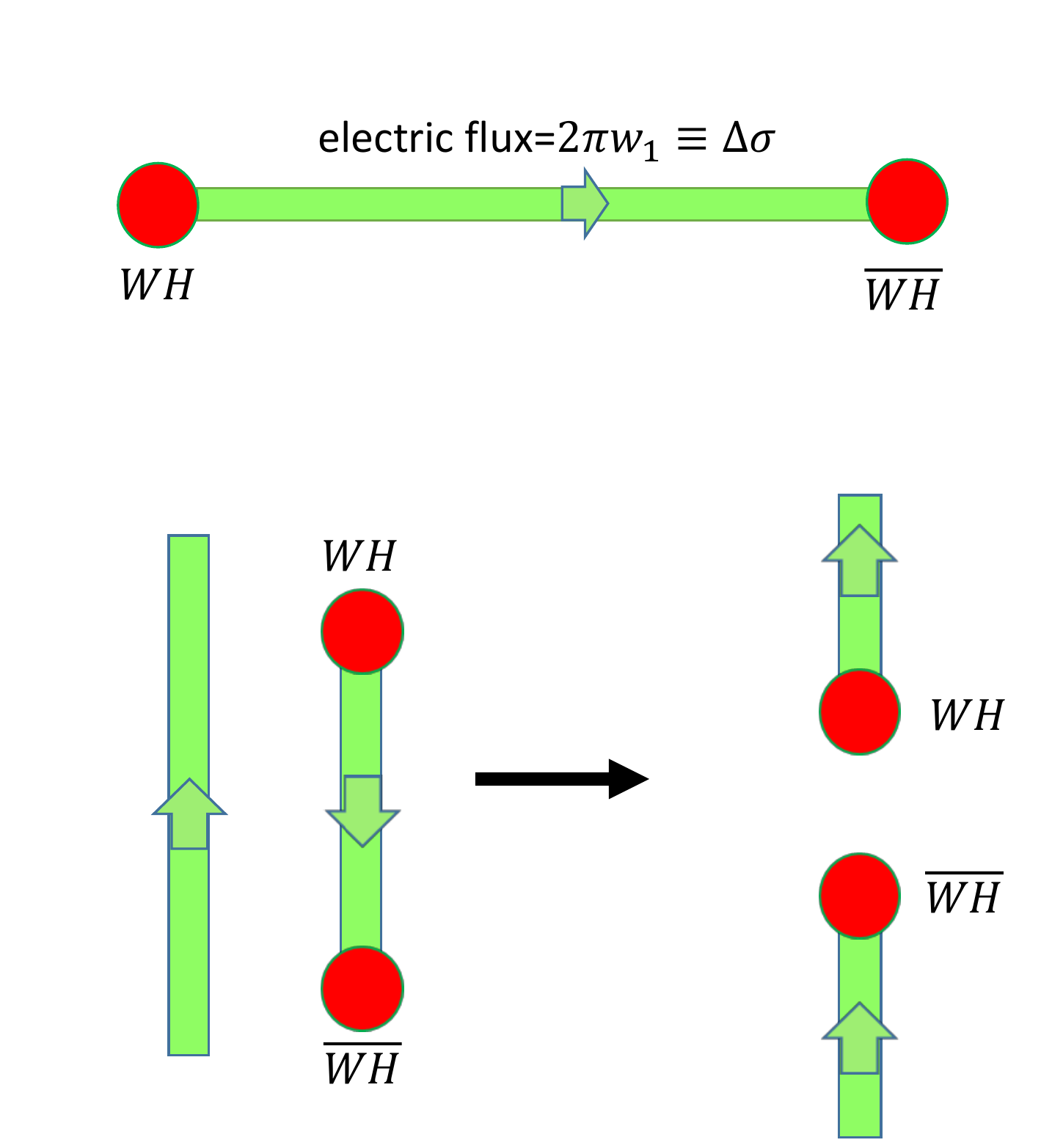} 
   \caption{{\it Top:} a string confining a local $WH$ probe in the $[SU(3)/\Z_3]_1$ theory is a domain wall like field configuration where $\pmb\sigma$ jumps by $\Delta\pmb \sigma = 2 \pi \pmb w_1$. The duality relation (\ref{v in terms of sigma}) shows that $\Delta\pmb \sigma$ is  the confined electric flux for quarks with charges in the highest weight of the fundamental. {\it Bottom}:  From left to right, a ``domain wall" interpolating between the $\pmb w_0$ and $\pmb w_1$  minima  is ``eaten" by  the creation of a $WH$--$\overline{WH}$ pair  connected by a confining string.}
   \label{fig:eatenwall}
\end{figure}

We now make some remarks on the vacuum structure we found:
\begin{enumerate}
 \item
First, we stress that the above counting of vacua is based on: {\it i.}) the understood confining dynamics of dYM at small-$L$ and {\it ii.}) the explicit $\hat{\gamma}^{(1,q)}$ action (\ref{centerk1}). 

Our counting of vacua  is consistent  with the heuristic picture for pure YM advocated in \cite{Aharony:2013hda}, as we review now. One begins with Seiberg-Witten (SW) theory (${\cal N}=2$ SYM softly broken to ${\cal N}=1$) $SU(N_c)$ theory on $\R^4$. SW theory has $N_c$ vacua where monopoles (one vacuum) or dyons ($N_c-1$ vacua) condense.  For an   $[SU(N_c)/\Z_{N_c}]_q$ gauge group SW theory has the same number of vacua on $\R^4$. However, $N_c-1$ of these vacua have area law for the genuine line operator $W^q H$ and only one vacuum has perimeter law. The perimeter law vacuum exhibits an unbroken $\Z_{N_c}$ emergent magnetic gauge symmetry (i.e.~the ``Higgs field" $W^q H$, really a line operator on $\R^4$, has charge unity, while the condensing objects have charge $N_c$). Upon compactification on $\R^3 \times \S^1$, the area law vacua persist, but the perimeter law vacuum is expected to split into $N_c$ distinct vacua, labeled by the expectation value of the $W^q H$ line operator winding around $\S^1$, which is now a local Higgs field.\footnote{We shall see that this counting, giving a total of $2N_c-1$ vacua for $[SU(N_c)/\Z_{N_c}]_q$ SYM with $N_c$-prime on $\R^3 \times \S^1$ is also valid for QCD(adj).} 

The relation to pure YM follows after turning on a small supersymmetry breaking gaugino mass, which selects, depending on its phase (as described in e.g.~\cite{AlvarezGaume:1996zr,Evans:1996hi}), one of the $N_c$ vacua on $\R^4$. For one of the $N_c$ theories with gauge group $[SU(N_c)/\Z_{N_c}]_q$,  this vacuum has perimeter law, while for the $N_c-1$ remaining ones  it has area law for the genuine line operators $W^q H$.\footnote{For an $SU(N_c)$ gauge group, there are only vacua with area law  for the genuine line operator $W$, hence one expects (after supersymmetry breaking) a unique vacuum for dYM on $\R^3 \times \S^1$, exactly as we found earlier in this section.} Upon compactification on $\R^3 \times \S^1$, one then expects that one of the $[SU(N_c)/\Z_{N_c}]_q$ theories (the one with perimeter law on $\R^4$) has $N_c$ vacua and the other $N_c-1$ theories have unique vacua. Further, if one assumes that this counting persists upon decoupling the gauginos and scalars  of SW theory, one arrives at a prediction for the number of vacua of pure YM on $\R^3 \times \S^1$. As our study shows, this counting is borne out by the dYM calculable dynamics.

 \item Second, we note that the vacuum structure can be understood using   the picture of confining strings on $\R^3 \times \S^1$ as  domain wall-like configurations in the noncompact $\R^3$, originated in \cite{Polyakov:1976fu}.  A domain wall-like configuration in the noncompact $\R^{3}$ can be either a confining string {\it or} a domain wall separating distinct vacua, but not both. Indeed, if a domain wall separating distinct vacua was also a confining string, one could imagine a process  (pictured on Fig.~\ref{fig:eatenwall}) whereby the domain wall would be ``eaten" by a pair production of the confined probes (presumed sufficiently heavy, but dynamical),  an event which contradicts  the existence of distinct vacua. Thus the multiplicity of ground states is directly correlated with the number of local probes with area law. In particular if there are no confined local probes, all domain wall like configurations should be true domain walls connecting distinct vacua.
  
 Consider for simplicity the $N_c=3$ case pictured on Fig.~\ref{fig:psu3vacua}. 
 
 For $q=0$, the  domain wall field configurations interpolating between $\pmb w_{0}, \pmb w_1$, and $\pmb w_{2}$ are true domain walls separating distinct vacua. That these are distinct vacua with true domain walls between them reflects  the fact that in this theory there is no area law for the genuine line operator $H$, i.e.~there are no confined local probes. Instead the expectation value  (i.e.~perimeter law) for the local (on $\R^3$) operator $H$ distinguishes  the three ground states.  
 
 For $q=1$ (or $2$), on the other hand, all three minima are identified. The domain wall field configurations interpolating between them are now confining strings. Indeed, the $W H$  (or $W^2 H$ for $q=2$) genuine line operators exhibit an area law on $\R^{3}$, determined by the tension of the appropriate ``domain walls" (between the different ``vacua" $\pmb w_{0,1,2}$). Recall that confinement on $\R^3$ is abelian  and the  precise map between the weights (charges) of the confined quarks and the ``domain wall" confining strings is, for dYM, simpler than the one for QCD(adj) from \cite{Anber:2015kea}. For example, a domain-wall configuration between the ``vacua" $\pmb w_1$ and $\pmb w_0$, i.e.~with ``monodromy" $\Delta \pmb \sigma = 2 \pi (\pmb w_1 - \pmb w_0)$ is a string confining fundamental quarks, whose weight is $\pmb w_1$---the electric part of the $WH$ operator for $q=1$.
 \end{enumerate}

 %%%%%%%%%%%%%%%%%%%%%%%%%%%%%%%%%
 \subsubsection{dYM with $\mathbf{[SU(N_c)/\Z_{N_c}]_q}$ for non-prime $\mathbf{N_c}$}
%%%%%%%%%%%%%%%%%%%%%%%%%%%%%%%%%%
 
 \label{dymnonprime}
 
 The modification from the discussion for prime $N_c$ is minimal. First, for $q=0$, there are $N_c$ ground states, as the vacuum identification is the same. For $q>0$ we still have the $k \leftrightarrow k+q$(mod $N_c$) vacuum identification  due to $ {\cal P}\pmb w_{k } +\pmb w_q = \pmb w_{k+q({\rm mod} N_c)} ({\rm mod}\; \Gamma_r)$. However, for gcd$(q,N_c) \ne 1$ the action of $\hat\gamma^{(1,q)}$ on the $N_c$ minima splits  into gcd$(q,N_c)$ orbits (each containing $N_c/$gcd$(q,N_c)$ minima), hence these theories  have gcd$(q,N_c)$  ground states. 
 
Physically, this split of the $N_c$ minima into orbits of the $\hat\gamma^{(1,q)}$ action reflects the fact the $[SU(N_c)/\Z_{N_c}]_q$-theory genuine line operator $(W^q H)^{N_c \over {\rm gcd}(q,N_c)}$ does not have area law as it has root-lattice charges and can be screened by $W$-boson pair creation (on $\R^4$ this holds in the appropriate vacuum, see below). The ``domain walls" between the $N_c/$gcd$(q,N_c)$ minima in each $\hat\gamma^{(1,q)}$-orbit are strings leading to area law for the genuine line operators $(W^q H)^{k}$, with $1\le k < N_c/$gcd$(q,N_c)$. 

The  simplest example is the $[SU(4)/\Z_4]_2$ theory where the two $\hat\gamma^{(1,2)}$ orbits of minima are ($\pmb w_0, \pmb w_2$) and ($\pmb w_1, \pmb w_3$). ``Domain walls" connecting the minima in each orbit are strings leading to area law for the $W^2 H$ genuine line operator. This is  clear from the fact that across such ``walls", $\Delta \pmb \sigma =2\pi( \pmb w_2 +$roots$)$, giving the correct confined electric flux for $N$-ality two representations  (recall the duality relation (\ref{v in terms of sigma})). On the other hand, the walls between the two sets of vacua have $\Delta \pmb \sigma =2\pi( \pmb w_1 +$roots$)$. They do not lead to area law for genuine line operators  and are true domain walls. On the other hand, the $[SU(4)/\Z_4]_{1 (3)}$ theories have unique ground state, implying that all domain walls are confining strings, leading to area law of the $W H$ ($W^3 H$) and its powers.

Finally, we note that the non-confined $(W^q H)^{N_c \over {\rm gcd}(q,N_c)}$  has nonzero magnetic $N$-ality and that the heuristic picture of \cite{Aharony:2013hda} also applies here. To see this, observe that the number of vacua we found corresponds to SW theory with a supersymmetry-breaking gaugino mass selecting the $\R^4$ vacuum with $(0,N_c)$  monopole (no electric charge) condensation. In this $\R^4$ vacuum, the $W^q H$ genuine line operator is confined, but $(W^q H)^{N_c \over {\rm gcd}(q,N_c)}$ has perimeter law. Hence there is  an emergent magnetic $\Z_{{\rm gcd}(q,N_c)}$ gauge symmetry, with unit charge for the $(W^q H)^{N_c \over {\rm gcd}(q,N_c)}$ genuine line operator and charge ${{\rm gcd}(q,N_c)}$ of the condensing $(0,N_c)$ monopole. The vacuum with an unbroken  $\Z_{{\rm gcd}(q,N_c)}$ symmetry on $\R^4$ is expected to split into ${\rm gcd}(q,N_c)$ vacua upon compactification,  consistent with our finding.
  
%%%%%%%%%%%%%%%%%%%%%%%%%%%%%%%%	
 \subsubsection{dYM with $\mathbf{[SU(N_c)/\Z_{k}]_q}$, with  $\mathbf{k k' = N_c}$}
%%%%%%%%%%%%%%%%%%%%%%%%%%%%%%%%
 \label{dymk}

Begin with the simplest case, that of an $SU(4)/\Z_2$ gauge group. The fundamental domain of $\pmb\sigma$, the group lattice $\Gamma_{SU(4)/\Z_2}$, is the lattice of all weights of $N$-ality $2$. For this theory, the identification of vacua is by (\ref{centerk2}) with  $k'=2$ and $q=0,1$ and the minima (\ref{dymminima}) within $\Gamma_r$ are 
 at $\pmb w_{0,1,2,3}$. Thus, for the $[SU(4)/\Z_2]_q$ theory, we have to identify $\pmb w_k \sim {\cal P}^2 \pmb w_k +  \pmb w_{q}$(mod $\Gamma_G$). The genuine line operators here are $(W^2, W^q H^2)$.

 For $[SU(4)/\Z_2]_0$ we thus find that 
 ($\pmb w_0, \pmb w_2$) as well as ($\pmb w_1, \pmb w_3$) are identified by $\Gamma_G$ shifts
 and there are two ground states. The domain wall configurations connecting minima within each orbit are strings responsible for the area law of the $W^2$ genuine line operator, while the walls between the two vacua (e.g.~with $\Delta \pmb\sigma = 2\pi \pmb w_1$) are genuine domain walls (neither $W$ nor $WH^2$ are genuine line operators here). The two vacua are distinguished by the vev of the genuine line operator $H^2$.
 
 On the other hand, for $[SU(4)/\Z_2]_1$, there is one $\hat\gamma^{(2,1)}$ orbit and a unique vacuum. All domain walls here are confining strings, reflecting the fact that both genuine line operators $W^2$ and $WH^2$ have area law. In particular the domain walls between $\pmb w_0$ and $\pmb w_1$ are now confining strings.
 
 It is easy to see that this pattern continues to the general case. 
 
For  $[SU(N_c)/\Z_k]_0$ theories, we find gcd$(N_c,k)$ vacua. Indeed the only genuine line operator with an area law is $W^k$, hence all minima among $\pmb w_0, \ldots \pmb w_{N_c-1}$ whose indices differ by $k$ (i.e.~by $N$-ality $k$) are identified. The ``domain walls" connecting them are strings leading to area law for the $W^k$ genuine line operator. There are exactly gcd$(N_c,k)$ unidentified vacua left, labeled by $\pmb w_0,\ldots \pmb w_{{\rm gcd}(N_c,k)}$. These are connected by genuine domain walls---no genuine line operators of such $N$-alities exist for the $q=0$ theory. 
   
 For the $[SU(N_c)/\Z_k]_{1\le q < k}$ theories, on the other hand, we have gcd$(N_c,q)$ minima, $\pmb w_0, \ldots$ $\pmb w_l $$\ldots  \pmb w_{{\rm gcd}(N_c,q)-1}$ not identified under $l \leftrightarrow (l+q)$(mod$N_c)$ (i.e.~these are representatives of the $\hat\gamma^{(k',q)}$ orbits). Imposing  identification by $N$-ality $k$ shifts does not further restrict the number of vacua as gcd$(N_c,q)<k$ for $q<k$.
 This is also consistent with the string/domain wall dichotomy as there are no genuine line operators among $(W^k, W^q H^{k'})$ with an area law and $N$-alities smaller than gcd$(N_c,q$) and the domain walls between these vacua are genuine.
 
%%%%%%%%%%%%%%%%%%%%%%%%%%
 \subsection{dYM on $\mathbf{\R^2 \times \S^1_\beta \times \S^1_L}$, Kramers-Wannier duality  and global structure}
 \label{thermal}
%%%%%%%%%%%%%%%%%%%%%%%%%%%%%

We now consider a further compactification on $\S^1_\beta$, with $\beta = 1/T$. We do this because the effective description of the thermal theory  in the low temperature regime $\beta \gg L$  of \cite{Simic:2010sv}  exhibits interesting duality properties, not much noted before, except for some remarks in  \cite{Anber:2011gn}. There is an interesting interplay with  the global structure of the gauge group which was not properly discussed  earlier \cite{Anber:2011gn, Anber:2012ig}.

 The dynamics relevant to the finite temperature theory is as follows. The  monopole-instanton gas (with constituents labeled by the affine roots of $SU(N_c)$) remains intact in the low temperature limit $\beta \gg L$ (recall that monopole-instanton core size is $L$). In addition, at finite temperature, the $W$-bosons, the lightest $N_c$ types of which have mass $1\over N_cL$, can also appear with Boltzmann probability. Ref.~\cite{Simic:2010sv} (see also the earlier work of \cite{Dunne:2000vp} on a similar description in the Polyakov model, and also the work \cite{Anber:2013xfa} for other perspective) argued that the thermal partition function of dYM reduces to a two-dimensional ``classical" electric-magnetic Coulomb gas of $W$-bosons and monopole-instantons and that this gas exhibits a deconfinement  phase transition at $T_c = {g^2 \over 4 \pi L}$. Qualitatively, at low-temperatures magnetic charges (the monopole-instantons) are dominant, causing screening of magnetic charge and confinement of electric charges. At high-temperatures, dominance of electric charges (the $W$ bosons) sets in, causing screening of electric charge and confinement of magnetic charge.

\begin{figure}[htbp] %  figure placement: here, top, bottom, or page
   \centering
   \includegraphics[width=1\textwidth]{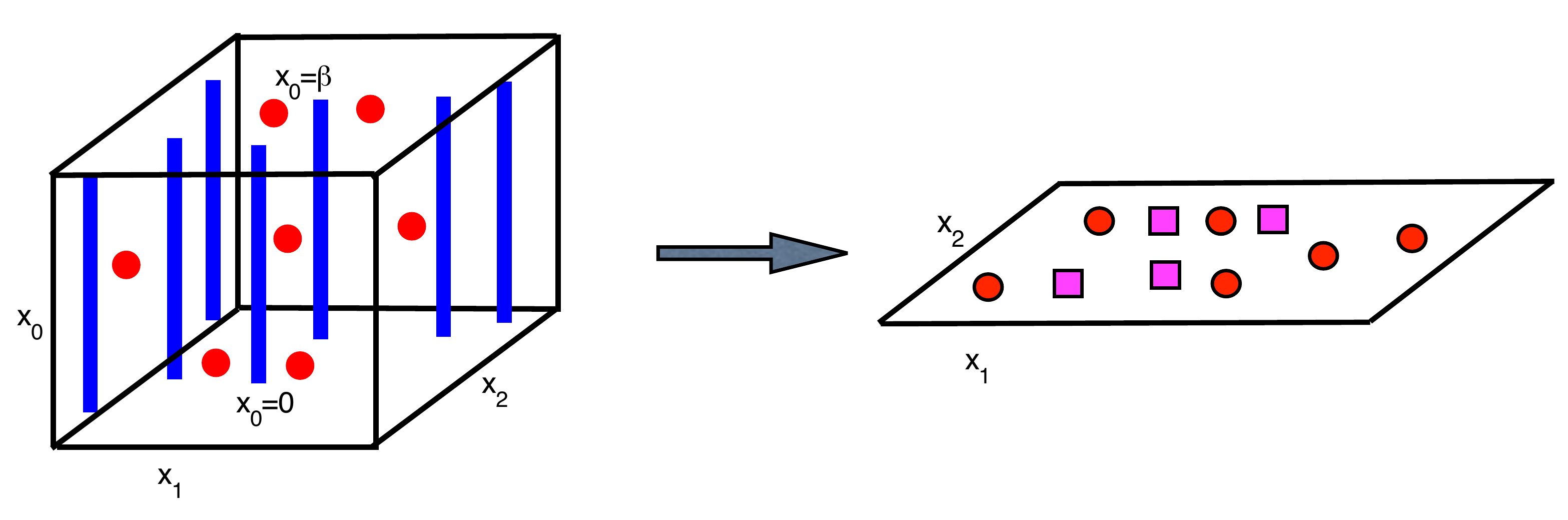} 
   \caption{A typical configuration in the gauge theory on $\R^2 \times \S^1_\beta$, with the much smaller $\S^1_L$ not shown.  Electric $W$-bosons  propagate along  static worldlines extending along $\S^1_\beta$ as shown on the picture. The   magnetic monopole-instantons, shown by dots, are localized in $\R^2$ and the Euclidean time direction and are extended along $\S^1_L$. Both gases are dilute in the  $m_W = { 1\over N_c L} \gg T$ regime. Further,  their separations are exponentially larger [this is not clear from the scale of the picture] than the   extent of the compact time direction, the inverse temperature $\beta$. The gas of monopole-instantons and $W$-bosons thus appears two dimensional and is described by the Coulomb-gas partition function  (\ref{sunpartition11}). Clearly, the duality (\ref{duality1}) exchanging  electric and magnetic objects emerges only in the 2D limit  as the two kinds of charges have distinct microscopic origin.  }
   \label{fig:2dgas}
\end{figure}

 Before we give the expression for the thermal partition function, on  Fig.~\ref{fig:2dgas} we  show a picture of a typical configuration of gauge theory objects contributing to the $\R^2 \times \S^1_\beta \times \S^1_L$  path integral. The  rationale for the dimensional reduction to (\ref{sunpartition11}) is also explained in the  caption. The description of the gauge theory by a dimensionally reduced partition function is valid for low temperatures, $m_W = { 1\over N_c L} \gg T$,  and  the usual $\Lambda_{\mbox{\scriptsize QCD}} L  N_c \ll 1$ condition for the validity of semiclassics is  assumed.
There are further corrections,  suppressed by  these two small parameters, to the dimensionally reduced partition function  (\ref{sunpartition11}), see \cite{Anber:2013doa} for a detailed discussion. 
 
Now, without much ado (see \cite{Simic:2010sv}, also \cite{Anber:2011gn} for the derivation), we write the partition function and explain the ingredients and notation in some detail:
\begin{eqnarray}
\label{sunpartition11}
\nonumber
Z &=& \sum\limits_{(N^i_{e \pm}, N^j_{m \pm} \geq0)}\sum\limits_{(   i\geq0,\; q^m_a=\pm1 )} \sum\limits_{( j\geq0,\; q^e_A=\pm1)}\frac{ y_m^{  \sum_{i}(N^i_{m +} + N^i_{m -})}  y_e^{  \sum_{i}(N^i_{e +} + N^i_{e -})} }{ \prod\limits_i  N^i_{m+}!N^i_{m-}!N^i_{e+}!N^i_{e-}! }   \int \prod_{a,i} d^2R_a^{i}\int \prod_{A,j}d^2R_A^{j}
\end{eqnarray}
\begin{eqnarray}
\nonumber 
&&\times\exp\left[{g^2 \over  4 \pi L T} \sum_{i\geq j}^{N_c}\sum_{A>B}^{N_e}q^e_A q^e_B\;  \pmb\alpha_i\cdot \pmb\alpha_j\ln |\vec R^i_A-\vec R_B^j|+\frac{{4 \pi L T}}{g^2}\sum_{i\geq j}^{N_c}\sum_{a>b}^{N_m}q^m_aq^m_b\; \pmb\alpha^*_i\cdot \pmb\alpha^*_j\ln |\vec R^i_a-\vec R_b^j| \right.\\
&&\qquad~~ \left.+ \;  i\;\sum_{i,j}^{N_c}\sum_{a,B}^{N_m,N_e}q^m_aq^e_B\; \pmb\alpha_j\cdot \pmb\alpha^*_i \;\Theta(\vec R^i_a-\vec R_B^j)    \right]\,. 
\end{eqnarray}
The dynamical objects in this 2D grand partition function are as follows. There are $N_c$ types of magnetically charged particles and anti-particles ($q^m=\pm 1$)---the magnetic monopole-instantons---labelled by their magnetic charges $\pmb \alpha^*_i$, $i=1,...,N_c$, the affine co-roots. There are also $N_c$ types of electrically charged particles and antiparticles ($q^e=\pm 1$)---the lightest degenerate $W$-bosons---labelled by their electric charges $\pmb \alpha_i$, $i=1,...,N_c$, the affine roots.\footnote{This is the one place in the paper where it is convenient to differentiate roots $\pmb \alpha_i$ (labeling electric charges) and co-roots $\pmb \alpha_i^*$ (labeling magnetic charges).}
The sums in (\ref{sunpartition11}) are  over all possible distributions and numbers of the electric and magnetic charges described above. The   magnetic and electric  fugacities are $y_m \sim {1\over L^{3} T} e^{- {8 \pi^2 \over g^2 N_c}}$ and $y_e \sim m_w T e^{- {m_W\over T}} = {T \over LN_c} e^{- {1 \over N_c L T}}$. The particles interact via: {\it i.}) 2D electric Coulomb law, with strength ${g^2 \over 4 \pi L T} \pmb \alpha_1 \cdot \pmb \alpha_2$ (the subscripts label the particles $1$ and $2$, rather than the first and second root), {\it ii.}) 2D magnetic Coulomb law, with strength ${4 \pi L T  \over g^2 } \pmb \alpha^*_1 \cdot \pmb \alpha^*_2$, and {\it iii.}) Aharonov-Bohm phase interactions, with exchange phases $\pmb \alpha_1 \cdot \pmb \alpha^*_2 \Theta_{12}$, where $\Theta_{12}$ is the angle between the $x$-axis and the vector from particle $1$ to particle $2$. 

    Having explained the physics behind the emergence of (\ref{sunpartition11}) as a description of the gauge theory on $\R^2 \times \S^1_\beta \times \S^1_L$, at $\beta \gg L$, we now note an interesting feature---the self-duality of the electric magnetic Coulomb gas. 
An inspection of Eq.~(\ref{sunpartition11}) shows that the effective theory is invariant under electric-magnetic duality (which we label by $\hat{S}$) acting as
 \beq
 \label{duality1}
\hat{S}: ~ (y_m, y_e) \rightarrow (y_e, y_m) ~,~~ (q^e \pmb\alpha_i, q^m \pmb\alpha_i^*)  \rightarrow (q^m \pmb \alpha_i^*, - q^e \pmb \alpha_i) ~, ~~{g^2 \over  4 \pi L T} \rightarrow \frac{{4 \pi L T}}{g^2}~,
 \eeq
 as well as an interchange of the coordinates of electric and magnetic charges.\footnote{We note that the partition function can be cast into the form of a self-dual sin-Gordon model, whose critical features have been studied in \cite{Lecheminant:2002va}; for related works see \cite{Kovchegov:2002vi,Lecheminant:2006hj,Anber:2012ig}.}
 Notice that (\ref{duality1}) acts as both electric-magnetic and high-T/low-T (Kramers-Wannier) duality. We stress again that we do not claim that (\ref{duality1}) is a fundamental (i.e.~all-scale) electric-magnetic duality in pure (d)YM
theory. Invariance under $\hat S$ is only a property of the long-distance effective theory of dYM on $\R^2 \times \S^1_\beta \times \S^1_L$ valid in the regime discussed above. Nonetheless, we shall see that with respect to the global structure of the theory,  (\ref{duality1}) has properties common with both Kramers-Wannier duality in the Ising model and strong-weak coupling duality in ${\cal N}=4$ SYM. We labeled (\ref{duality1}) $\hat{S}$ to underlie similarities with the latter case.\footnote{One notable distinction is that our $\hat{S}$ holds only for gauge theory $\theta$ angle $0$ or $2 \pi$. For nonzero $\theta$, phases appear in the fugacities of various monopole-instantons (see \cite{Unsal:2012zj}) but not in the $W$-boson fugacities. Notice that while these  $\theta$-dependent phases can be thought of as the analogue of the Witten effect for monopole-instantons (in the Euclidean sense of \cite{Kapustin:2005py}) they do not lead to electric charge of the monopole-instantons---as these are instantons with worldlines around $\S^1_L$, one obtains instead, in addition for the $\theta$-dependent phases shown in (\ref{dympotential}), a $\theta$-dependent $\pmb v_3$ (or $\pmb \phi$) charge. This charge is irrelevant for the dynamics because $\pmb\phi$ is gapped. }

 Before we discuss global structure, let us study the observables in  the effective theory (\ref{sunpartition11}). Since $Z$ describes a system of electric ($\pmb \alpha_i$) and magnetic ($\pmb \alpha^*_i$) charges, the  natural observables are correlation functions of external electric (of weights $\pmb \nu_e$)  and magnetic (of weights $\pmb \nu_m$) charges as a function of their separations. 
 In order to not introduce new formalism (see e.g.~\cite{Kadanoff:1978ve,Kovchegov:2002vi,Lecheminant:2002va}), as we will only need the results and a physical picture, we define the probes  via (\ref{sunpartition11}). Let us introduce, as in (\ref{WandT}), the fundamental Wilson and 't Hooft loops $W(\pmb w_1, \vec{r}) = ``e^{i \pmb w_1 \cdot \pmb\phi(\vec{r})}"$ and $H(\pmb w_1^*,\vec{r}) = ``e^{-i \pmb w_1^* \cdot \pmb \sigma(\vec{r})}"$, where $\vec{r} \in \R^2$ and the quotation marks appear because $\pmb \phi, \pmb\sigma$ are not  variables appearing in (\ref{sunpartition11}).  
 We define the operators via their correlation functions. For example, the two point function of $H$  and its antiparticle $\bar{H}$ (whose charge is $- \pmb w_1^*$) is  defined as the insertion of two external probe magnetic charges into (\ref{sunpartition11})
 \begin{eqnarray}
\label{sunpartition2}
\nonumber
\langle H(\pmb w_1^*, \vec{0}) \bar{H} (\pmb w_1^*, \vec{r}) \rangle &=&\bigg\langle \exp\left[
 \frac{{4 \pi L T}}{g^2}\left( - \pmb w^*_1\cdot \pmb w_1^* \ln |\vec r| + \sum_{i}^{N_c}\sum_{a}^{N_m}q^m_a \; \pmb w_1^* \cdot \pmb\alpha^*_i\ln |\vec R^i_a| + \ldots \right) \right.\\
&&\qquad~~ \left.+ \;  i\;\sum_{j}^{N_c}\sum_{ B}^{N_e} q^e_B\; \pmb\alpha_j\cdot \pmb w_1^*  \;\Theta(-\vec R_B^j)    + \ldots \right]\,\bigg\rangle~.
\end{eqnarray}

 It is easier to explain the physics  than to write down all terms or all correlators. The expectation value in (\ref{sunpartition2}) is taken with $Z$ from (\ref{sunpartition11}). The terms in the exponent on the top line are the magnetic Coulomb attraction between the two external charges and the interaction of the charge at $\vec{0}$ with all magnetic charges in the gas (the interaction between the charge at $\vec{r}$ and the magnetic charges in the gas is shown by dots). The bottom line shows the Aharonov-Bohm phase between the charge at $\vec{0}$ and the electric charges in the gas (again, omitting the phases for the charge at $\vec{r}$). 
 It is clear now that to define arbitrary correlation functions of $W(\pmb w_1, \vec{r})$'s and $H(\pmb w_1^*,\vec{r})$'s one simply has to keep track of all interactions between the external charges and between the external charges and the particles in the gas and take an expectation value using the grand partition function (\ref{sunpartition11}). Similarly, one can define correlation functions of the more general dyonic operator ${\cal D}(\pmb \nu_e, \pmb \nu_m, \vec{r},\theta)$
of (\ref{operators1}). Notice also that, as in the gauge theory, $H(\pmb w_1^*)$ and $W(\pmb w_1)$ are not mutually local with respect to each other (the Aharonov-Bohm phase interaction between them would be $e^{i \pmb w_1^* \cdot \pmb w_1 \Theta}$, which would change by a $\Z_{N_c}$ phase upon $\Theta \rightarrow \Theta + 2 \pi$). 
  
In our further remarks on the global structure, for brevity,  we shall explicitly consider the $su(2)$ case only. We also  drop the $\pmb w_1$ argument in $H$ and $W$  (the questions that arise from the observation of (\ref{duality1}) and their resolution are similar for the higher-rank cases). One finds, upon studying correlation functions using various dual representations of the Coulomb gas \cite{Simic:2010sv}  that at $r\rightarrow \infty$
\begin{equation}
\label{thooftthermal}
\langle H(r) \bar{H}(0)\rangle\big\vert_{r\rightarrow \infty}  = \left\{ \begin{array}{cl}   e^{ - {\sigma r \over T}}, & {\rm thus} \;  \langle H \rangle = 0 \;  {\rm for} \; T> T_c= {g^2 \over 4 \pi L}~,\\
 1,   & {\rm thus} \; \langle H \rangle = \pm 1  \; {\rm for} \; T< T_c ~,  \end{array}   \right.
\end{equation}
and 
\begin{equation}\label{wilsonthermal}
\langle W(r) \bar{W}(0)\rangle\big\vert_{r\rightarrow \infty}  = \left\{ \begin{array}{cl}   1, &   {\rm thus} \;  \langle W \rangle = \pm 1 \; ~ {\rm for} \; T> T_c= {g^2 \over 4 \pi L}~,\\
 e^{ - {\sigma r \over T}},  &  {\rm thus} \;  \langle W \rangle =  0 \; ~ ~ {\rm for} \; T< T_c ~. \end{array}   \right.
\end{equation}
   The question that arises is the consistency of these results  with the global structure of the gauge group.  
    For an $SU(2)$ gauge group, the genuine line operator is $W$. In the $T<T_c$ confining phase  there is a unique ground state $\langle W\rangle =0$, as per Section \ref{dymprime} and from (\ref{wilsonthermal}). At $T>T_c$, it is well known from thermal field theory that there are two, labelled by the expectation value of the fundamental Polyakov loop $W$ wrapped around $\S^1_\beta$ and breaking the zero-form $\Z_2$ center symmetry. This is also seen in (\ref{wilsonthermal}). A puzzle, similar to the one asked for the Ising model in \cite{Kapustin:2014gua} arises: since the number of ground states of an $SU(2)$ theory on the two sides of the Kramers-Wannier duality (\ref{duality1}) is different, the effective long-distance description (\ref{sunpartition11}) can not be self dual. 
    
    The resolution, also similar to \cite{Kapustin:2014gua}, is that the high-$T$ dual of the $SU(2)$ theory is an $SU(2)$ theory coupled to a discrete topological field theory, or, equivalently, an $SO(3)_+$ theory. 
To argue for this, consider the $SO(3)_+$ gauge theory, where the genuine line operator is $H$. At $T<T_c$, $H$ wrapped around $\S^1_L$  has a vev breaking the $\Z_2$-magnetic center symmetry and there are two vacua, as described in Section \ref{dymprime} (and is also seen in (\ref{thooftthermal})). This is the $\hat{S}$ dual of the high-$T$ phase of the $SU(2)$ theory. At $T>T_c$, on the other hand, there is a unique ground state as $H$ has an area law in the deconfined phase\footnote{For a study of 't Hooft loops in  thermal gauge theory, see~\cite{KorthalsAltes:1999xb}.}  because monopoles are confined in the electric plasma phase, as per (\ref{thooftthermal}). This is the $\hat{S}$ dual of the low-$T$ phase of the $SU(2)$ theory. Thus $\hat S$-duality of the effective theory (\ref{sunpartition11}) acts by interchanging $SU(2)$ with $SO(3)_+$, and  $H$ with $W$. 
  \begin{figure}[htbp] %  figure placement: here, top, bottom, or page
   \centering
   \includegraphics[width=1
   \textwidth]{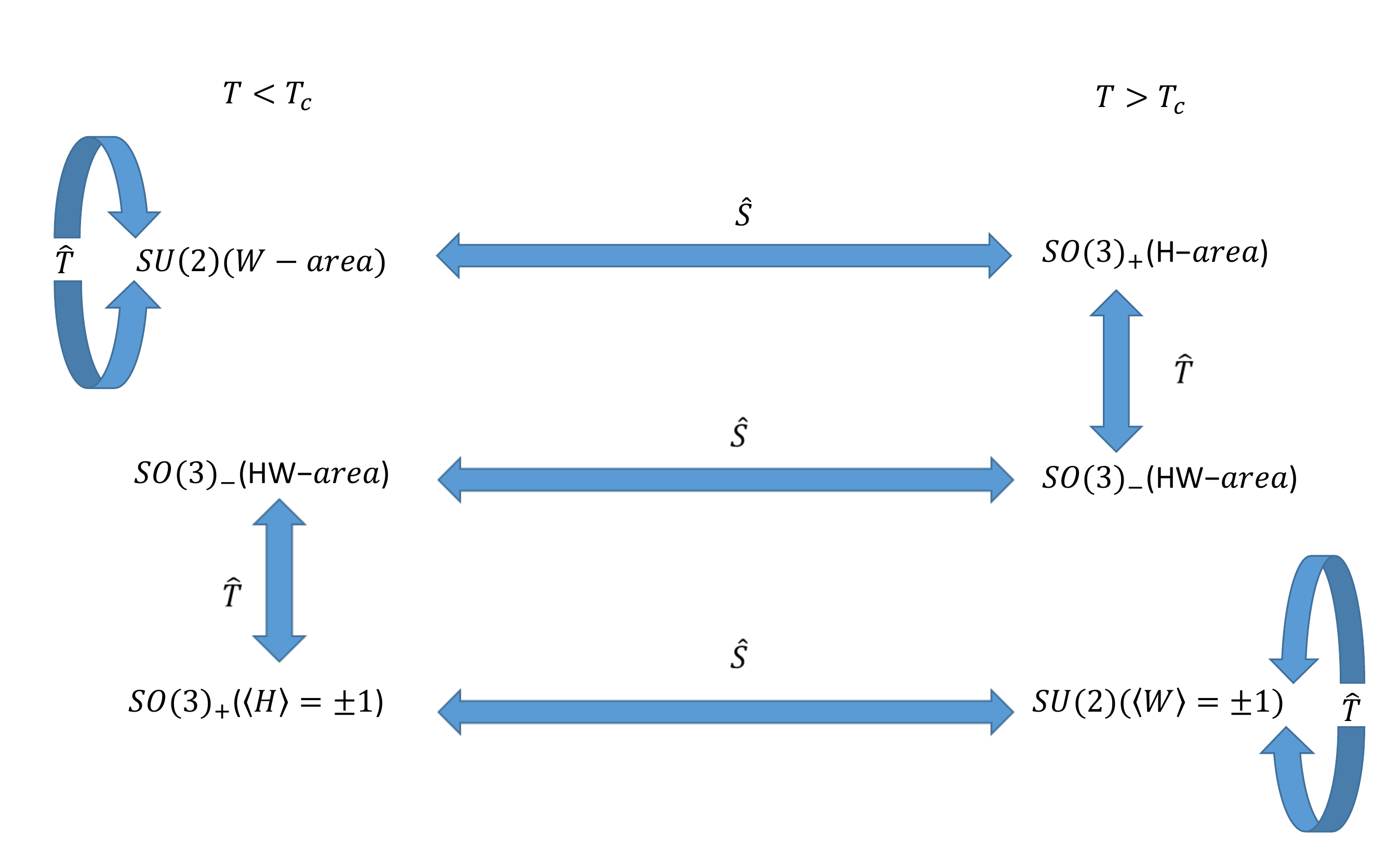} 
   \caption{The action of the Kramers-Wannier duality of the effective theory (\ref{sunpartition11}) on gauge theory observables.   $\hat{S}$ of Eq.~(\ref{duality1}) interchanges theories with different global structure. While the action of $\hat{S}$ and $\hat{T}$ is superficially similar to that in $N=4$ SYM, our $\hat{S}$  duality only holds for $\theta = 0$(mod $2 \pi$). }
   \label{fig:thermal}
\end{figure}

For the $SO(3)_-$ gauge theory, the genuine line operator is $WH$. At $T<T_c$, as already described, there is a unique ground state corresponding to the fact that $WH$ (its electric component) is confined in the monopole plasma. 
At $T> T_c$, there is also a unique ground  state as the magnetic component of $WH$ is confined in the $W$-boson plasma. We conclude that $SO(3)_-$ is self dual with respect to $\hat{S}$ with  the genuine line operator $WH$  mapped to itself. 
  \begin{figure}[htbp] %  figure placement: here, top, bottom, or page
   \centering
   \includegraphics[width=1\textwidth]{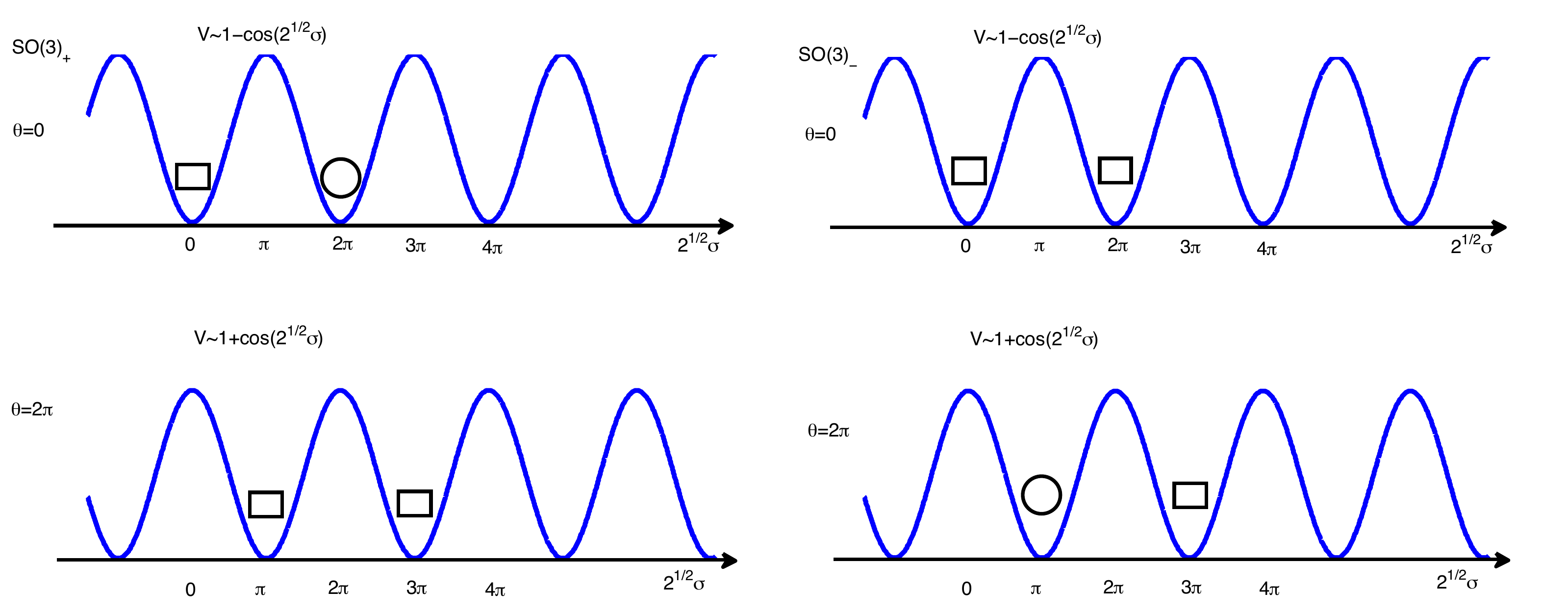} 
   \caption{The potential (\ref{dympotential}) for $N_c=2$ for $\theta = 0$ ({\it top}) and $\theta=2\pi$ ({\it bottom}) as a function of $\sqrt{2} \sigma$. The domain of $ \sqrt 2\sigma$ is $ \sqrt 2\sigma\sim\sqrt 2 \sigma+4\pi$. {\it Left:}~{\bf  $\mathbf{SO(3)_+}$ dYM theory}:  obtained by the identification $\sqrt 2\sigma\rightarrow -\sqrt 2 \sigma$. On the top figure, $\theta = 0$, the  two  minima  at $0$ and $2 \pi$ are distinct, not  identified by $\Z_2$ and $4\pi$-periodicity (one is indicated by a square and the other by a circle). For $\theta = 2 \pi$, the two  minima at $\pi$ and $3 \pi$ are identified under the $\Z_2$ and $4 \pi$ periodicity (hence both are indicated by a square). Thus,  this is now a theory with a single vacuum and confining strings instead of domain walls, as per  Section \ref{dymprime}, i.e.~the $SO(3)_-$ theory. {\it Right:}~{\bf $\mathbf{SO(3)_-}$ dYM theory}: obtained by the identification $\sqrt 2\sigma\rightarrow -\sqrt 2\sigma+2\pi$.  For $\theta = 0$, the two minima are identified, but for $\theta = 2 \pi$ they are distinct, indicating the absence of confining strings; this is  thus  the $SO(3)_+$ theory.    }
   \label{fig:so3potentials}
\end{figure}

Thus, the picture that emerges is that the action of the $\hat S$ duality (\ref{duality1}) in the effective theory (\ref{sunpartition11})  is very similar to the action of $S$-duality in $N=4$ SYM, as we show on Fig.~\ref{fig:thermal}. 
The $\hat{T}$ transformation represents a $\theta$-angle shift by $2 \pi$ which exchanges the $SO(3)_\pm$ theories
and leaves the $SU(2)$ theory invariant. The fact that  $SO(3)_\pm$ theories are interchanged by a $2\pi$ shift of $\theta$ also follows by studying the minima of the potential (\ref{dympotential}) in the $\Gamma_r$ fundamental domain for $\theta = 0$ vs. $\theta = 2\pi$. For $N_c=2$, the potential (\ref{dympotential}) is $V(\sigma, \theta)\sim 2 - 2 \cos {\theta\over 2} \; \cos \sqrt{2} \sigma$, using $  \alpha_1 = -   \alpha_2 = \sqrt{2}$. In the $SU(2)$ theory, the $\Gamma_w$ fundamental domain is $\sigma \sim \sigma + {2 \pi/\sqrt{2}}$, as $ w_1 = 1/\sqrt{2}$. We observe that the potential has a unique minimum within the fundamental domain regardless of the value of $\theta$, and so the $SU(2)$ theory has a unique ground state (except at $\theta = \pi$, see \cite{Unsal:2012zj}). On the other hand, in the $SO(3)_\pm$ theories, we have periodicity in the twice larger $\Gamma_r$:  $\sigma \sim \sigma + 2 \sqrt{2} \pi$. Further,  for $SO(3)_+$ we have  the identification $\sigma \rightarrow - \sigma$ (the    action of $\cal{P}$ for $su(2)$) and,  for $SO(3)_-$: $\sigma \rightarrow - \sigma + {2 \pi/\sqrt{2}}$. An inspection of the potentials on Fig.~\ref{fig:so3potentials}, plotted  for $\theta=0$ and $2 \pi$, shows if the $\theta = 0$ theory has  one ground state, the $\theta = 2 \pi$ has two and vise versa.

%%%%%%%%%%%%%%%%%%%%%%%%%%%%%%%%%%%%%%%%%%%%%%
\subsection{QCD(adj)}
%%%%%%%%%%%%%%%%%%%%%%%%%%%%%%%%%%%%%%%%%%%%%%
\label{adj}

 According to (\ref{qcdadjminima}), we have the minima $\langle \pmb \sigma \rangle_k = {2 \pi k \pmb \rho \over N_c}, ~k=0,...N_c-1$ (modulo arbitrary $\Gamma_w$ shifts). For an $SU(N_c)$ gauge group, the fundamental domain is $\Gamma_w$ itself, hence there are $N_c$ ground states related by the broken chiral $\Z_{N_c}$ symmetry.
  Next, we follow the same strategy as in dYM. We shall be brief and less general and only consider $N_c= 2,3,4$.\footnote{This is because, while  the combinatorics of    identification of  the minima (\ref{qcdadjminima}) in the case of QCD(adj)   is manageable and can potentially be automated, as opposed to the dYM case, we have not found an efficient way to treat all $N_c$ and $k$. }  These three classes of theories provide examples of all cases considered in dYM.
  
%%%%%%%%%%%%%%%%%%%%%%%%%%%%%%%%%%%%%%%%%%%%%%%
\subsubsection{Theories with $\mathbf{su(2)}$ algebra}
%%%%%%%%%%%%%%%%%%%%%%%%%%%%%%%%%%%%%%%%%%%%%%%
\label{adjsu2}

\begin{figure}[htbp] %  figure placement: here, top, bottom, or page
   \centering
   \includegraphics[width=.9\textwidth]{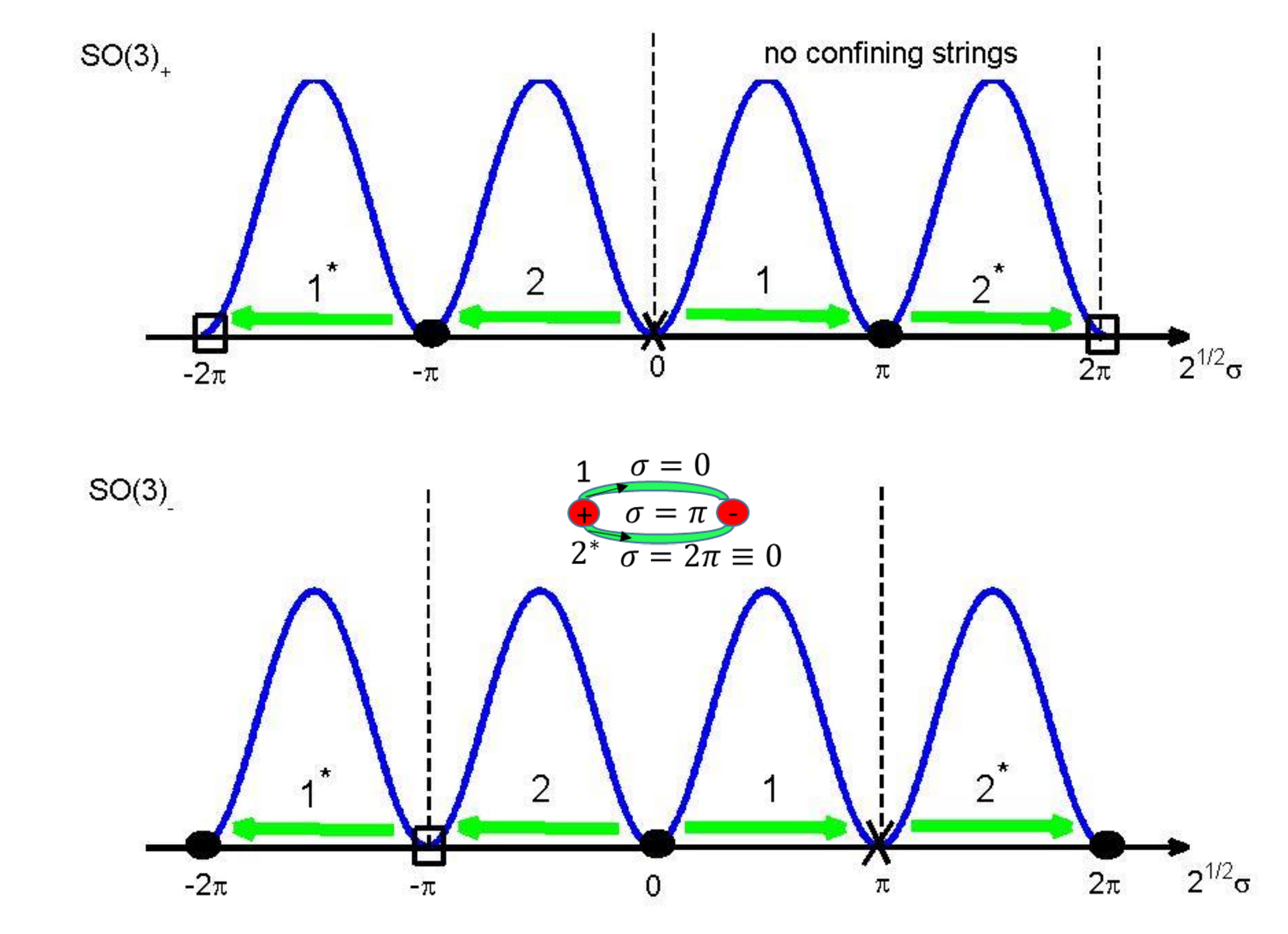} 
   \caption{{\it Top:}~{\bf $\mathbf{SO(3)_+}$ QCD(adj)}: The vacua are at $\sqrt 2 \sigma = 0, \pi, 2 \pi$ (this follows from the vacuum identification  $\sqrt 2 \sigma \sim \sqrt 2 \sigma + 4\pi$ and $\sqrt 2 \sigma \sim -  \sqrt 2 \sigma$, with a fundamental domain denoted by vertical dashed lines). It is not possible to construct confining string configurations, which are now necessarily made out of two domain walls (in order to carry the right electric flux), in any of the three vacua. The domain walls labeled by $1$ and $2$ are, in the case of SYM, the two known $SU(2)$ BPS domain walls. {\it Bottom:}~{\bf $\mathbf{SO(3)_-}$ QCD(adj)}:  the vacua are at $\sqrt 2\sigma = - \pi, 0, \pi$ (the identification is $\sqrt 2 \sigma \sim \sqrt 2 \sigma + 4\pi$ and $\sqrt 2 \sigma \sim -  \sqrt 2 \sigma + 2 \pi$). Composite strings  confining the electric part of the genuine line operator $WH$ are allowed in every vacuum (a confining string in the $\sigma = 0$ vacuum is pictured, see also Figure~\ref{fig:doublestring}). }
   \label{fig:so3adjoint}
\end{figure}

We begin by illustrating the simplest example: theories with gauge group $SO(3)$. This case can be worked out explicitly and relatively briefly. We shall use it to illustrate the main points and to connect with the study of $SO(3)$ supersymmetric theories    \cite{Aharony:2013kma}.\footnote{\label{f11}
The weights of the fundamental and adjoint (i.e. the nonzero roots $\pmb \alpha$) representations are given by (in this simplest case it is easier to revert to an $r$-component basis) $
\pmb \nu=\pm\frac{1}{\sqrt{2}}\,,\quad \pmb \alpha=\pm\sqrt{2}\,.$}
 In this case  the magnetic weights have to obey (\ref{diracrelation}) with $\pmb \nu_e$ in the root lattice, hence $\pmb \nu_m$ is in the weight lattice. Now, with $k=2$ and $k'=1$, we see from (\ref{choices}) that there are two   choices of commuting dyonic operators in this case, given by $(1, H) \sim (W^2, H)$ and $(1, WH)$, respectively.
More explicitly, in  the $(1, H)$ case, called $SO(3)_+$,   the lowest charge probes are purely magnetic ones with weights of the fundamental representation. In the other, $SO(3)_-$  case,  the lowest charge probe is dyonic.
The $SO(3)_-$ and $SO(3)_+$ theories are also labeled by $[SU(2)/\Z_2]_0$ and $[SU(2)/\Z_2]_1$, respectively. This classification of the probes is exactly as in dYM. In this simple case, it is easier to plot the potential (\ref{qcdadjpotential}), which after using the root form footnote \ref{f11}, up to a constant, is  $V(\sigma) = 1 - \cos 2 \sqrt 2 \sigma$, plotted on Fig.~\ref{fig:so3adjoint} as a function of $\sqrt 2 \sigma$, a variable with  periodicity $4 \pi$. 
 
 For the $SO(3)_+$ theory, it is easy to see from the identifications given on the figure that there are three distinct vacua, at $\sqrt 2 \sigma = 0, \pi, 2 \pi$ (indicated by different symbols; the fundamental domain of $\sqrt 2 \sigma$ is now the segment $[0,2\pi]$). The three vacua are distinguished by the broken $\Z_2$ magnetic center symmetry, with order parameter $H$ (or Re($e^{i \pmb w_1 \cdot \pmb \sigma}) = \cos ({\sqrt 2 \sigma \over 2})$).  
  For $SO(3)_-$, there are also three vacua, at $\sqrt 2 \sigma = -\pi, 0, \pi$ (the fundamental domain is now $[-\pi, \pi]$). The three vacua are distinguished by the expectation value of the (wrapped on $\S^1_L$) $WH$ operator,\footnote{This is $\sin {\sqrt 2 \sigma \over 2} \sim {\rm Re}( e^{i (\pmb \phi + \pmb \sigma) \cdot \pmb w_1})$ (the real part accounts for the Weyl reflection in (\ref{centerk2}), and $e^{i \langle \pmb \phi \rangle \cdot \pmb w_1} = i$).} while $WH$  in $\R^2$ has area law due to confinement of its electric part.

\begin{figure}[htbp] %  figure placement: here, top, bottom, or page
   \centering
   \includegraphics[width=.7\textwidth]{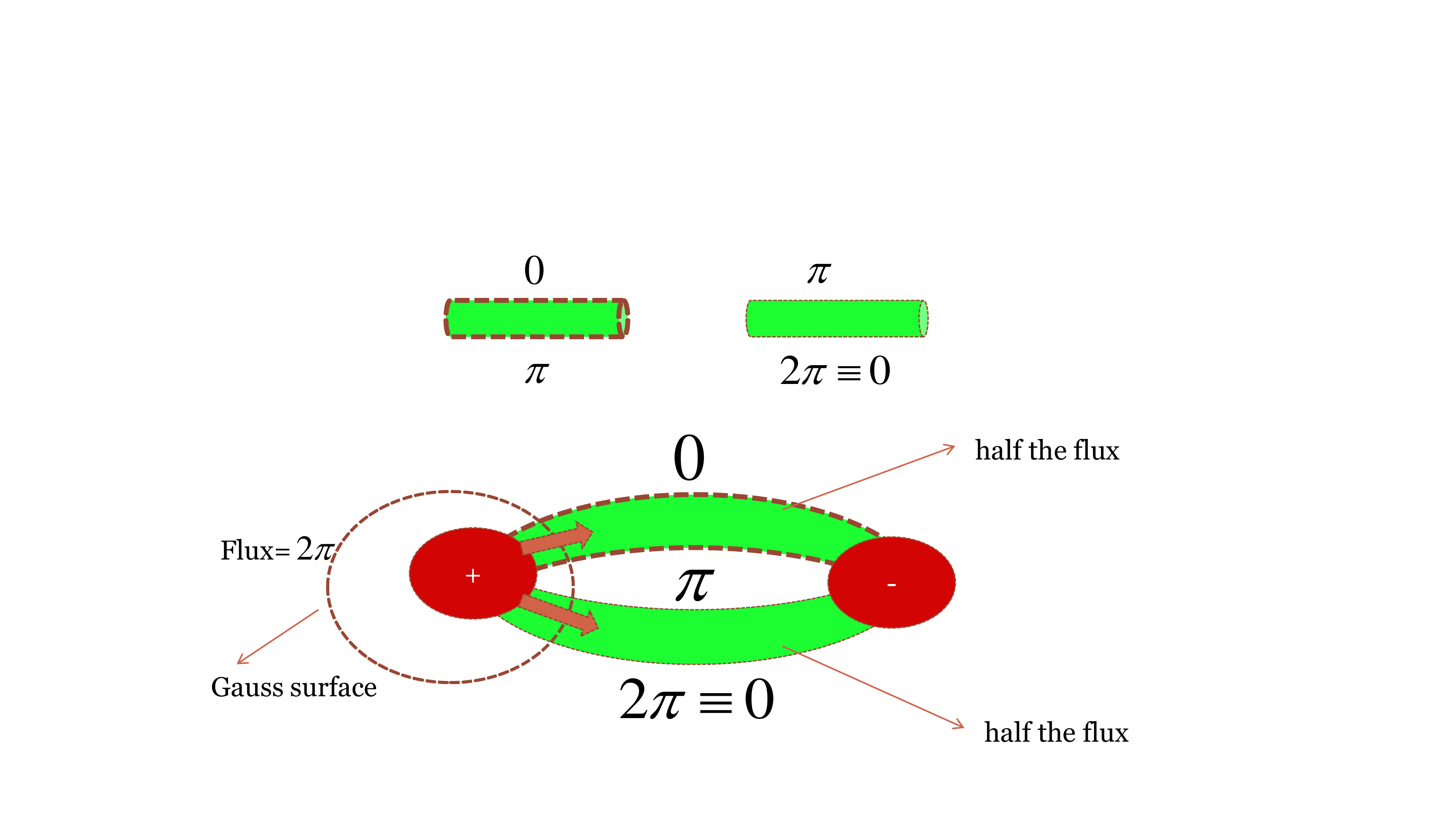} 
   \caption{{\it Top:}~The two distinct domain wall configurations in $su(2)$ QCD(adj), interpolating between the minima  with $\sqrt 2\sigma = 0$ (labeled by $1$ on Fig.~\ref{fig:so3adjoint}) and   $\sqrt 2\sigma = \pi$, and between $\sqrt 2\sigma = \pi $ and   $\sqrt 2\sigma = 2\pi \equiv 0$ (labeled by $2^*$ on Fig.~\ref{fig:so3adjoint}). Each of them carries half the electric flux of a fundamental charge. {\it Bottom:}~The double string confining fundamental charges in  the $SU(2)$ and $SO(3)_-$ QCD(adj) theories, shown here in the $\sigma = 0 = 2 \pi$ vacuum. It is a simple exercise to show that such configurations are allowed in all two vacua of the $SU(2)$ theory and all three vacua of the $SO(3)_-$ theory. On the other hand, the vacua identification in $SO(3)_+$ does not permit this configuration, as the $\sqrt 2\sigma =0$ and $\sqrt 2\sigma =2 \pi$ vacua are distinct (this holds in all three vacua), consistent with the absence of local confined probes.  }
   \label{fig:doublestring}
\end{figure}

The absence/presence of area law in these theories can also be understood  using our understanding of confining strings \cite{Anber:2015kea}. For the $SO(3)_+$ theory, from general arguments, we already know that there are no local probes with area law. To see this from the point of view of confining strings, note that the main difference from  dYM  is that in QCD(adj) the domain walls carry electric flux that can only confine half a quark (this is because the magnetic bions, whose ``condensation" is responsible for the confining potential have twice the magnetic charge of fundamental monopole-instantons, see Fig.~\ref{fig:doublestring}). Nonetheless, it is easy to see that the identification of vacua for $SO(3)_+$ does not allow any quark-like (dyonic or not) confined probe. This follows from considering the domain walls, denoted by $1$ and $2$ on the figure (they carry opposite electric flux, each equal to the flux of half quark, and are both BPS in the case of supersymmetry). A fundamental quark/antiquark  probe can only be confined by a configuration of a $1$-wall and a $2$-antiwall. However, such a configuration is impossible to arrange in any of the vacua of the $SO(3)_+$ theory, because all vacua connected by walls $1$ and $2$ are distinct.
 On the other hand, for the $SO(3)_-$ theory, confining configurations between quark/antiquarks are possible in all vacua: this is illustrated on the bottom figure, where such a configuration embedded in the vacuum $\sigma=0$ is shown.  

Finally we note that the counting of vacua and the   identification under the gauged center symmetry are the ones already given in \cite{Aharony:2013kma} for the supersymmetric $n_f=1$ case. We found that that the number of vacua of each of the $SO(3)_\pm$ theories is 3. This is in accord with the Witten index calculations for $SO(3)$ theories (for $n_f=1$) \cite{Witten:2000nv} and with the splitting of vacua argument of  \cite{Aharony:2013hda}  for the SYM case, reviewed in Section \ref{dymprime}.
%%%%%%%%%%%%%%%%%%%%%%%%%%%%%%%%%%%%%%%%%%%%%%%
\subsubsection{Theories with $\mathbf{su(3)}$ algebra}
%%%%%%%%%%%%%%%%%%%%%%%%%%%%%%%%%%%%%%%%%%%%%%%%
 \label{adjsu3}
 
In this Section, we consider QCD(adj) with  $su(3)$  algebra. We have three different theories that we label as  $\left[SU(3)/Z_3\right]_{0}$,  $\left[SU(3)/Z_3\right]_{1}$, and  $\left[SU(3)/Z_3\right]_{2}$. For each theory, the set of the compatible dyonic probes are $(1,H)$, $(1, WH)$ and $(1, W^2 H)$. %
The extrema of ${\cal W}$ (\ref{superpotential1}) are located at
\begin{eqnarray}\label{su3vacua}
\sigma_1^0=\frac{2\pi \sqrt 2}{3}\left(n_1+\frac{n_2}{2}\right)\,,\quad
\sigma_2^0=\frac{2\pi}{\sqrt{6}}n_2\,,
\end{eqnarray}
and $n_1,n_2 \in \mathbb Z$.  For the $SU(3)$ group, the fundamental domain of $\pmb \sigma$ is the weight lattice $\Gamma_w$ with basis vectors
$
\pmb w_1=\left(\frac{1}{\sqrt{2}},\frac{1}{\sqrt{6}}\right)$, $\pmb w_2=\left(0,\sqrt{\frac{2}{3}}\right).$
In this case, we have $3$ vacua which can be chosen to be $\{(n_1,n_2)=(0,0),(1,0),(2,0)\}.$
For the $\left[SU(3)/\mathbb Z_3\right]_{0,1,2}$ theories, the fundamental domain of $\pmb \sigma$ is the root lattice $\Gamma_r$. Hence, we find that there are $9$ vacua (\ref{su3vacua}) (the tripling is expected, since $\Gamma_w/\Gamma_r = \Z_3$) in the fundamental domain, given by the pairs\footnote{The results of this Section can be obtained geometrically from Fig.~\ref{fig:symsu3} (showing the vacua of $su(3)$ QCD(adj)  in $\Gamma_r$) upon an identification of the vacua under the action of  (\ref{centerk2}), recalling that $\cal{P}$ is a counterclockwise $\pi/3$ rotation around the origin.}
\begin{eqnarray}
\{(n_1,n_2)=(-2,2),(-1,1),(-1,2),(0,0),(0,1),(0,2),(1,0),(1,1),(2,0)\}\,.
\end{eqnarray}
 In order to avoid notational clutter, we just use these ordered pairs  to label the vacua. The theories $\left[SU(3)/\mathbb Z_3\right]_{0,1,2}$ are obtained from the $su(3)$ algebra by moding by the center, and hence gauging away the center symmetry amounts to the  identification (this is (\ref{centerk2}) written for this case):
\begin{eqnarray}
\nonumber
\sigma_1 &\rightarrow& -\frac{1}{2}\sigma_1-\frac{\sqrt{3}}{2}\sigma_2+2\pi k_1 w_1^1+2\pi k_2 w_2^1\,,\\
\sigma_2 &\rightarrow& \frac{\sqrt 3}{2}\sigma_1-\frac{1}{2}\sigma_2+2\pi k_1 w_1^2+2\pi k_2 w_2^2\,,
\label{center identification for su3}
\end{eqnarray}
where $k_1,k_2 \in \mathbb Z$. The three choices of gauged center correspond to taking $k_1 =0, k_2 =0$, 
$k_1 = 1, k_2 =0$, and $k_1 = 0, k_2 = 1$.
Choosing $k_1=0, k_2=0$, we find that under  (\ref{center identification for su3}) we have the following identification of the vacua:
\begin{eqnarray}
\nonumber
[SU(3)/\Z_3]_0:~~&&(0,0)\leftrightarrow(0,0)\,,\quad(1,1)\leftrightarrow(1,1)\,,\quad(-1,2)\leftrightarrow(-1,2)\,,\\
&&(-2,2)\leftrightarrow(0,1)\leftrightarrow(2,0)\,,\quad(-1,1)\leftrightarrow(0,2)\leftrightarrow(1,0)\,.
\label{case I su3}
\end{eqnarray}
Choosing $k_1=1, k_2=0$ we have 
\begin{eqnarray}
\nonumber
[SU(3)/\Z_3]_1:~~&&(2,0)\leftrightarrow(2,0)\,,\quad(0,1)\leftrightarrow(0,1)\,,\quad(-2,2)\leftrightarrow(-2,2)\,,\\
&&(-1,1)\leftrightarrow(1,0)\leftrightarrow(0,2)\,,\quad (-1,2)\leftrightarrow(0,0)\leftrightarrow(1,1)\,,
\label{case II su3}
\end{eqnarray}
and for $k_1=0, k_2=1$ we have 
\begin{eqnarray}
\nonumber
[SU(3)/\Z_3]_2:~~&&(-1,1)\leftrightarrow(-1,1)\,,\quad(0,2)\leftrightarrow(0,2)\,,\quad(1,0)\leftrightarrow(1,0)\,,\\
&&(-2,2)\leftrightarrow(2,0)\leftrightarrow(0,1)\,,\quad(-1,2)\leftrightarrow(1,1)\leftrightarrow(0,0)\,.
\label {case III su3}
\end{eqnarray}
 The number of the different vacua of each theory is $5$. This is exactly the number of vacua (for $N=3$) in a $[SU(N)/\mathbb Z_N]_{k}$ supersymmetric theory  on $\mathbb R^3 \times \mathbb S^1$, which is given by $\sum_{k=1}^N \mbox{gcd}(N,k)$, and $\mbox{gcd}$ stands for the greatest  common divisor. Thus, for prime values of $N$ the number of vacua is $2N-1$.
 
 In \cite{Aharony:2013hda} this result is obtained from knowledge of the magnetic charges of the condensed objects in the corresponding Seiberg-Witten theory on $\R^4$. This information is not available for QCD(adj) with $n_f>1$, where we instead rely only on a study of the unbroken gauge symmetries,  our detailed knowledge of the semiclassical  small-$L$ dynamics in QCD(adj), and consistency of the long-distance theory.

%%%%%%%%%%%%%%%%%%%%%%%%%%%%%%%%%%%%%%%%%%%%%%%%%%%%%%%%%%%%%%%
\subsubsection{Theories with $\mathbf{su(4)}$ algebra}
%%%%%%%%%%%%%%%%%%%%%%%%%%%%%%%%%%%%%%%%%%%%%%%%%%%%%%%%%%%%%%%%
\label{adjsu4}

As a final example, we consider QCD(adj) with an $su(4)$ algebra. Unlike the previous two cases where the center groups have a prime number of elements, the center symmetry of $su(4)$ is $\mathbb Z_4$ with non-prime numbers of elements. Thus, one obtains theories with distinct global structures by modding the group $SU(4)$ either by $\mathbb Z_4$ or by its subgroup $\mathbb Z_2$. Theories with an $su(4)$ algebra thus provide the QCD(adj) analogues of all cases considered for dYM in Sections~\ref{dymprime}, \ref{dymnonprime}, and \ref{dymk}.
The details of the calculations for QCD(adj) with $su(4)$ are given in Appendix \ref{su4appendix} and only the results will be reviewed here. To the best of our knowledge, the results for $su(4)$ SYM vacua on $\R^3\times \S^1$ with different global structure  are new (but they fit the pattern of vacua splitting upon compactification of softly-broken Seiberg-Witten theory).

Modding by $\Z_4$, there are four different theories that we label as  $[SU(4)/\Z_4]_{0}$, $[SU(4)/\Z_4]_{1}$, $[SU(4)/\Z_4]_{2}$, and $[SU(4)/\Z_4]_{3}$ which admit probes $(1,H)$, $(1,WH)$, $(1,WH^2)$, and $(1,WH^3)$, respectively. Each of these theories have $8$ distinct vacua. For SYM, this is in agreement  with the splitting of vacua picture \cite{Aharony:2013hda} (discussed in Section~\ref{dymprime}), which leads to  $\sum_{k=1}^{N_c}\mbox{gcd}(N,k)$ ($=8$, for $N=4$) vacua on $\R^3 \times \S^1$.

Modding $SU(4)$ by $\mathbb Z_2$, there are two theories  $[SU(4)/\Z_2]_{0}$
and  $[SU(4)/\Z_2]_{1}$ which respectively admit the probes $(W^2,H^2)$ and $(W^2,WH^2)$. The theory $[SU(4)/\Z_2]_{0}$ has $8$ distinct vacua, while $[SU(4)/\Z_2]_{1}$ has $4$ vacua. For SYM, this is consistent with the splitting of vacua picture: $H^2$ has perimeter law with a magnetic $\Z_2$ one-form gauge symmetry emerging in all four vacua on $\R^4$, each of which splits upon $\R^3 \times \S^1$ compactification yielding $8$ vacua,  while $WH^2$ has area law in each of the vacua on $\R^4$ giving $4$ vacua in the compactified theory (the other nontrivial operator common to the two theories, $W^2$ has area law). 

Evidently, as in the $su(2)$ and $su(3)$ cases presented earlier, the pattern of vacua for QCD(adj) with different global structure is, in each case, the same as in SYM and consistent with the emergence of  magnetic gauge symmetries in the softly-broken Seiberg-Witten theory on $\R^4$, despite the fact that there is no such $\R^4$ picture for QCD(adj). 

\acknowledgments
 
We would like to thank Aleksey Cherman, Markus Dierigl, Tin Sulejmanpasic, and Mithat \"Unsal for discussions related to the topic of this paper. EP   acknowledges support by an NSERC Discovery Grant. MA  acknowledges support by the Swiss National Science Foundation.

%\newpage

\appendix

%%%%%%%%%%%%%%%%%%%%%%%%%%%%%%%%%%%%%%%%%%%%%%%%%%%%%%%%%%%%%%%%
\section{The fundamental domain of the dual photon $\pmb\sigma$}
\label{appendixsigma}
%%%%%%%%%%%%%%%%%%%%%%%%%%%%%%%%%%%%%%%%%%%%%%%%%%%%%%%%%%%%%%%

As described in the main text, the fundamental domain of $\pmb \phi$ in the $\tilde{G}$ theory is such that $\pmb \phi \sim \pmb \phi+2\pi\pmb\alpha_k$, ($k=1,...,N_c-1$ for $SU(N_c)$), see (\ref{phiperiodicity1}, \ref{weylchamber}).  In the $\tilde{G}/K$ theory, the domain of $\pmb \phi$ is further reduced upon gauging the center subgroup $K$ acting as in (\ref{zenphi}).  In contrast, 
the fundamental domain of the dual photon field $\pmb \sigma$, is extended (rather then reduced) compared to the $\tilde{G}=SU(N_c)$ theory,  by permitting only a subset of all electric representations in the $\tilde{G}/K$ theory. The fundamental domain $\pmb \sigma$ is enhanced, from the unit cell of the finer weight lattice in the $\tilde{G}$ theory, to the unit cell of the coarser group lattice in the $\tilde{G}/K$ theory. Below, we give a canonical formalism derivation of this well-known result.

In order to determine the fundamental period of the dual photon field $\pmb \sigma$, we compactify the spatial directions, $x,y$, over a tow-torus $\mathbb T_2$, and use Gau\ss' law,  the quantization of magnetic flux on $\mathbb T_2$, and the duality (\ref{v in terms of sigma}) to find that the period of $\pmb \sigma$.  
We begin with the Wilson loop given by:
\begin{eqnarray}
{\cal W}_{xy}=\exp\left[i \oint_c  dl_i v_i\right]\,,
\label{Wilson loop in 3D}
\end{eqnarray} 
where the contour $c$ lies in the $x-y$ plane, or in other words on the torus surface. Using Stoke's theorem the line integral above can be written as
\begin{eqnarray}
{\cal W}_{xy}=\exp\left[i \oint_c  dl_i v^i\right]=\exp\left[i\int_{\Sigma \subset \mathbb T_2} ds B^3\right]=\exp\left[-i\int_{\Sigma_o} ds B^3\right]\,,
\end{eqnarray} 
where $B^3=v^{12}$ is the magnetic field in $2+1$ D, $\Sigma$ is the interior surface enclosed by $c$, while $\Sigma_o$ is the exterior or complementary surface, i.e. $\Sigma_o=\mathbb T_2-\Sigma$. The last equality results from the fact that the line integral is equivalent, by Stoke's theorem, to the integral over the internal and external areas enclosed by the loop.  Hence, we find the Dirac quantization condition
\begin{eqnarray}
\exp\left[i\int_{\mathbb T_2} ds B^3\right]=1\,,\mbox{or}\quad \int_{\mathbb T_2} ds B^3=2\pi n\,, \quad n \in Z\,.
\label{quantization condition}
\end{eqnarray}
The Wilson loop (\ref{Wilson loop in 3D}) measures the magnetic field probed by an electric charge that belongs to a representation ${\cal R}$, and  hence we have
\begin{eqnarray}
B^3= \pmb B^3 \cdot \pmb H_{{\cal R}_e}\,,
\end{eqnarray}
where $\pmb H_{{\cal R}_e}$ are the Cartan generators of the electric group in representation ${\cal R}$. 
Using (\ref{v in terms of sigma}) the magnetic field $\pmb B$ on the torus can be expressed in terms of the fields $\pmb \sigma$ and $\pmb \phi$:
\begin{eqnarray}
\pmb B^3=\pmb v^{12}=\frac{g^2}{8\pi^2 R}\left(\dot{\pmb \sigma}+\frac{\theta}{2\pi}\dot{\pmb \phi}\right)\,,
\label{B sigma and phi}
\end{eqnarray}
where $\dot{\pmb \sigma}\equiv \partial_t {\pmb \sigma}$ and $R=L/(2\pi)$. In addition, since we have a compact space, $\mathbb T_2$, we can ignore all higher modes of $\pmb \sigma$ and $\pmb \phi$ keeping only the zero modes $\pmb\phi_0$ and $\pmb\sigma_0$. Hence, the action (\ref{total action of the system}) takes the form ($A_{\mathbb T_2}$ is the area of the torus)
\begin{eqnarray}
S=\frac{A_{\mathbb T_2}}{2\pi R}\int dt \left[ \frac{1}{g^2}(\dot{\pmb \phi}_0)^2+\frac{g^2}{16\pi^2}\left(\dot{\pmb \sigma_0}+\frac{\theta}{2\pi}\dot{\pmb \phi_0}\right)^2\right]\,,
\label{zero mode action}
\end{eqnarray}  
with the equations of motion implying that the momenta are conserved, i.e.
\begin{eqnarray}
\nonumber
\frac{1}{g^2}\dot{\pmb \phi_0}+\frac{g^2\theta}{32\pi^3}\left(\dot{\pmb \sigma_0}+\frac{\theta}{2\pi }\dot{\pmb \phi_0}\right)&=&\pmb C\,,\\
\dot{\pmb \sigma_0}+\frac{\theta}{2\pi}\dot{\pmb \phi_0}&=&\pmb U\,,
\end{eqnarray}
where $\pmb U$ and $\pmb C$ are constants of motion. Using the second equation above and (\ref{B sigma and phi}) we find
\begin{eqnarray}
\pmb B^3=\frac{g^2}{8\pi^2 R}\; \pmb U\,,
\end{eqnarray} 
and hence the magnetic field is constant on the torus. The allowed values of $\pmb U$ are determined using the Dirac quantization condition (\ref{quantization condition}):
\begin{eqnarray}
\frac{g^2}{8\pi^2 R}A_{\mathbb T_2}\pmb U\cdot \pmb H_{{\cal R}_e}=2\pi (n_1,n_2,...n_{\mbox{\scriptsize dim}{\cal R}_e})\,.
\label{dual2}
\end{eqnarray}
 If the gauge group is $G$, the weights of all faithful representations ${\cal{R}}_e$ form the group lattice $\Gamma_G$. Then, equation (\ref{dual2}) implies that 
 \begin{equation}
 \label{dual3}
{ 16 \pi^3 R \over g^2 A_{\mathbb T_2}} \; \pmb  U \subset \Gamma_G^*,~ {\rm equivalently} ~~ { A_{\mathbb T_2} \over 2 \pi} \; \pmb B^3\subset \Gamma_G^*~,
 \end{equation}
where $\Gamma_G^*$ is the lattice dual to  $\Gamma_G$. 
The two extreme examples are $G = \tilde{G}$, i.e.~$\Gamma_G = \Gamma_w$, $\Gamma_G^* = \Gamma_{\alpha^*}$, where we find
\begin{equation}
\label{Bforcover}
\pmb B^3=\frac{2\pi}{A_{\mathbb T_2}}\sum_{a=1}^{r}n_a \pmb\alpha^*_a\,, ~ G = \tilde{G}
\end{equation}
and $G=\tilde{G}/C$, i.e. $\Gamma_G = \Gamma_r$, $\Gamma_G^* = \Gamma_{w^*}$, when (\ref{dual3}) implies that
\begin{equation}
\label{Bforadjoint}
\pmb B^3=\frac{2\pi}{A_{\mathbb T_2}}\sum_{a=1}^{r}n_a \pmb w^*_a\;, ~ G = {\rm \tilde{G}}/C\,.\end{equation}
Here $\tilde G$ is the covering group and $C$ is its center ($\pmb\alpha^*$ are the dual roots, $\pmb  w ^*$ are the dual weights and  $\{n_a\}$ are integers). In the general case, intermediate between (\ref{Bforcover}) and (\ref{Bforadjoint}), we have to replace the dual roots/weights in  above with the basis vectors of $\Gamma_G^*$, the lattice dual to the group lattice:
\begin{equation}
\label{Bgeneral}
\pmb B^3=\frac{2\pi}{A_{\mathbb T_2}}\sum_{a=1}^{r}n_a \pmb g^*_a\;, ~ \pmb g^*_a \cdot \pmb g_b = \delta_{ab}, ~{\rm for \; any} ~ \pmb g_b \in  \Gamma_G ~. \end{equation}

Further, from the action (\ref{total action of the system}) we find the  momenta conjugate to the fields $\pmb\sigma$ and $\pmb \phi$
\begin{eqnarray}
\nonumber
\pmb \Pi_\sigma&=&\frac{\delta S}{\delta \dot{\pmb \sigma}}=\frac{g^2 }{16 \pi^3R}\left(\dot{\pmb \sigma}+\frac{\theta}{2\pi}\dot{\pmb \phi}\right),\\
\pmb \Pi_\phi&=&\frac{\delta S}{\delta \dot{\pmb \phi}}=\frac{1}{\pi R g^2}\dot{\pmb \phi}+\frac{g^2}{32\pi^4 R}\theta\left(\dot{\pmb \sigma}+\frac{\theta}{2\pi}\dot{\pmb\phi}\right)\,.
\end{eqnarray}
In the case of compactifying the $x$$-$$y$ plane over the torus, (\ref{zero mode action}), we have
\begin{eqnarray}
\nonumber
\pmb \Pi_{\sigma_0}&=&\frac{g^2  A_{\mathbb T_2} }{16 \pi^3R}\left(\dot{\pmb \sigma}_0+\frac{\theta}{2\pi}\dot{\pmb \phi}_0\right)=\frac{g^2 A_{\mathbb T_2}}{16 \pi^3R}\; \pmb U=\frac{A_{\mathbb T_2}}{2\pi}\;\pmb B^3 =  \left\{\begin{array}{c} \sum_{a=1}^{r}n_a \pmb\alpha^*_a\,,\mbox{for}\,G=\tilde G \\ \sum_{a=1}^{r}n_a \pmb w^*_a\,,\mbox{for}\;\,G=\tilde G/C\end{array}\,,\right.\\\label{momenta}
\pmb \Pi_{\phi_0}&=&\frac{A_{\mathbb T_2}}{\pi R g^2}\; \dot{\pmb \phi_0}+\frac{A_{\mathbb T_2}g^2}{32\pi^4 R}\theta\left(\dot{\pmb \sigma_0}+\frac{\theta}{2\pi}\dot{\pmb\phi_0}\right) = \, \frac{A_{\mathbb T_2}}{\pi R } \;\pmb C~,
\end{eqnarray}
and the total Hamiltonian of the system reads
\begin{eqnarray}
H=\frac{8\pi^3 R}{g^2 A_{\mathbb T_2}}\; \pmb{\Pi}_{\sigma_0}^2+\frac{\pi g^2 R}{2 A_{\mathbb T_2}}\left(\pmb \Pi_{\phi_0}-\frac{\theta}{2\pi}\pmb \Pi_{\sigma_0}\right)^2\,.
\end{eqnarray}
In order to determine the period of $\pmb \sigma_0$ we can set $\theta=0$ and ignore $\pmb \phi_0$. Then,  using (\ref{momenta}) and (\ref{Bgeneral}) we find that the energy of the field $\pmb \sigma_0$ is, for general $G$:
\begin{eqnarray}
H_{\sigma_0}=\frac{8\pi^3 R }{g^2A_{\mathbb T_2}}\; \pmb \Pi_{\sigma_0}\cdot\pmb \Pi_{\sigma_0}=\frac{8\pi^3 R }{g^2A_{\mathbb T_2}}\left[ \sum_{a=1}^r n_a\pmb g_a^*\right]^2~.
\label{energy for sigma}
\end{eqnarray}
This energy can also be obtained by promoting the field $\pmb\sigma_0$ to an operator $\pmb\sigma_0\rightarrow \hat{\pmb\sigma}_0$ and $\pmb\Pi_{\sigma_0}\rightarrow \pmb{\hat\Pi}_{\sigma_0}=-i\pmb{\partial}_{\sigma_0}$. Thus, the quantum mechanical Hamiltonian reads
\begin{eqnarray}
\hat H_{\sigma_0}=-\frac{8\pi^3 R }{g^2A_{\mathbb T_2}}\; \pmb{\partial}_{\sigma_0}\cdot \pmb{\partial}_{\sigma_0}\,.
\end{eqnarray}
The wave function that gives the correct energy (\ref{energy for sigma}) is 
\begin{eqnarray}
\psi= \exp\left[i\pmb\sigma_0\cdot \sum_{a=1}^r n_a \pmb g^*_a \right] 
\end{eqnarray}
For single valued $\psi$ we demand that $\psi$ changes by a trivial phase as $\pmb\sigma_0\rightarrow \pmb\sigma_0+\pmb\Lambda$. Thus, we have $\pmb\Lambda =2\pi \pmb  g_a$  
(recall that $\pmb g_a = \pmb w_a$ for $G=\tilde G$, and $\pmb \alpha_a$ for $G = \tilde G/C$). Thus, the fundamental domain of $\pmb \sigma$ is the group lattice.

\section{Derivation of the line operators on $\R^3 \times \S^1_L$}
\label{operatorsappendix}

\subsection{The 't Hooft operator in the canonical formalism}
We begin with two   remarks. First, for our purposes, it  is   more convenient  to use the Hilbert space representation of 't Hooft line operators \cite{'tHooft:1977hy}, rather than their definition based on prescribed monopole singularities  of the gauge field configurations   in the path integral; see Witten's lecture in vol.~II of \cite{Deligne:1999qp} for an introduction. Second, following \cite{Kapustin:2005py}, we shall use the term ``'t Hooft operator"  to denote operators that are more general than the ones originally introduced  in \cite{'tHooft:1977hy}, as this is commonly done in the current literature (see also Footnote~\ref{thooftnote}).
 
In a canonical Hilbert space representation, as usual most convenient in $A_0=0$ gauge, the action of the four dimensional 't Hooft operator can be described as the creation of an infinitely thin magnetic flux line along a spacelike curve  $\partial \Sigma$, the boundary of a two-surface 
$\Sigma$, as will be made explicit below. This picture is dual to the one that can be applied to the Wilson loop, which can be thought of as creating an electric flux along $\partial \Sigma$.  
More explicitly, the four dimensional 't Hooft loop operator is determined by choosing a constant vector, or  ``magnetic weight"  $\pmb \nu_{m}$, an $r$-component vector which we shall not specify yet. The 't Hooft loop operator  can then be written in the form:
\begin{eqnarray}\label{T1}
{\cal T}^{4D}(\pmb \nu, \Sigma)=\exp\left[ i 2\pi  \pmb \nu_{m}\cdot \int_\Sigma d^2s\; n_i \; \pmb  \Pi^{i\; 4D} \right]\,,
\end{eqnarray}
where $i$$=$$1,2,3$, $n_i$ is the unit normal to the surface $\Sigma$, assumed orientable, with boundary $\partial \Sigma$, and $\pmb \Pi^{i \;4D}$ is the four dimensional canonical momentum (essentially, the electric field operator  if the $\theta$ angle vanishes; thus one can think of (\ref{T1}) as measuring the electric flux through $\Sigma$). Notice that despite the appearance of a constant Lie-algebra valued vector in (\ref{T1}), the 't Hooft loop operator maps physical states into physical states.\footnote{See \cite{'tHooft:1977hy,KorthalsAltes:1999xb,Reinhardt:2002mb} in the Hilbert space formalism and   \cite{Gomis:2009ir} within the Euclidean path integral definition of ${\cal T}^{4D}$. We do not dwell on this  here, as we study dYM and QCD(adj) in the  dynamically abelianized regime.}

Using the canonical commutation relations  $[\Pi_{A}^{i\; 4D}(\vec{x}), v_{j \; B}(\vec{y})] = - i \delta_{AB} \delta_{j}^i \delta^{(3)}(\vec{x} - \vec{y})$, where $A,B$ are Lie-algebra indices (we use $a,b$ below denote their restriction to the Cartan subalgebra), one finds that the action of (\ref{T1}) on the canonical coordinate $v_{i \; A}(\vec{y})$ is to shift it by an amount given in the last term below:\footnote{In an abuse of notation, we do not put hats over operators, hoping that the distinction between operators and $c$-functions is evident in each case.}
\beq
\label{transform}
{\cal T}^{4D}(\pmb \nu, \Sigma) \; v_{i \; A} (\vec{y}) \; {\cal T}^{4D}(\pmb \nu, \Sigma)^\dagger &=& v_{i \; A} (\vec{y}) + 2 \pi \delta_{Aa}  \;\nu_{m \;a} \int_\Sigma d^2 s\; n_i \delta^{(3)} (\vec{x}_\Sigma - \vec{y}) \nonumber~, \\
&\equiv& v_{i \; A} (\vec{y}) + 2 \pi \delta_{Aa} \; \nu_{m \;a} \; A_i (\vec{y})
\eeq
where $\vec{x}_\Sigma \in \R^3$ denotes a point on $\Sigma$, the last line defines the $c$-number $A_i(\vec{y})$, and we used  $\nu_{m \; a}$ for the $a$-th component of the magnetic weight $\pmb \nu_m$.  
The shift of the operator $v_{i \; a}$ induced by the action of the 't Hooft operator ${\cal T}^{4D}$, proportional to $A_i (\vec{y}) =  \int_\Sigma ds\; n_i \delta^{(3)} (\vec{x}_\Sigma - \vec{y})$, can  be easily seen to correspond to the field of an infinitely thin unit magnetic flux line (vortex) along the boundary of $\Sigma$. 
To show this, let us calculate the circulation of $A_i$ along a closed contour $C$, $\oint_C d y^i  A_i (\vec{y})$. This is  equal  to the flux of the  magnetic field, $\vec{B}_{A} = \vec \nabla \times \vec{A}$, through a surface $S$ such that $\partial S = C$, via the chain of identities:
\beq
\label{t3} 
\int_S d^2 \vec{s} \cdot  \vec{B}_A = \oint_{C} d y^i  A_i (\vec{y}) =  \oint_{C = \partial S}  d y^i  \int_\Sigma d^2s\; n_i \delta^{(3)} (\vec{x}_\Sigma - \vec{y}) \equiv   I(C, \Sigma)~,
\eeq
which shows that $\int_S d^2 \vec{s} \cdot  \vec{B}_A$ equals\footnote{If $C$ does not intersect $\Sigma$, the argument of the delta function has no support and the integral vanishes. If $C$ intersects $\Sigma$ once, the integral is $\pm 1$ depending on the direction of $C$, etc. (to see this one can choose local coordinates near the intersection point of $C$ with $\Sigma$ such that, e.g. $\Sigma$ is in the $xy$ plane, then $dy^i n_i = \pm dz$ and the result follows).}  the intersection number $I(C,\Sigma)$ of $C$ and $\Sigma$. Thus an arbitrarily small  $C$ that winds once around the boundary of $\Sigma$  has $I=1$. Thus, we conclude that $A_i$ in (\ref{transform}) indeed corresponds to the field of an infinitely thin unit magnetic flux line. 
The chain of arguments from (\ref{T1}) to (\ref{t3})  proves  the assertion that the action of the 't Hooft operator (\ref{T1}) on a ``position eigenstate" (an eigenstate of $v_{i \; a}$) shifts its eigenvalue by an amount representing the creation of an infinitely thin magnetic flux line (a vortex) along the boundary of $\Sigma$. 

 Let us also stress that only the location of the boundary of $\Sigma$ is essential. Consider the difference between the action of two operators   with the same boundary $\partial \Sigma$, but different choice of surfaces $\Sigma_1$ and $\Sigma_2$, on the canonical coordinate, i.e. the action of ${\cal T}^{4D\;\dagger} (\pmb \nu_m, \Sigma_2) {\cal T}^{4D}(\pmb \nu_m, \Sigma_1)$. The result  is also given by the last line in  (\ref{transform}), but now  $A_i (\vec{y}) =  \oint_{\Sigma'} d^2 s\; n_i \delta^{(3)} (\vec{x}_{\Sigma'} - \vec{y})$, where  $\Sigma'$  is  a closed surface, the union of $\Sigma_1$ and $\Sigma_2$, and $n_i$ is the unit  outward  normal to the surface (the joining of the two surfaces has to be sufficiently smooth near $\partial \Sigma$). As we shall see, $A_i (\vec{y})$ is a total derivative, hence a gauge transformation. Consider the integral $\omega(\Sigma', \vec{y})  = \oint_{\Sigma'} d^2 s n^i \partial_i^{y} {1\over |\vec{x}_{\Sigma'} -  \vec{y}|}$, which, thus defined, is either $4\pi$ or $0$, depending on whether $\vec{y}$ is inside or outside $\Sigma'$. This follows from noticing that $\omega(\Sigma', \vec{y}) $ is proportional to the flux of the electric field of a pointlike charge at $\vec{y}$ through the closed surface $\Sigma'$. Then, the gradient\footnote{\label{gaugefootnote}Heuristically, this is because  $\omega(\Sigma', \vec{y})$, being piecewise constant, changes only for $\vec{y}$ at the surface. Further, its gradient is along the normal to $\Sigma'$ and is negative if $n^i$ is the outward normal, as the function decreases stepwise from $4 \pi$ to $0$ upon $\vec{y}$ crossing to the outside. The derivative is thus proportional to $- 4\pi$ times a delta function of the normal component of $\vec{y}$. To see that the formula given in the text correctly reproduces this,  choose coordinates with origin at  a point on the surface $\vec{x}_{\Sigma'}^0 = (0,0,0)$. Then, we  have $\vec{y}=(x_1,x_2,x_3)$, s.t. $x_3$  is the normal direction near $\vec{x}_{\Sigma'}^0$ and $x_1,x_2$ are the tangential directions to $\Sigma'$ (we also have $\vec{x}_{\Sigma'} = (x_1^{\Sigma'}, x_2^{\Sigma'}, x_3^{\Sigma'}$) near $\vec{x}_{\Sigma'}^0$). Then, the only nonvanishing component of $A_i$ is $A_3 = \delta({x_3})$, since the surface integral over the  surface removes two of the delta functions, thus $\partial_3\omega(\Sigma',\vec{y}) = - 4 \pi A_3 = - 4 \pi \delta(x_3)$, the expected result.} of $\omega(\Sigma', \vec{y})$ is   $\partial_i^{y} \omega(\Sigma', \vec{y}) = - 4 \pi A_i(\vec{y})$. Hence, the surface $\Sigma$ can be moved around by gauge transformations while keeping its boundary $\partial \Sigma$ fixed  and its location is not essential. Further, for probes obeying (\ref{magneticweights}), the surface is not only topological, but unobservable.

The magnetic flux of the vortex line along $\partial \Sigma$, as follows from (\ref{transform}, \ref{t3}), is 
\begin{equation}
\label{flux}
\Phi = 2 \pi \pmb \nu_m \cdot \pmb H_{\cal{R}}~,
\end{equation} where $\pmb H_{{\cal{R}}}$ are the Cartan generators in a representation ${\cal{R}}$ of the gauge group with weights $\pmb \nu_{\cal R}$. Flux quantization requires that 
\beq
\label{magneticweights}
e^{i \Phi} = {\rm id}_G~~\Leftrightarrow~~\pmb \nu_m \cdot \pmb \nu_{\cal{R}} \in \Z~,
\eeq
where ${\cal{R}}$ is any faithful representation of the gauge group $G$  and id$_G$ denotes the identity in $G$. This condition ensures   that operators in faithful representations of $G$, e.g. the ones corresponding to the representations of (possibly heavy) local fields in the theory, are single-valued around the vortex and their correlation functions with ${\cal T}^{4D}$ are well defined. Gauge invariant Wilson loops of  probes in all 
  $\cal R$ obeying  (\ref{magneticweights}) will also be local with respect to the 't Hooft operator, see (\ref{algebra1}), (\ref{commutator}). 
Eq.~(\ref{magneticweights}) is also equivalent to the GNO magnetic charge quantization condition -- one way to visualize this would be to imagine that the vortex line ends on a monopole so that $\Phi$ becomes its GNO flux  \cite{Goddard:1976qe}.\footnote{\label{thooftnote}
To connect (\ref{T1}) with the original definition of the 't Hooft loop \cite{'tHooft:1977hy}, notice that if one takes the magnetic weight $\pmb \nu_m$  to be a  weight of the fundamental representation, a non-dynamical electric probe in the fundamental representation  will not be single valued around the vortex,  since it will detect fractional flux (\ref{flux}) with eigenvalues   $\Phi_i  = {2 \pi \over N} k_i$, $k_i \in \Z$. The fractional flux occurs  because weights of the fundamental representation obey $\pmb \nu_i \cdot \pmb \nu_j = \delta_{ij} - {1 \over N_c}$ ($i,j=1,\ldots N_c$) and both the magnetic weight and the weights of ${\cal R}$ are now taken to be weights of the fundamental representation. This fractional flux is usually called ``'t Hooft flux", the corresponding vortex line---a ``center vortex", and the corresponding ${\cal T}^{4D}$---a ``center vortex creation" operator. In the modern terminology, one of the Wilson/'t Hooft operators introduced \cite{'tHooft:1977hy} is a surface rather than a genuine line operator, as they do not obey the GNO condition.}

Next, we find the 't Hooft operator in the long distance theory on $\R^3 \times \S^1$, beginning with (\ref{T1}). Recall that the four dimensional Lagrangian, restricted to Cartan subalgebra components, is given by
\begin{eqnarray}
\label{l4d}
{\cal L}_{4D}=-\frac{1}{2g^2}\pmb v_{mn}^2+\frac{\theta}{16\pi^2}\tilde{\pmb v}_{mn}\pmb v^{mn}=\frac{1}{g^2}\left( \pmb E_i\pmb E_i-\pmb B_i\pmb B_i \right)-\frac{\theta}{4\pi^2}\pmb E_i \cdot \pmb B_i\,,
\end{eqnarray}
where $\pmb E_i=\pmb v_{i0}=\partial_i \pmb v_0- \dot{\pmb v_i}$, $\pmb B_i=\frac{1}{2}\epsilon_{ijk}\pmb v_{jk}$. Then, the conjugate momenta to the field $\pmb v_i$ are given by 
\begin{eqnarray}
\pmb\Pi_i^{4D}=\frac{\partial {\cal L}_{4D}}{\partial \dot{\pmb v}_i }=\frac{2}{g^2}\left(-\partial_i \pmb v_0+ \dot{\pmb v_i}\right)+\frac{\theta}{4\pi^2}\pmb B_i\,.
\end{eqnarray}
Further, using (\ref{v in terms of sigma}), $\pmb B_{1}=\frac{\partial_{2} \pmb\phi}{2\pi R}$, $\pmb B_{2}=-\frac{\partial_{1} \pmb\phi}{2\pi R}$, and $\pmb E_3=-\frac{\partial_t\pmb \phi}{2\pi R}$, we obtain
\begin{eqnarray}
\nonumber
\pmb \Pi_1^{4D}&=&-\frac{1}{4\pi^2 R}\partial_2 \pmb \sigma \,~, ~~
\pmb \Pi_2^{4D} = \frac{1}{4\pi^2 R}\partial_1 \pmb \sigma \,,\\
\pmb \Pi_3^{4D}&=&\frac{1}{\pi g^2 R}\partial_t \pmb \phi+\frac{\theta g^2}{32\pi^4 R}\left(\partial_t \pmb \sigma +\frac{\theta}{2\pi}\partial_t \pmb\phi\right)\,.
\end{eqnarray}

We first take  $n_i=n_3=\hat e_z$, such that the surface $\Sigma$ lies in the $x-y$ plane (hence we denote it by $\Sigma_{xy}$), to  find that (\ref{T1}) becomes
\begin{eqnarray}
{\cal T}^{4D} (\pmb \nu_m, \Sigma_{xy})=\exp\left[ i2\int_{\Sigma_{xy}} ds \left[\frac{1}{g^2 R}\partial_t \pmb \phi+\frac{\theta g^2}{32\pi^3 R}\left(\partial_t \pmb \sigma +\frac{\theta}{2\pi}\partial_t \pmb\phi\right) \right] \cdot \pmb \nu_m \right]\,.
\label{four D 't hhoft 3}
\end{eqnarray}
In fact,  one can use the second equation in (\ref{momenta}) to rewrite (\ref{four D 't hhoft 3}) as, omitting the $4D$ superscript from now on:
\begin{eqnarray}
{\cal T}  (\pmb \nu_m, \Sigma_{xy})=\exp\left[i2\pi\int_{\Sigma_{xy}} ds \; \pmb\Pi_{\phi}\cdot \pmb \nu_m \right]\,.
\label{t4}
\end{eqnarray}
 
Now, if we take $n_i=n_x=\hat e_x$ we find that the surface $\Sigma$ is closed surface in the $y-z$ plane, which wraps around the compact dimension $z$.  When projected onto the noncompact space, $\R^{2}$,  this surface is a line with two end points. Hence, since we shall be working in the long-distance theory, we shall denote $\Sigma$ in this case as
 $\Sigma_{\vec{r}_1, \vec{r}_2}$ where $\vec{r} \in \R^2$ (for the above example, we have $\vec{r}_1 = (x, y_1)$, $\vec{r}_2 = (x, y_2)$).
Then  we find
\begin{eqnarray}
\nonumber 
{\cal T} (\pmb \nu_m, \Sigma_{(x,y_1),(x,y_2)})&=&\exp \left[-i  \int_{z=0}^{2\pi R} dz \int_{y=y_1}^{y=y_2} dy\frac{1}{2\pi R}\partial_2 \pmb \sigma \cdot \pmb \nu_m \right]  \nonumber \\
&=&e^{-i\left(\pmb \sigma(x,y_2) -\pmb \sigma(x,y_1)\right) \cdot \pmb \nu_m }\,.
\label{thooftwrapped}
\end{eqnarray}
The last expression shows that  the  't Hooft operator corresponding to a surface wrapped around the compact direction   depends on the position of the initial and final points in $\R^2$ (this is the remnant of it being, generally, a surface operator).  A true local operator should not depend on two points. One can consider the local 't Hooft operator, with $\vec{r} \in \R^2$:
\begin{eqnarray}
\label{t5}
{\cal T}(\pmb \nu_m,  \vec{r})=e^{- i\pmb \sigma(\vec{r}) \cdot \pmb \nu_m}\,,
\end{eqnarray}
which can be thought of as (\ref{thooftwrapped}) with one of the points taken to infinity.

%%%%%%%%%%%%%%%%%%%%%%%%%%%%%%%%%%%%%%%%%%%%%%%%%
\subsection{The  Wilson operator }
%%%%%%%%%%%%%%%%%%%%%%%%%%%%%%%%%%%%%%%%%%%%%%%%%%%%

The Wilson operators can be similarly defined. As usual, they are specified by choosing a representation ${\cal R}_e$ of the gauge group, with Cartan generators  $\pmb H_{{\cal R}_e}$. By Gau\ss' law, they can be thought of as measuring the magnetic flux through  a surface. For a loop lying in the $xy$-plane, we have, following the same steps as above, dimensionally reducing and replacing the magnetic field $\vec{\pmb B}$ by its dual via (\ref{v in terms of sigma})
\begin{eqnarray}
\label{wilsonloop1}
{\cal W}({\cal R}_e, \Sigma_{xy})=\exp\left[i  \int_{\partial \Sigma_{xy}} dl^i \pmb v_i  \cdot \pmb H_{{\cal R}_e} \right]\,=  \exp\left[i 2\pi\int_{\Sigma_{xy}} ds \; \pmb\Pi_{\sigma}(x,y)\cdot \pmb H_{{\cal R}_e} \right]\,.
\end{eqnarray}
In addition to the Wilson loop defined above, which measures the magnetic field in the $x-y$ plane probed by an electric charge,  one can also define the Wilson loop that measures the magnetic field in the $y-z$ or $x-z$ planes which wraps the $\mathbb S^1$ circle. The new Wilson loop operator is given by, using the same notation for the loop as for the 't Hooft operator (\ref{thooftwrapped})
\begin{eqnarray}
{\cal W}({\cal R}_e,\Sigma_{\vec{r}_1,\vec{r}_2})&=&\exp\left[i \int_{y_1}^{y_2} dy \int_0^{2\pi R}dz \; \pmb B_1\cdot \pmb H_{{\cal R}_e} \right] \nonumber = e^{i \left( \pmb \phi(x,y_2)-\pmb \phi(x,y_1)\right)\cdot \pmb H_{{\cal R}_e} }\,,
\label{wilson23}
\end{eqnarray}   
with $\vec{r}_1 = (x, y_1)$, $\vec{r}_2 = (x, y_2)$.
The operator ${\cal W}({\cal R}_e,\Sigma_{\vec{r}_1,\vec{r}_2})$ is not a local operator in the dimensionally reduced theory, as it depends on two points. One can also define the local operator, similar to (\ref{t5})
\begin{eqnarray}
\label{wilson24}
{\cal W}({\cal R}_e,  \vec{r})=e^{i\pmb \phi(\vec{r})\cdot \pmb H_{{\cal R}_e} }\,, \end{eqnarray}
and one can think of it as the limit of (\ref{wilson23}) when one of the points is taken to infinity.

%%%%%%%%%%%%%%%%%%%%%%%%%%%%%%%%%%%%%%%%%%%%%%%%%%%%%%%%%%%%%%%%%%%%%%%%%%%%%%%%%%%%
\subsection{The Wilson and 't Hooft operators commutation relations}
%%%%%%%%%%%%%%%%%%%%%%%%%%%%%%%%%%%%%%%%%%%%%%%%%%%%%%%%%%%%%%%%%%%%%%%%%%%%%%%%%%%%%

The basic set of operators we shall study are ${\cal T}(\pmb \nu_m, \vec{r})$ of
 (\ref{t5}), ${\cal T}  (\pmb \nu_m, \Sigma_{xy})$ of (\ref{t4}), ${\cal W}({\cal R}_e,  \vec{r})$ of (\ref{wilson24}), and ${\cal W}({\cal R}_e, \Sigma_{xy})$ of (\ref{wilsonloop1}). It is clear from their definitions that there are two pairs among the four operators that may not commute. Consider one pair, 
  ${\cal T}(\pmb \nu_m, \vec{r})$ and ${\cal W}({\cal R}_e, \Sigma_{xy})$.
From the canonical commutation relation 
$
\left[ \Pi_{i\sigma}(\vec{r}), \sigma_{j}(\vec{r}^{\; \prime}) \right]=-i\delta^{(2)}(\vec{r} - \vec{r}^{\; \prime})\delta_{ij},
$
we find
\begin{eqnarray}
2\pi\int_{\Sigma_{xy}} ds \left[\pmb\Pi_\sigma(\vec{r}_{\Sigma_{xy}})\cdot \pmb H_{{\cal R}_e},\pmb \sigma(
\vec{r}^{\; \prime}) \cdot \pmb \nu_m \right] =-2\pi i\ell(\Sigma_{xy}, \vec{r}^{\; \prime}) \; \pmb H_{{\cal R}_e}\cdot \pmb \nu_m \,,
\end{eqnarray}
where $\ell$ is the linking number between the position of the 't Hooft operator and the Wilson loop which can be either $0$ or $1$ ($1$ if $\vec{r}'$ is inside $\Sigma_{xy}$ and $0$ otherwise).
Hence, from the Baker-Campbell-Hausdorff formula we obtain
\begin{eqnarray}
\label{algebra1}
 {\cal W}({\cal R}_e, \Sigma_{xy}) \; {\cal T}(\pmb \nu_m, \vec{r}) \;=e^{2\pi i\ell(\Sigma_{xy}, \vec{r})  \pmb H_{{\cal R}_e}\cdot \pmb \nu_m}{\cal T}(\pmb \nu_m, \vec{r}) \; {\cal W}({\cal R}_e, \Sigma_{xy})\end{eqnarray}
The other two operators,  ${\cal W}({\cal R}_e, \vec{r})$ and  ${\cal T}(\pmb \nu_m, \Sigma_{xy})$, obey  a similar relation.

We now recall the terminology of \cite{Aharony:2013hda}. 
 If the GNO condition (\ref{magneticweights}) is satisfied for $\pmb \nu_m$ and   ${\cal R}_e$, then we have $ \pmb \nu_{{\cal R}_e} \cdot \pmb \nu_m =n$, $n\in \Z$, where $\pmb \nu_{{\cal R}_e}$ are the 
weights of the representation ${\cal R}_e$ of the gauge group (the eigenvalues of $\pmb H_{{\cal R}_e}$). Thus the phase in (\ref{algebra1}) vanishes and the Wilson and 't Hooft operators commute. We shall call these operators ``genuine line operators". 
On the other hand, for operators that do not obey his condition there is a nontrivial phase (such as the original one defined by 't Hooft, where there is a $\Z_N$ phase) and there must be a physical significance to the surface attached to the operators. We shall call these operators ``surface operators" (for theories without dynamical fundamental fields, only the topology of the surface matters).

%%%%%%%%%%%%%%%%%%%%%%%%%%%%%%%%%%%%%%%%%%%%%%%%%%%%%%%%%%%%%%%%%%%%%%%%%
\subsection{Including dyonic operators,  $\pmb \theta$ angle, and Witten effect}
\label{witteneffect}
%%%%%%%%%%%%%%%%%%%%%%%%%%%%%%%%%%%%%%%%%%%%%%%%%%%%%%%%%%%%%%%%%%%%

For the study of the ground states of QCD(adj), it will be of interest to consider general dyonic, or Wilson-'t Hooft operators, as they will be essential in distinguishing theories with different choices of gauge group. 
Dyonic operators can be defined in the nonabelian case as a product of the 't Hooft operator (\ref{T1}) with a Wilson operator  along the same  $\partial \Sigma$:
\beq
\label{d1}
 {\cal D}^{4D}(\pmb\nu_m, {\cal R}({\pmb \nu_m}), \Sigma) = e^{ i 2\pi  \pmb \nu_{m}\cdot \int_\Sigma d^2s\; n_i \; \pmb  \Pi^{i\; 4D}} \times {\rm Tr}_{{\cal R}({\pmb \nu_m})}  {\cal P} e^{ i \oint_{\partial \Sigma} v_i  d l^i}~.
\eeq
There are some subtleties: as before, the 't Hooft operator is labeled by a magnetic weight $\pmb \nu_m$, but the Wilson operator is taken in a representation of the stabilizer subgroup of the magnetic weight  \cite{Kapustin:2005py}.
Thus, the magnetic flux along $\partial \Sigma$  due to the action of 't Hooft loop operator and the electric flux  due to the Wilson loop commute. In our abelianized long-distance theory this is manifestly true.
 Focusing on  the abelian case from now on, 
 we  define the four dimensional Wilson-'t Hooft operators as
\beq\label{t6}
  {\cal D}^{4D}(\pmb\nu_m, \pmb\nu_e, \Sigma) = {\cal T}^{4D}(\pmb \nu_m, \Sigma)\; {\cal W}(\pmb\nu_e, \Sigma_{xy})
  \eeq
  where 
  \beq
  {\cal W}(\pmb\nu_e, \Sigma )=\exp \left[i  \int_{\partial \Sigma} dl^i \pmb v_i  \cdot \pmb \nu_e \right] \,=  \exp\left[i 2\pi\int_{\Sigma} ds n_i \; \pmb \Pi^i(x,y)\cdot \pmb \nu_e \right]\,
\eeq
is one of the eigenvalues of (\ref{wilsonloop1}). 
As we did in the previous sections, we break the four dimensional operator into two classes of operators by choosing either $n_i=n_3$ or $n_i=n_1$. The resulting operators, corresponding to loops wrapped around $\S^1$ (combine (\ref{t5}) with (\ref{wilson24}))  or in the noncompact directions (combine (\ref{t4}) with (\ref{wilsonloop1})), are  
\begin{eqnarray}
\label{operators11}
{\cal D}(\pmb \nu_e, \pmb \nu_m, \vec{r})&=&\exp\left[-i \pmb \sigma(\vec{r})\cdot\pmb\nu_m +i \pmb \phi(\vec{r}) \cdot \pmb \nu_e \right]\,,\\ \nonumber
{\cal D}(\pmb \nu_e, \pmb \nu_m, \Sigma_{xy})&=&\exp\left[i 2\pi \int_{\Sigma_{xy}} d^2 s \left\{ \pmb\Pi_\phi \cdot \pmb\nu_m+\pmb \Pi_\sigma \cdot \pmb \nu_e  \right\} \right]\,.
\end{eqnarray} 
In the case of ${\cal D}(\pmb \nu_e, \pmb \nu_m, \vec{r})$ we have taken one of the ``constituent" operators to infinity, as already done in (\ref{t5}), (\ref{wilson24}). 
The nontrivial commutation relation of the dyonic operators (\ref{operators11}) is easily seen to be Eq.~(\ref{diracrelation}) from  the main text.

Finally, we comment on the Witten effect in the canonical formalism. In the (Euclidean) path integral definition of the line operators, the 't Hooft loop is defined as a boundary condition, imposed on the fields one integrates over,   on a thin ``tube" around the loop. The Witten effect for a 't Hooft loop in this formulation arises from a surface term coming from the ${\theta}$ term, see \cite{Kapustin:2005py}. In the canonical formalism, on the other hand, 
 we have from (\ref{transform}) that the 't Hooft loop, acting (for simplicity) on the vacuum state, using a  field-eigenstate basis, creates a thin magnetic vortex line, explicitly
 \beq
 \label{state1}
   {\cal T}^{4D}(\pmb \nu, \Sigma)|0 \rangle = | 2 \pi \delta_{Aa}  \nu_{m \;a}   A_i\rangle, ~~~ A_i (\vec{y}) =  \int_\Sigma ds\; n_i \delta^{(3)} (\vec{x}_\Sigma - \vec{y})
   \eeq
    where $A_i$ was shown to be the vector potential of a thin magnetic vortex, along $\partial \Sigma$ and of magnetic flux $\pmb \Phi = 2 \pi \pmb \nu_m$, see (\ref{flux}) and note that all  notation is the same as around Eq.~(\ref{transform}). In the presence of a nonzero $\theta$ angle, the creation of a magnetic flux is accompanied by the creation of electric flux: from (\ref{l4d}), the Hamiltonian is $H= \int d^3 x ( {g^2 \over 4} (\pmb\Pi_i - {\theta \over 4 \pi^2} \pmb B_i)^2 + {1 \over g^2} \pmb B_i^2)$ and the electric field is $\pmb E_i = - {g^2 \over 2}(\pmb \Pi_i -  {\theta \over 4 \pi^2} \pmb B_i)$. Thus, since $\pmb \Pi_i$ commutes with ${\cal T}^{4D}$, it is easily seen that the state (\ref{state1}) also carries electric flux along $\partial \Sigma$, proportional to $\theta$ times the magnetic flux, i.e.~the Witten effect.
    
     More explicitly, the electric flux carried by the state (\ref{state1}), measured in an arbitrary direction $\pmb \nu$, is\footnote{The factor of $g^2/2$ is due to our normalization of charge, see beginning of Section \ref{sigmadomain}.}
    \beq
    \label{magnflux}
    \pmb \nu \cdot \pmb \Phi_{\pmb E} \equiv \int\limits_S d^2 \vec{s} \; \vec{\pmb E} \cdot \pmb \nu= {g^2 \over 2} {\theta \over 4 \pi^2}\; \pmb \Phi\cdot \pmb \nu =  {g^2 \over 2} {\theta \over 2 \pi}\; \pmb \nu_m \cdot \pmb \nu ,
    \eeq 
   where $S$ is a small open surface intersecting the vortex $\partial \Sigma$ and we used (\ref{flux}). The flux (\ref{magnflux}) is the same as the electric flux of a state obtained by applying a  Wilson loop  along $\partial \Sigma$, with a $\theta$-dependent noninteger charge, i.e.  
    \beq
    \label{wilson2}
    e^{- i  {\theta \over 2 \pi} \; \pmb\nu_m \cdot \oint_{\partial\Sigma}  \pmb v_i d x^i} \vert 0 \rangle ,
    \eeq which is an eigenstate of electric flux with  
        \beq
      \pmb \nu \cdot \pmb \Phi_{\pmb E} = - {g^2\over 2} \int_S d^2 \vec{s} \; \pmb \nu \cdot \left[ \vec{\pmb \Pi}, -  i  {\theta \over 2 \pi} \; \pmb\nu_m \cdot \oint_{\partial\Sigma}  \pmb v_i d x^i  \right] = {g^2\over 2} {\theta \over 2 \pi}   \; \pmb\nu_m \cdot \pmb \nu. \eeq
      As the above discussion shows, changing the $\theta$ angle by $2\pi$ makes 't Hooft operators, which create magnetic flux, become Wilson-'t Hooft dyonic operators, which create both electric and magnetic fluxes. 
 Finally, note that the same reasoning applies to the operators (\ref{operators11}) in our long-distance theory on $\R^3 \times \S^1$. Consider for example the state created by $e^{- i \pmb w_1 \cdot \pmb \sigma}$ acting on the vacuum, an eigenstate of magnetic flux wrapped on $\S^1_L$. That this state, at nonzero $\theta$,   also carries electric flux along $\S^1_L$ follows from recalling (see the discussion  after  (\ref{l4d})) that $\pmb E_3$$=$$- \partial_t \pmb \phi/(2 \pi R)$ and, from (\ref{momenta}), that $- {2\over g^2} \pmb E_3 = \pmb \Pi_{\pmb\phi} - {\theta \over 2\pi} \pmb \Pi_{\pmb \sigma}$.

%%%%%%%%%%%%%%%%%%%%%%%%%%%%%%%%%%%%%
\section{QCD(adj) with $\mathbf{su(4)}$ algebra}
\label{su4appendix}
%%%%%%%%%%%%%%%%%%%%%%%%%%%%%%%%%%%%%

In this appendix, we consider QCD(adj) with an $su(4)$ algebra. Unlike $su(2)$ and $su(3)$ where the center groups have a prime number of elements, the center symmetry of $su(4)$ is $\mathbb Z_4$ with non-prime numbers of elements. Thus, one obtains theories with distinct global structures by modding the group $SU(4)$ either by $\mathbb Z_4$ or by its subgroup $\mathbb Z_2$.  
For  $SU(4)/\mathbb Z_4$,  the allowed dyonic probes can be classified into $4$ mutually non-local operators, $(1,H)$, $(1, W H)$, $(1,W^2H)$ and $(1, W^3H)$, while for  $SU(4)/\Z_2$ they are $(W^2, H^2)$ and $(W^2, WH^2)$.
 
 The fundamental  weight vectors are $\pmb w_1=\left(\frac{1}{\sqrt 2}, \frac{1}{\sqrt 6}, \frac{1}{2\sqrt 3}\right)$, $\pmb w_2=\left(0, \sqrt{\frac{2}{3}}, \frac{1}{\sqrt 3}\right) $, $\pmb w_3=\left(0, 0, \frac{\sqrt 3}{2}\right)$ and 
the global minima of the potential are located at
\begin{eqnarray}
\sigma_1^{0}&=&\frac{\pi}{4}\sqrt{2}(n_1+2n_2+3n_3)\, ,~
\sigma_2^0=\frac{\pi}{4}\sqrt{\frac{2}{3}}(3n_1+6n_2+n_3)\, ,~ \sigma_3^0=\frac{\pi}{2\sqrt{3}}(3n_1-n_3)\,.
\label{global minima in su4}
\end{eqnarray}
where $n_1,n_2,n_3\in \mathbb Z$. For $SU(4)$ group, there are $4$ global minima within the fundamental domain of $\pmb \sigma$ which is bounded by the weight vectors. The number of the global minima increases as we mod by the center symmetry $\mathbb Z_4$ and its subgroup $\mathbb Z_2$ as we show below. 

\begin{enumerate}

\item $SU(4)/{\mathbb Z_4}$ 

The domain of $\pmb \sigma$ is $\Gamma_w$. 
 Hence, there are $16$ vacua given by
\begin{eqnarray}
\nonumber
\
\{(n_1,n_2,n_3)&=&(0, 0, 0), (1, -2, 1), (1, -2, 2), (1, -2, 3), (1, -1, 1), (1, -1, 2), (2, -3, 2)\\
\nonumber
&&, (2, -3, 3), (2, -2, 1), (2, -2, 2), (2, -2, 3), (3, -4, 2), (3, -4, 3), (3, -4, 4)\\
 &&,(3, -3, 2), (3, -3, 3)\}.
\end{eqnarray}  
To reduce the notational clutter, these vacua will be numbered from $1$ to $16$ according their position in the above list. 

Not all the vacua are distinct: under the $\mathbb Z_4$ center symmetry identification, Eq.~(\ref{centerk2}) with $k'=1$, we have
\begin{eqnarray}
\nonumber
\sigma_1 &\rightarrow& -\frac{1}{2}\sigma_1-\frac{1}{2\sqrt{3}}\sigma_2-\sqrt{\frac{2}{3}}\sigma_3+2\pi\sum_{a=1}^3k_a w_a^{(1)}\,, \\
\sigma_2&\rightarrow& \frac{\sqrt{3}}{2}\sigma_1-\frac{1}{6}\sigma_2-\frac{\sqrt{2}}{3}\sigma_3+2\pi\sum_{a=1}^3k_a w_a^{(2)}\,,\\
\sigma_3 &\rightarrow& \frac{4}{3\sqrt{2}}\sigma_2-\frac{1}{3}\sigma_3+2\pi\sum_{a=1}^3k_a w_a^{(3)}\,,\nonumber
\label{the Z4  transformation}
\end{eqnarray}
where $k_a$ are integers. 

For $k_1=k_2=k_3=0$ (a $2 \pi \pmb w_1$ shift on the r.h.s.~of (\ref{centerk2})) we obtain the vacua identification
\begin{eqnarray}
\nonumber
&&4 \leftrightarrow 7\,,\quad5 \leftrightarrow 16\,,\quad1 \leftrightarrow 1\,,\quad 2 \leftrightarrow 2\,,\quad 10 \leftrightarrow 10\,,\quad13 \leftrightarrow 13\,,\\
&&3 \leftrightarrow 11 \leftrightarrow 14 \leftrightarrow 15\,, \quad 6\leftrightarrow 12 \leftrightarrow 9 \leftrightarrow 8\,.
\label{1 theory su4}
\end{eqnarray}
For $k_1=1,k_2=k_3=0$ (a $2 \pi \pmb w_2$ shift on the r.h.s.~of (\ref{centerk2}))we obtain
\begin{eqnarray}
\nonumber
&&11 \leftrightarrow 15\,,\quad 3 \leftrightarrow 14\,,\quad 6 \leftrightarrow 6\,,\quad 8 \leftrightarrow 8\,,\quad 9 \leftrightarrow 9\,,\quad 12 \leftrightarrow 12\,,\\
&&1 \leftrightarrow 2 \leftrightarrow 10 \leftrightarrow 13 \,,\quad4 \leftrightarrow 16 \leftrightarrow 7 \leftrightarrow 5\,.
\label{2 theory su4}
\end{eqnarray}
For $k_1=0,k_2=1,k_3=0$ (a $2 \pi \pmb w_3$ shift on the r.h.s.~of (\ref{centerk2}))we obtain
\begin{eqnarray}
\nonumber
&&1 \leftrightarrow 10\,, \quad
2 \leftrightarrow 13\,, \quad
4 \leftrightarrow 4\,, \quad
5 \leftrightarrow 5\,,\quad
7 \leftrightarrow 7\,,\quad
16 \leftrightarrow 16\,,\\
&&3 \leftrightarrow 15 \leftrightarrow 14 \leftrightarrow 11 \,,\quad 6 \leftrightarrow 8 \leftrightarrow 9 \leftrightarrow 12\,.
\label{3 theory su4}
\end{eqnarray}
For $k_1=k_2=0,k_3=1$ we obtain
\begin{eqnarray}
\nonumber
&&6 \leftrightarrow 9\,,\quad
8 \leftrightarrow 12\,,\quad 2 \leftrightarrow 2\,,
\quad 3 \leftrightarrow 3\,,\quad 14 \leftrightarrow 14\,, \quad
15 \leftrightarrow 15\,,\\
&&1 \leftrightarrow 13 \leftrightarrow 10 \leftrightarrow 2 \,,
\quad 4 \leftrightarrow 5 \leftrightarrow 7 \leftrightarrow 16\,.
\label{4 theory su4}
\end{eqnarray}
These $4$ different choices correspond to $\left[SU(4)/Z_4\right]_{0,1,2,3}$. Each of these theories have $8$ distinct vacua as shown above, in agreement with  the Witten index result  $\sum_{k=1}^{N_c} {\rm gcd}(N,k)$ for this case from \cite{Aharony:2013hda}.

 \item $SU(4)/\mathbb Z_2$

In this case, the $\mathbb Z_2$ center symmetry acts on $\pmb \sigma$ as Eq.~(\ref{centerk2}) with $k'=2$ (the $\mathbb Z_2$ transformation below is obtained from (\ref{the Z4  transformation}) by applying the permutation operation  in (\ref{the Z4  transformation}) twice):
\begin{eqnarray}
\left[\begin{array}{c}\sigma_1\\\sigma_2\\\sigma_3 \end{array}\right]\rightarrow A \left[\begin{array}{c}\sigma_1\\\sigma_2\\\sigma_3 \end{array}\right]+2\pi\sum_{a=1}^3  k_a\pmb\omega_a^T\,,\quad A= \left[\begin{array}{ccc} 0&\frac{-1}{\sqrt 3} & \sqrt{\frac{2}{3}}\\-\frac{1}{\sqrt 3} &-\frac{2}{3}&-\frac{\sqrt 2}{3}\\ \sqrt{\frac{2}{3}}&-\frac{\sqrt 2}{3}&-\frac{1}{3} \end{array} \right]\,.
\end{eqnarray}
By modding by $\mathbb Z_2$ we obtain a coarser lattice compared to the weight lattice. This lattice, called the group lattice, is still finer than the root lattice. The fundamental domain of $\pmb \sigma$ is the group lattice of $SU(4)/\Z_2$. The group lattice is generated by the following  vectors $\pmb \kappa_1, \pmb \kappa_2, \pmb \kappa_3$:
\begin{eqnarray}
\nonumber
\pmb\kappa_1&\equiv&\pmb\omega_2=\left(0,\sqrt{\frac{2}{3}},\frac{1}{\sqrt{3}}\right)\,,\quad \pmb\kappa_2\equiv\pmb\omega_1+\pmb\omega_3=\left(\frac{1}{\sqrt 2}, \frac{1}{\sqrt 6}, \frac{2}{\sqrt 3} \right)\,,\\
\pmb\kappa_3&\equiv&\pmb\omega_1-\pmb\omega_3=\left(\frac{1}{\sqrt{2}},\frac{1}{\sqrt 6},-\frac{1}{\sqrt 3}\right)\,.
\end{eqnarray}

 To make the analysis easier, we define the reciprocal vectors $\{\pmb{\cal C}_i\}$ such that 
 $\pmb{\cal C}_i\cdot \pmb\kappa_j=\delta_{ij}$, from which we can solve for $\pmb{\cal C}_{1,2,3}$:
\begin{eqnarray}
\pmb{\cal C}_1=\left(-\frac{1}{\sqrt 2},\sqrt{\frac{3}{2}},0\right)\,,\quad
\pmb{\cal C}_2=\left(\frac{1}{\sqrt 2},-\frac{1}{\sqrt 6},\frac{1}{\sqrt 3}\right)\,,\quad
\pmb{\cal C}_3=\left(\frac{1}{\sqrt 2},\frac{1}{\sqrt 6},-\frac{1}{\sqrt 3}\right)\,.
\end{eqnarray}
In order to further simplify our analysis, we define the new coordinates $\tilde{\pmb\sigma}$:
\begin{eqnarray}
\tilde {\pmb\sigma}_1={\cal C}_1\cdot \pmb \sigma\,,\quad \tilde {\pmb\sigma}_2={\cal C}_2\cdot \pmb \sigma\,,\quad\tilde {\pmb\sigma}_1={\cal C}_1\cdot \pmb \sigma\,.
\end{eqnarray}
Thus, we can write the following linear transformation between $\pmb \sigma$ and $\tilde{\pmb \sigma}$:
\begin{eqnarray}
\left[\begin{array}{c} \tilde\sigma_1\\\tilde\sigma_2\\\tilde\sigma_3 \end{array}\right]=T\left[\begin{array}{c} \sigma_1\\\sigma_2\\\sigma_3 \end{array}\right]\,,\quad T=\left[\begin{array}{ccc}-\frac{1}{\sqrt 2} &\sqrt{\frac{3}{2}} &0\\\frac{1}{\sqrt 2} &-\frac{1}{\sqrt 6} & \frac{1}{\sqrt 3}\\ \frac{1}{\sqrt{2}}& \frac{1}{\sqrt 6} &-\frac{1}{\sqrt 3} \end{array}\right]\,.
\label{sigma coordinates transformation}
\end{eqnarray}
These new coordinates rectify the fundamental domain of the group lattice such that this lattice is bounded by the unit vectors $(1,0,0),(0,1,0),(0,0,1)$. In terms of the new coordinates, we find that the the $\mathbb Z_2$ center symmetry acts as
\begin{eqnarray}
\left[\begin{array}{c}\tilde\sigma_1\\\tilde \sigma_2\\\tilde\sigma_3 \end{array}\right]\rightarrow TAT^{-1} \left[\begin{array}{c}\tilde\sigma_1\\\tilde\sigma_2\\\tilde\sigma_3 \end{array}\right]+2\pi\sum_{a=1}^3  k_a T\pmb\omega_a^T\,,\quad TAT^{-1}= \left[\begin{array}{ccc} -1&-2 & 0\\0 &1&0\\ 0&0&-1 \end{array} \right]\,,
\end{eqnarray}
where
\begin{eqnarray}
T\pmb\omega_1^T=\left(0,\frac{1}{2},\frac{1}{2}\right)^T\,,\quad
T\pmb\omega_2^T=\left(1,0,0\right)^T\,,\quad
T\pmb\omega_3^T=\left(0,\frac{1}{2},-\frac{1}{2}\right)^T\,.
\end{eqnarray}

Next,  we express the global  minima of the potential ${\cal W}$ as given by  (\ref{global minima in su4})  in terms of the new coordinates $\tilde{\pmb \sigma}$
\begin{eqnarray}
\tilde \sigma_1^0=\frac{\pi}{2}(n_1+2n_2-n_3)\,,\quad
\tilde \sigma_2^0=\frac{\pi}{2}(n_1+n_3)\,,\quad
\tilde \sigma_3^0=\pi (n_2+n_3)\,.
\end{eqnarray}
The minima in the fundamental domain of $\tilde{\pmb \sigma}$ are give by  
\begin{eqnarray}
\nonumber
(n_1,n_2,n_3)=\{(0,0,0),(0,1,0),(1,0,0),(1,0,1),(1,1,0),(2,0,0),(2,0,1),(3,0,0)\}\,,\\
\end{eqnarray}
which is half the number of the minima in the case of $SU(4)/\mathbb Z_4$. We will label these vacua with numbers from $1$ to $8$.
Under this center identification we have the following theories:

For $k_1=k_2=k_3=0$
\begin{eqnarray}
\nonumber
&&1 \leftrightarrow 1\,,\quad 2\leftrightarrow 2\,,\quad 3 \leftrightarrow 3\,,\quad 4\leftrightarrow 4\\
&&5 \leftrightarrow 5\,,\quad 6 \leftrightarrow 6\,,\quad7 \leftrightarrow 7\,,\quad 8 \leftrightarrow 8\,.
\label{su4 z3 1}
\end{eqnarray}
Therefore, the center symmetry transformation acts trivially on the minima and we end up having $8$ distinct vacua in this theory. This theory is $\left[SU(4)/\mathbb Z_2\right]_0$.

For $k_1=1,k_2=k_3=0$
\begin{eqnarray}
1 \leftrightarrow 4\,,\quad 2\leftrightarrow 6\,,\quad
3 \leftrightarrow 7\,,\quad
5\leftrightarrow 8\,.
\label{su4 z3 2}
\end{eqnarray}
Thus, in this theory we have only four vacuua. This vacua identification corresponds to the $\left[SU(4)/\mathbb Z_2\right]_1$  theory. One can also check that all other values of $k_a$ do not give new theories.

\end{enumerate}

 \bibliography{Global_Structure_JHEP}

\bibliographystyle{JHEP}

\end{document}